\documentclass[aps,prb,twocolumn,showpacs,eqsecnum]{revtex4}
\usepackage{makeidx}
\usepackage{amsmath}
\usepackage{amsfonts}
\usepackage{amssymb}
\usepackage{graphicx}
\usepackage{graphics}
\usepackage{color}
\usepackage{colordvi}
\usepackage{dcolumn}
\usepackage{bm}
\usepackage{hyperref}
\usepackage{simplewick}

\begin{document}

\title{Describing systems of interacting fermions by boson models:\\
exact mapping in arbitrary dimension and applications}
\author{K. B. Efetov${}^{1}$}
\author{C. P\'{e}pin${}^{2}$}
\author{H. Meier${}^{1}$}
\affiliation{$^{1}$Theoretische Physik III, Ruhr-Universit\"{a}t Bochum, 44780 Bochum,
Germany\\
$^{2}$IPhT, CEA-Saclay, L'Orme des Merisiers, 91191 Gif-sur-Yvette, France }
\date{\today }

\begin{abstract}
We develop a new method that allows us to map models of interacting fermions
onto bosonic models describing collective excitations in an arbitrary
dimension. This mapping becomes exact in the thermodynamic limit in the
presence of a bath. The boson models can be written either in the form of a
model of non-interacting bosons in a fluctuating auxiliary field or in the
form of a superfield theory of interacting bosons. We show how one can study
the latter version using perturbation theory. Using the developed
diagrammatic technique we compared the first two orders of perturbation
theory with the corresponding results for the original fermion model and
found a perfect agreement. As concerns the former representation, we suggest
a scheme that may be suitable for Monte Carlo simulations and demonstrate
that it is free of the fermionic sign problem. We discuss in details the
properties of the bosonic representation and argue that there should not be
any obstacles preventing from an efficient computation.
\end{abstract}

\pacs{71.10.Ay, 71.10.Pm, 75.40.Cx}
\maketitle

\section{Introduction}

Many body electron systems with interaction traditionally attract a strong
attention. Numerous interesting phenomena of the condensed matter physics
originate from the electron-electron, electron-phonon and other types of the
interaction and do not exist in the ideal Fermi gas of the electrons.

Generally, models with interactions cannot be solved exactly in the space of
dimensionality $d>1$ and one has to use approximate schemes of calculation.
The most developed method is the perturbation theory with respect to the
interaction. Within this theory one starts with the ideal Fermi gas and
calculates corrections to physical quantities assuming that the interaction
is weak. Sometimes it is sufficient to compute only the first several orders
of this perturbation theory, but very often in order to capture important
physics, one has to sum certain series.

The formalism of the perturbation theory for the interacting electron gas is
well developed (see, e.g., Ref.~\onlinecite{agd}). Each term of the
perturbation theory is represented diagrammatically, which enables one to
carry out quite complicated calculations. Many interesting phenomena like,
e.g., superconductivity are not seen in the lowest orders of the
diagrammatic expansion and show up after summation of certain ladder
diagrams. Consideration of chain diagrams is necessary in the microscopic
theory of the Landau Fermi liquid and there are many examples of this type.

At the same time, such sequences of diagrams correspond to physically
well-defined collective bosonic excitations. In the Fermi liquid theory, the
chain diagrams describe particle density and spin excitations that can be
spoken of as bosonic quasiparticles. From this and many other examples,
where a well-defined physical quasiparticle is represented by an infinite
sequence of diagrams, one can conclude that the conventional diagrammatic
technique is not always the most convenient technique for the description of
interacting many body systems.

The inconveniences of the conventional perturbative methods are especially
evident in situations when the interaction between the quasiparticles is
important. This happens in one-dimensional systems, $d=1,$ but models in
higher dimensions, $d>1$, used for describing high $T_{c}$ superconducting
cuprates\cite{lee}, heavy fermions\cite{norman}, etc., can hardly be
successfully studied by conventional diagrammatic approaches either. These
difficulties call for constructing different approaches dealing with
collective excitations rather than with single particles.

Such an approach called \textit{bosonization }is well known for one
dimensional ($1D$) systems. A huge number of publications are devoted to it
(see, for a review, Refs.~\onlinecite{tsvelik,giamarchi}). Attempts to
bosonize higher dimensional fermionic systems have been undertaken in the
past starting from the work by Luther \cite{luther} where the Fermi surface
of a special form (square or cubic) was considered. This idea was further
developed by Haldane\cite{haldane} who suggested to patch the Fermi surface
by a large number of segments and continued in a number of publications\cite%
{houghton1,neto,kopietz,kopietzb,khveshchenko,khveshchenko1,castellani}. All
these bosonization schemes are based on the assumption that only small
momenta are transferred by the electron-electron interaction such that,
after any scattering event, the electron remains in the same segment of the
Fermi surface. This case is, of course, relevant to a long range interaction
but a generalization to an arbitrary interaction and arbitrary momentum
transfer is hardly possible within this scheme. At the end one can reproduce
results of the random phase approximation (RPA) that can be interpreted in
terms of quasiparticles but the interaction between these quasiparticles
cannot be taken into account properly.

Another method based on quasiclassical equations has been developed recently
for an arbitrary interaction\cite{aleiner}. Using this approach anomalous
contributions to the specific heat and magnetic spin susceptibility\cite%
{schwiete} have been calculated in all dimensions and new logarithmic
contributions were found. In $1D$, the results obtained are in agreement
with those known from renormalization group considerations\cite{dl} and
exact solutions for spin chains\cite{affleck,lukyanov}. They have also been
confirmed in $1D$ by the conventional diagrammatic expansion\cite{cms} where
the first non-vanishing logarithmic contribution was calculated. However,
the same diagrammatic method did not exactly reproduce \cite{chubukov} the
logarithmic contributions to $\delta C/T^{d}$ for $d>1$, where $\delta C$ is
the anomalous correction to the specific heat. This discrepancy has been
attributed to the observation that not all the effects of Fermi surface
curvature have been taken into account in Ref.~\onlinecite{aleiner}.

One can conclude from this discussion that the problem of the bosonization
in $d>1$ has not been completely solved yet but such a method is very
desirable because it would be a new tool for analytical calculations.
Remarkably, fermionic models are shown to be problematic for numerical
computations as well. The powerful Monte Carlo (MC) method suffers from the
famous fermionic sign problem\cite%
{blankenbecler,hirsch,linden,dosantos,troyer}. The MC method implies that
the partition function can be represented as a sum of terms with positive
probabilities and in this case, the computation time is proportional to a
power of the size of the system. This makes the MC method very robust
compared to other approaches. However, the fermion determinant is negative
for some paths and, actually, its average sign becomes zero in the sampling
process. As a result, the computation time grows exponentially with the size
of the system or inverse temperature and the advantages of MC get lost.

We are not able to list here all publications where the sign problem was
discussed but (except special cases like electron systems with attraction,
systems with repulsion but with half filling and some others) the solution
has not been found. Moreover, according to Ref.~\onlinecite{troyer}, the
sign problem should be NP-hard \cite{cook}, which means that its resolution
is almost impossible. We are not aware about any mathematically rigorous
proof of the NP-hardness of the fermionic sign problem but it is anyway
clear that it is the main problem for MC simulations.

This problem does not arise in (non-frustrated) bosonic systems, though.
Therefore, it is quite natural to think about a possibility of circumventing
the sign problem by mapping the initial fermionic model onto a bosonic one.
Of course, it is impossible to convert real fermions into real bosons but
one can think of recasting the interacting part of the fermionic action in
the language of collective bosonic excitations. This leads us again to the
idea of the bosonization.

In this paper, we suggest a new bosonization scheme that allows one to map
in the fermionic model to a model describing bosonic excitations \textit{%
exactly }in the sense that\textit{\ }the mapping works in any dimension at
any temperature and for arbitrary interactions. It can be written in a form
of a model of non-interacting bosons in an effective Hubbard-Stratonovich
(HS) field. The transformation from the fermionic system to the bosonic one
corresponds to using in quantum mechanics the density matrix instead of the
wave functions. We argue that the representation of the model in terms of
bosons in the external field should be free of the negative sign problem and
can be used for MC simulations.

It should be emphasized, though, that for finite systems used in the MC
simulations the new bosonized model is not exactly equivalent to the initial
fermionic one. In order to derive the boson model, we need a certain
regularization and a bath attached to the system. This is necessary to avoid
some uncertainties in the bosonized expressions. Only after the
regularization is carried out we obtain a bosonic model free of the sign
problem. Completely exact mapping of the fermionic model onto the bosonic
one is impossible because, technically, several electron Green functions
with coinciding Matsubara frequencies and momenta cannot be converted into
bosonic excitations. If we did only identical transformations we would not
be able to get rid of the sign problem. At the same time, one can see from
the derivation that considering a sufficiently large system with the bath
any desired precision can be achieved.

Following an alternative route one can average over the HS field before
making any approximation. This goal is achieved using the famous BRST
(Becci-Rouet-Stora-Tyutin) transformation based on the introduction of
superfields\cite{brst}. This allows us to represent the partition function
in such a form that the average over the HS field can be done immediately.
After the averaging one comes to a model of bosons with quadratic, cubic and
quartic interactions. We show how this model can be studied with the help of
perturbation theory in the interaction terms and compare the first orders
with the corresponding terms of the conventional perturbation theory
demonstrating a full agreement.

The paper is organized as follows:

In Section \ref{model}, we formulate the model of interacting fermions,
decouple the interaction with the help of the HS transformation and map the
model onto the model of bosons in the external field. We introduce a
regularization that allows us to avoid singularities in the equation of
motion.

In Section \ref{pt}, we introduce superfields, average over the HS field and
derive the model of interacting bosons. After that we develop a diagrammatic
technique for the superfield theory describing the bosonic excitations and
demonstrate in the first orders in the interaction its agreement with the
conventional perturbation theory.

Section \ref{mc} is devoted to the discussion of the bosonization from the
point of view of MC simulations. We discuss in detail how the negative sign
contributions originate in the fermionic formulation and why they do not
exist in the bosonic regularized model. Properties of the partition
functions are considered in the complex plane of the electron-electron
interaction, which helps to understand in details what happens in the
procedure of the bosonization. We derive formulas directly suitable for
numerical simulations and give an example of numerical convergence in a case
of static fields.

We discuss the obtained results in Section \ref{dis}.

The main results of this paper have already been presented in a shorter
publication \cite{epm}.

\section{Fermionic model and bosonization\label{model}}

\subsection{Reduction to fermions in the auxiliary field}

Our method of the bosonization is quite general and is applicable to a broad
class of models. Therefore we start our considerations with a general model
of interacting fermions on a lattice of an arbitrary dimension. We write the
Hamiltonian $\hat{H}$ of the system as follows
\begin{equation}
\hat{H}=\hat{H}_{0}+\hat{H}_{int},  \label{a1}
\end{equation}%
where $\hat{H}_{0}$ is the bare part describing non-interacting electrons,
\begin{equation}
\hat{H}_{0}=-\sum_{r,r^{\prime },\sigma ,\sigma ^{\prime }}(t_{r,r^{\prime
}}+\mu \delta _{r,r^{\prime }})c_{r\sigma }^{+}c_{r^{\prime }\sigma }
\label{a2}
\end{equation}%
with a chemical potential $\mu $, a hopping amplitude $t_{r,r^{\prime }}$
between the sites $r$ and $r^{\prime }$ and the spin $\sigma =+1,-1$.

The term $\hat{H}_{int}$ describes the electron-electron interaction,
\begin{equation}
\hat{H}_{int}=\frac{1}{2}\sum_{r,r^{\prime },\sigma ,\sigma ^{\prime
}}V_{r,r^{\prime }}c_{r\sigma }^{+}c_{r^{\prime }\sigma ^{\prime
}}^{+}c_{r\sigma ^{\prime }}c_{r\sigma }.  \label{a3}
\end{equation}

The partition function $Z$ can obtained from $\hat{H}$ the Hamiltonian $\hat{%
H}$ using the standard relation%
\begin{equation}
Z=\mathrm{Tr \exp }\left( -\beta \hat{H}\right) ,\quad \beta =1/T,
\label{a4}
\end{equation}%
where $T$ is temperature.

In order to proceed with the mapping of the fermion model onto a boson on,
we will decouple the interaction term $\hat{H}_{int}$ by integration over an
auxiliary HS field. In principle, the decoupling can be performed in
different ways and we should choose one of them. The choice can be different
depending on the problem studied. In this work we are mostly interested in
systems with an on-site repulsion. We also keep in mind a possibility of
using the method for MC computations. The latter desire enforces us to use
real HS fields $\phi _{r}$ depending only on one coordinate $r$.

At the same time, it would be advantageous to keep an arbitrary sign of the
off-diagonal interaction $V_{r,r^{\prime }\text{ }}.$ The on-site attraction
is less interesting for us because MC schemes can be free of sign in this
case \cite{dosantos} anyway. The bosonization procedure can be carried out
also in this case with minimal changes but we do not do this here.

In order to carry out the HS transformation with the real fields $\phi _{r}$
we split the matrix~$V_{r,r^{\prime }}$ into its diagonal $V_{r,r}$ and
off-diagonal $\tilde{V}_{r,r^{\prime }}$ parts,
\begin{equation}
V_{r,r^{\prime }}=\tilde{V}_{r,r^{\prime }}+V_{r,r}\delta _{r,r^{\prime
}},\quad \tilde{V}_{r,r}=0 .  \label{a4a}
\end{equation}

Assuming that~$\tilde{V}_{r,r^{\prime }}$ is bounded from above, we can add
a sufficiently large constant~$\bar{V}$ to~$-\tilde{V}_{r,r^{\prime }}$,
ensuring the matrix~%
\begin{equation}
V_{r,r^{\prime }}^{\left( 1\right) }=-\tilde{V}_{r,r^{\prime }}+\bar{V}%
\delta _{r,r^{\prime }}  \label{a4b}
\end{equation}
to be positive definite. (Its Fourier transform $V^{\left( 1\right)}_\mathbf{%
q}$ should be positive for all momenta $\mathbf{q}$.) Introducing the
coupling constant
\begin{equation}
V_{0}=V_{r,r}+\bar{V} \ ,  \label{a4c}
\end{equation}
assuming it is site independent, and using the fermionic commutation
relations we rewrite the interaction term $\hat{H}_{int}$ as%
\begin{eqnarray}
\hat{H}_{int} &=&\hat{H}_{int}^{\left( 0\right) }+\hat{H}_{int}^{\left(
1\right) },  \label{a5} \\
\hat{H}_{int}^{\left( 0\right) } &=&-\frac{V_{0}}{2}\sum_{r}\left(
c_{r+}^{+}c_{r+}-c_{r-}^{+}c_{r-}\right) ^{2}  \notag \\
\hat{H}_{int}^{\left( 1\right) } &=&-\frac{1}{2}\sum_{r,r^{\prime },\sigma
,\sigma ^{\prime }}V_{r,r^{\prime }}^{\left( 1\right) }c_{r\sigma
}^{+}c_{r\sigma }c_{r^{\prime }\sigma ^{\prime }}^{+}c_{r^{\prime }\sigma
^{\prime }}  \notag
\end{eqnarray}%
while the chemical potential $\mu $ should be replaced by $\tilde{\mu}=\mu
-\left( V_{0}+\bar{V}\right) /2$.

Due a somewhat arbitrary choice of the constant $\bar{V}$ the coupling
constants $V_{0}$ and $V^{\left( 1\right) }$ are also not uniquely defined.
However, this does not create any problems.

As the interaction term $\hat{H}_{int}$ does not commute with $\hat{H}_{0}$
the HS transformation can be performed first subdividing the interval $%
\left( 0,\beta \right) $ into segments of the length $\Delta =\beta /N$ with
$N\gg 1$ and then integrating over the auxiliary field for each $\tau
_{l}=\Delta \left( l-1/2\right) ,$ $l=1,2...N$.

In this Section we are interested in analytical calculations and therefore
we write all formulas in the continuous limit ($\Delta \rightarrow 0$) with
respect to the \textquotedblleft imaginary time" $\tau $. Transformations
for the discrete time (finite $\Delta $) will be performed in Section \ref%
{mc}$.$

We rewrite the partition function $Z$, Eq.~(\ref{a4}), as%
\begin{equation}
Z=\mathrm{Tr}_{\bar c, c}\Big[\exp\left( -\beta \hat{H}_{0}\right) T_{\tau
}\exp \Big(-\int_{0}^{\beta }\hat{H}_{int}\left( \tau \right) d\tau \Big)%
\Big]  \label{a8}
\end{equation}%
where $T_{\tau }$ is the time ordering operator (the time grows from the
right to the left) and $\hat{H}_{int}\left( \tau \right) $ is the operator
in the interaction representation,%
\begin{equation}
\hat{H}_{int}\left( \tau \right) =\exp \left( \hat{H}_{0}\tau \right) \hat{H}%
_{int}\exp \left( -\hat{H}_{0}\tau \right) .  \label{a9}
\end{equation}%
The term $\hat{H}_{int}^{\left( 0\right) }$ can be decoupled as follows%
\begin{eqnarray}
&&T_{\tau }\exp \Big(-\int_{0}^{\beta }\hat{H}_{int}^{\left( 0\right)
}\left( \tau \right) d\tau \Big)  \label{a11} \\
&=&\int T_{\tau }\exp \Big(\sum_{r,\sigma }\int_{0}^{\beta }\sigma \phi
_{r}^{\left( 0\right) }\left( \tau \right) \bar{c}_{r\sigma }\left( \tau
\right) c_{r\sigma }\left( \tau \right) d\tau \Big)  \notag \\
&&\times W_{0}\big[ \phi ^{\left( 0\right) }\big] D\phi ^{\left( 0\right) }%
\Big(\int_{0}^{\beta }W_{0}\big[ \phi ^{\left( 0\right) }\big] D\phi
^{\left( 0\right) }\Big)^{-1},  \notag
\end{eqnarray}%
\begin{equation}
W_{0}\big[ \phi ^{\left( 0\right) }\big] =\exp \Big(-\frac{1}{2V_{0}}%
\sum_{r}\int_{0}^{\beta }\left( \phi _{r}^{\left( 0\right) }\left( \tau
\right) \right) ^{2}d\tau \Big)  \label{a12}
\end{equation}%
where $c\left( \tau \right) $ and $\bar{c}\left( \tau \right) $ are creation
and annihilation operators in the interaction representation [introduced
analogously to Eq.~(\ref{a9})], $\phi _{r}^{\left( 0\right) }\left( \tau
\right) $ is a real field with the bosonic periodicity, $\phi _{r}^{\left(
0\right) }\left( \tau \right) =\phi _{r}^{\left( 0\right) }\left( \tau
+\beta \right) $.

Before decoupling the term $\hat{H}_{int}^{\left( 1\right) },$ Eq.~(\ref{a5}%
), we rewrite it in the form
\begin{eqnarray}
\hat{H}_{int}^{\left( 1\right) } &=&-\frac{1}{2}\sum_{r,r^{\prime },\sigma
,\sigma ^{\prime }}V_{r,r^{\prime }}^{\left( 1\right) }\left( c_{r\sigma
}^{+}c_{r\sigma }-n\right) \left( c_{r^{\prime }\sigma ^{\prime
}}^{+}c_{r^{\prime }\sigma ^{\prime }}-n\right)  \notag \\
&&-2n\sum_{r}V^{\left( 1\right) }c_{r\sigma }^{+}c_{r\sigma }+\frac{N_{d}}{2}%
V^{\left( 1\right) }(2n)^{2},  \label{a13}
\end{eqnarray}%
where $V^{\left( 1\right) }=\sum_{r^{\prime }}V_{r,r^{\prime }}^{\left(
1\right) }$, $n=\left\langle c_{r\sigma }^{+}c_{r\sigma }\right\rangle $ is
the fermion density per spin direction, and $N_{d}$ is the number of sites
in the system.

The second term in Eq.~(\ref{a13}) renormalizes once more the chemical
potential, such that now it equals%
\begin{equation}
\mu ^{\prime }=\mu -\frac{1}{2}\left( V_{0}+\bar{V}\right) +2nV^{\left(
1\right) }.  \label{a14}
\end{equation}%
The last term in Eq.~(\ref{a13}), $(2n)^{2}V^{\left( 1\right) }N_{d}/2$, is
a trivial contribution to the thermodynamic potential~$\Omega $.

The fermion density $n$ should be calculated using the shifted chemical
potential $\mu ^{\prime }$, Eq.~(\ref{a14}), and, thus solving a
self-consistency equation. We assume that this has been done and consider
the first term in $\hat{H}_{int}^{\left( 1\right) }$ in Eq.~(\ref{a13})
decoupling it by integration over another field $\phi _{r}^{\left( 1\right)
}\left( \tau \right) $,%
\begin{eqnarray}
&&T_{\tau }\exp \Big(-\int_{0}^{\beta }\hat{H}_{int}^{\left( 1\right)
}\left( \tau \right) d\tau \Big)  \label{a15} \\
&=&\int T_{\tau }\exp \Big(\sum_{r,\sigma }\int_{0}^{\beta }\phi
_{r}^{\left( 1\right) }\left( \tau \right) \left( \bar{c}_{r\sigma }\left(
\tau \right) c_{r\sigma }\left( \tau \right) -n\right) d\tau \Big)  \notag \\
&&\times W_{1}\big[\phi ^{\left( 1\right) }\big]D\phi ^{\left( 1\right) }%
\Big[\int_{0}^{\beta }W_{1}\big[\phi ^{\left( 1\right) }\big]D\phi ^{\left(
1\right) }\Big]^{-1}  \notag
\end{eqnarray}%
where%
\begin{equation}
W_{1}\big[\phi ^{\left( 1\right) }\big]=\exp \Big(-\frac{1}{2}%
\sum_{r,r^{\prime }}\int_{0}^{\beta }\phi _{r}^{\left( 1\right) }\left( \tau
\right) \left( V^{\left( 1\right) }\right) _{r,r^{\prime }}^{-1}\phi
_{r^{\prime }}^{\left( 1\right) }\left( \tau \right) d\tau \Big)  \label{a16}
\end{equation}%
and $\phi _{r}^{\left( 1\right) }\left( \tau \right) $ is also a real
periodic field, $\phi _{r}^{\left( 1\right) }\left( \tau \right) =\phi
_{r}^{\left( 1\right) }\left( \tau +\beta \right) $. The matrix $\left(
V^{\left( 1\right) }\right) _{r,r^{\prime }}^{-1}$ is a matrix inverse to
the matrix $V_{r,r^{\prime }}^{\left( 1\right) }$. The integral over $\phi
_{r}^{\left( 1\right) }\left( \tau \right) $ in Eqs.~(\ref{a15}, \ref{a16})
converges because $\left( V^{\left( 1\right) }\right) _{r,r^{\prime }}^{-1}$
is positive definite.

Introducing the field
\begin{equation}
\phi _{r\sigma }\left( \tau \right) =\phi _{r}^{\left( 0\right) }\left( \tau
\right) +\sigma \phi _{r}^{\left( 1\right) }\left( \tau \right)  \label{a17}
\end{equation}%
we write finally the partition function $Z$, Eq.~(\ref{a8}), in the form%
\begin{equation}
Z=\frac{\int Z\left[ \phi \right] W\left[ \phi \right] D\phi }{\int W\left[
\phi \right] D\phi },\quad W\left[ \phi \right] =W_{0}\big[\phi ^{\left(
0\right) }\big]W_{1}\big[\phi ^{\left( 1\right) }\big]  \label{a10}
\end{equation}%
with $Z\left[ \phi \right] $ equal to%
\begin{eqnarray}
&&Z\left[ \phi \right] =\mathrm{Tr}_{\bar{c},c}\Big[\mathrm{\exp }\left(
-\beta \hat{H}_{0}\right) T_{\tau }\exp \Big(-\int_{0}^{\beta }\hat{H}%
_{int}^{\left( \phi \right) }\left( \tau \right) d\tau \Big)\Big],  \notag \\
&&\hat{H}_{int}^{\left( \phi \right) }\left( \tau \right) =\sum_{r,\sigma
}\sigma \phi _{r\sigma }\left( \tau \right) \left[ \bar{c}_{r\sigma }\left(
\tau \right) c_{r\sigma }\left( \tau \right) -n\right] .  \label{a18}
\end{eqnarray}%
and the Hamiltonian $\hat{H}_{0}$ now given by
\begin{equation}
\hat{H}_{0}=\sum_{r,r^{\prime },\sigma ,\sigma ^{\prime }}(-t_{r,r^{\prime
}}-\delta _{r,r^{\prime }}\mu ^{\prime })c_{r\sigma }^{+}c_{r^{\prime
}\sigma },  \label{a18c}
\end{equation}%
with the modified chemical potential $\mu ^{\prime }$ from Eq.~(\ref{a14}).

We calculate the trace over the fermionic operators $c,c^{+}$ introducing an
additional variable $0\leq u\leq 1$, replacing $\phi _{r\sigma }\left( \tau
\right) $ by $u\phi _{r\sigma }\left( \tau \right) $ in Eq.~(\ref{a18}) and
writing $Z\left[ \phi \right] $ in the form%
\begin{eqnarray}
Z\left[ \phi \right] &=&Z_{0}\exp \Big[\sum_{r,\sigma }\int_{0}^{\beta
}\int_{0}^{1}\sigma \phi _{r\sigma }\left( \tau \right)  \label{a19} \\
&&\times \left( G_{r,r,\sigma }^{\left( u\phi \right) }\left( \tau ,\tau
+0\right) -G_{r,r,\sigma }^{\left( 0\right) }\left( \tau ,\tau +0\right)
\right) dud\tau \Big]  \notag
\end{eqnarray}%
In Eq.~(\ref{a19}), $Z_{0}$ is the partition function of the ideal Fermi
gas,
\begin{equation}
Z_{0}=\prod_{k,\sigma }\left( 1+\exp \left( -\frac{\epsilon _{k}}{T}\right)
\right)  \label{a19a}
\end{equation}%
where $\epsilon _{k}$ are the eigenvalues of the Hamiltonian $\hat{H}_{0}$,
Eq.~(\ref{a18c}), and
\begin{equation}
G_{r,r^{\prime },\sigma }^{\left( u\phi \right) }\left( \tau ,\tau ^{\prime
}\right) =-\frac{\left\langle T_{\tau }c_{r\sigma }\left( \tau \right) \bar{c%
}_{r^{\prime }\sigma }\left( \tau ^{\prime }\right) \exp \left(
-\int_{0}^{\beta }\hat{H}_{int}^{\left( u\phi \right) }d\tau \right)
\right\rangle _{0}}{\left\langle T_{\tau }\exp \left( -\int_{0}^{\beta }\hat{%
H}_{int}^{\left( u\phi \right) }d\tau \right) \right\rangle _{0}}
\label{a20}
\end{equation}%
is the electron Green function in the field $\phi _{r\sigma }\left( \tau
\right) $. In Eq.~(\ref{a19a}), the operators $c,\bar{c}$ and $\hat{H}%
_{int}^{\left( u\phi \right) }\left( \tau \right) $ are written in the
interaction representation using $\hat{H}_{0}$, Eq.~(\ref{a18c}), as the
bare Hamiltonian and $G_{r,r^{\prime },\sigma }^{\left( 0\right) }\left(
\tau ,\tau ^{\prime }\right) $ is the Green function of the ideal Fermi gas
described by the Hamiltonian $\hat{H}_{0}$. The symbol $\left\langle
...\right\rangle _{0}$ in Eq.~(\ref{a20}) stands for the Gibbs averaging
over the states of the Hamiltonian $\hat{H}_{0}$, Eq.~(\ref{a18c}).

The Green function $G_{r,r;\sigma }^{\left( u\phi \right) }\left( \tau ,\tau
^{\prime }\right) ,$ Eq.~(\ref{a20}), satisfies the following equation%
\begin{align}
& \left( -\frac{\partial }{\partial \tau }-\hat{h}_{r\sigma }\left[ u\phi
\left( \tau \right) \right] \right) G_{r,r^{\prime };\sigma }^{\left( u\phi
\right) }\left( \tau ,\tau ^{\prime }\right) =\delta _{r,r^{\prime }}\delta
\left( \tau -\tau ^{\prime }\right) ,  \notag \\
& \hat{h}_{r\sigma }\left[ u\phi \left( \tau \right) \right] =\hat{%
\varepsilon}_{r}-\mu ^{\prime }-\sigma u\phi _{r\sigma }\left( \tau \right)
\label{a21}
\end{align}%
where $\hat{\varepsilon}_{r}f_{r}\equiv -\sum_{r^{\prime }}t_{r,r^{\prime
}}f_{r^{\prime }}$ for an arbitrary function $f_{r}$.

A conjugated equation can be written as%
\begin{equation}
\left( \frac{\partial }{\partial \tau ^{\prime }}-\hat{h}_{r^{\prime }\sigma
}\left[ u\phi \left( \tau ^{\prime }\right) \right] \right) G_{r,r^{\prime
};\sigma }^{\left( u\phi \right) }\left( \tau ,\tau ^{\prime }\right)
=\delta _{r,r^{\prime }}\delta \left( \tau -\tau ^{\prime }\right) .
\label{a22}
\end{equation}%
The electron Green function $G_{r,r;\sigma }^{\left( u\phi \right) }\left(
\tau ,\tau ^{\prime }\right) $, Eq.~(\ref{a20}), satisfies the fermionic
boundary conditions%
\begin{equation*}
G_{r,r^{\prime };\sigma }^{\left( u\phi \right) }\left( \tau ,\tau ^{\prime
}\right) =-G_{r,r^{\prime };\sigma }^{\left( u\phi \right) }\left( \tau
+\beta ,\tau ^{\prime }\right) =-G_{r,r^{\prime };\sigma }^{\left( u\phi
\right) }\left( \tau ,\tau ^{\prime }+\beta \right) .
\end{equation*}

Eqs.~(\ref{a19}, \ref{a19a}, \ref{a21}, \ref{a22}) can serve as the starting
point of our bosonization scheme. In the next subsection, we derive
equations for the bosonic excitations in an external time dependent HS field.

\subsection{Equations for bosons in the fluctuating field}

\label{subsecIIb}

\subsubsection{General equations.}

According to the results of the previous subsection one can calculate the
partition function $Z$ solving Eq.~(\ref{a21}) or (\ref{a22}) for a given
configuration of the field $\phi _{r\sigma }\left( \tau \right) $,
substitute the solution into Eq.~(\ref{a19}), thus obtaining the functional $%
Z\left[ \phi \right] $, $\,$and then calculate $Z$ using Eq.~(\ref{a10}).
Actually, solving Eqs.~(\ref{a21}, \ref{a22}) one obtains more information
than necessary because the Green function $G_{r,r^{\prime };\sigma }^{\left(
u\phi \right) }\left( \tau ,\tau ^{\prime }\right) $ that should be obtained
from this equations depends on two times, $\tau $ and $\tau ^{\prime },$
whereas these time should be taken equal in Eq.~(\ref{a19}).

It is not difficult to derive closed equations for the Green function $%
G_{r,r^{\prime };\sigma }^{\left( u\phi \right) }\left( \tau ,\tau +0\right)
$ entering Eq.~(\ref{a19}). Subtracting Eq.~(\ref{a21}) from Eq.~(\ref{a22})
and putting $\tau ^{\prime }=\tau +0$, we obtain%
\begin{equation}
\left( \frac{\partial }{\partial \tau }-\hat{h}_{r\sigma }\left[ u\phi
\left( \tau \right) \right] +\hat{h}_{r^{\prime }\sigma }\left[ u\phi \left(
\tau \right) \right] \right) G_{r,r^{\prime };\sigma }^{\left( u\phi \right)
}\left( \tau ,\tau +0\right) =0  \label{a22a}
\end{equation}%
We write here equal times in the functions $\phi _{r}\left( \tau \right) $
because we assume that they are continuous. This is not so for the Green
functions $G_{r,r^{\prime };\sigma }^{\left( u\phi \right) }\left( \tau
,\tau ^{\prime }\right) $ because they have a jump at equal times. For
discontinuous functions $\phi _{r}\left( \tau \right) $ one would have to
take slightly different times $\tau $ and $\tau +0$, too.

It follows from Eq.~(\ref{a22a}) that
\begin{equation}
\frac{\partial }{\partial \tau }\sum_{r}G_{r,r;\sigma }^{\left( u\phi
\right) }\left( \tau ,\tau +0\right) =0  \label{a22ab}
\end{equation}%
which shows that the total number of particles in the system of the
non-interacting fermions in the auxiliary HS field $\phi _{r}\left( \tau
\right) $ does not depend on time. To avoid confusion, we remind the reader
that we started from the grand canonical formulation for the model of the
interacting particles, Eqs.~(\ref{a1}, \ref{a4}), and writing about the
particle conservation now we do not mean that we change the formulation of
our initial model.

For subsequent manipulations, it is convenient to introduce a new function
\begin{equation}
A_{r,r^{\prime }}\left( z\right) =G_{r,r^{\prime },\sigma }^{\left( 0\right)
}\left( \tau ,\tau +0\right) -G_{r,r^{\prime };\sigma }^{\left( u\phi
\right) }\left( \tau ,\tau +0\right)  \label{a22ac}
\end{equation}%
where $z=\left( \tau ,\sigma ,u\right) $.

The function $A_{r,r^{\prime }}\left( z\right) $ has the property%
\begin{equation}
\frac{\partial }{\partial \tau }\sum_{r}A_{r,r}\left( z\right) =0,
\label{a22c}
\end{equation}%
is periodic,
\begin{equation}
A_{r,r^{\prime }}\left( \tau ,\sigma ,u\right) =A_{r,r^{\prime }}\left( \tau
+\beta ,\sigma ,u\right) ,  \label{a23a}
\end{equation}%
and, hence, describes bosons.

With this function, the functional $Z\left[ \phi \right] $ takes a simple
form%
\begin{equation}
Z\left[ \phi \right] =Z_{0}\exp \Big[-\sum_{r,\sigma }\int_{0}^{\beta
}\int_{0}^{1}\sigma \phi _{r\sigma }\left( \tau \right) A_{r,r}\left(
z\right) dud\tau \Big]  \label{a24}
\end{equation}%
Eq.~(\ref{a24}) is a reformulation of (\ref{a19}) in terms of $%
A_{r,r^{\prime }}\left( z\right) $, Eq.~(\ref{a22ac}).

Having written Eq.~(\ref{a22a}) we can derive a closed equation for the
function $A_{r,r}\left( z\right) ,$ Eq.~(\ref{a22ac}). Our procedure is very
similar to deriving the kinetic equation starting from equations for Green
functions \cite{keldysh}.

We rewrite Eq.~(\ref{a22a}) at $u=0$, and subtract it from Eq.~(\ref{a22a}).
Using the definition of $A_{r,r^{\prime }}\left( z\right) $, we come to the
following equation for this function
\begin{align}
\frac{\partial }{\partial \tau }A_{r,r^{\prime }}\left( z\right)
+&M_{r,r^{\prime }}A_{r,r^{\prime }}\left( z\right) =-u\sigma \Phi
_{r,r^{\prime };\sigma }\left( \tau \right) n_{r,r^{\prime },\sigma },
\notag \\
M_{r,r^{\prime }}\left( z\right) &=\hat{\varepsilon}_{r}-\hat{\varepsilon}%
_{r^{\prime }}-u\sigma \Phi _{r,r^{\prime };\sigma }\left( \tau \right) ,%
\text{ }  \label{a25} \\
\Phi _{r,r^{\prime };\sigma }\left( \tau \right) &=\phi _{r\sigma }\left(
\tau \right) -\phi _{r^{\prime }\sigma }\left( \tau \right)  \notag
\end{align}%
where $n_{r,r^{\prime }}=G_{r,r^{\prime }}^{\left( 0\right) }\left( \tau
,\tau +0\right) $ is the Fermi distribution function of the ideal gas in the
coordinate representation%
\begin{eqnarray}
n_{r,r^{\prime }} &=&\int n_{\mathbf{p\sigma }}e^{i\mathbf{p}(r-r^{\prime })}%
\frac{d\mathbf{p}}{(2\pi )^{d}},  \label{a25a} \\
n_{\mathbf{p}} &=&\left[ \exp \left\{ \beta \left( \varepsilon _{\mathbf{p}%
}-\mu ^{\prime }\right) \right\} +1\right] ^{-1}.  \notag
\end{eqnarray}

The solution of Eq.~(\ref{a25}), as it stands, is not unique. One can easily
see that there is a non-physical solution $A_{r,r^{\prime }}\left( z\right)
=n_{r,r^{\prime }}$ for any $\phi _{r\sigma }\left( \tau \right) $.
Moreover, for static fields $\phi _{r\sigma }$, one can add to the solution
of Eq.~(\ref{a25}) a combination $A_{r,r^{\prime }}^{\left( 0\right) }$ of
the form
\begin{equation}
A_{r,r^{\prime }}^{\left( 0\right) }=\sum_{k}C_{k}v_{r}^{k}v_{r^{\prime
}}^{k}  \label{a25b}
\end{equation}%
where $v_{r}^{k}$ are eigenfunctions of the operator $\hat{h}_{r}\left[
u\phi \right] $ and $C_{k}$ are time independent coefficients.

There can be less trivial solutions of the homogeneous equation
corresponding to Eq.~(\ref{a25}) when $\phi _{r\sigma }\left( \tau \right) $
is a discontinuous function of time with large jumps. Such solutions
correspond to poles in the Green functions $G_{r,r^{\prime }}\left( \tau
,\tau +0\right) $ and should be treated very carefully. The sign problem in
the MC simulations appears as a result of the poles in the integrals over $u$
in Eqs.~(\ref{a24}) which arise when the fields $\phi _{r\sigma }\left( \tau
\right) $ are discontinuous in time. We discuss this question in Section \ref%
{mc}.

It is clear that the ambiguity in solving Eq.~(\ref{a25}) following from the
existence of solutions of the homogeneous equation creates problems in both
numerical and analytical treatment. This means that a procedure fixing the
proper solutions of Eq.~(\ref{a25}) is necessary and we introduce it below.

\subsubsection{Regularization of the bosonized model.}

\begin{figure}[t]
\includegraphics{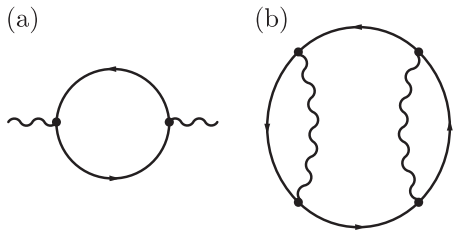}
\caption{Simple electronic diagrams.}
\label{reg}
\end{figure}
Seeking for the unique correct solution of Eq.~(\ref{a25}) we are guided by
the conventional diagrammatic technique for fermions which is well defined
and all corresponding diagrams can be calculated at least in principle. In
order to visualize how we come to solving Eq.~(\ref{a25}) instead of summing
the conventional diagrams we consider the simplest loop represented in Fig. %
\ref{reg} (a). The expression corresponding to this diagram can be written as%
\begin{eqnarray}
&&T\sum_{\varepsilon ,\mathbf{p}}\frac{1}{i\varepsilon -\varepsilon _{%
\mathbf{p}}^{\prime }}\frac{1}{i\varepsilon +i\omega -\varepsilon _{\mathbf{%
p+q}}^{\prime }}  \notag \\
&=&\sum_{\mathbf{p}}\frac{n_{\mathbf{p}}-n_{\mathbf{p+q}}}{i\omega
-\varepsilon _{\mathbf{p+q}}^{\prime }+\varepsilon _{\mathbf{p}}^{\prime }}
\label{a25bc}
\end{eqnarray}%
where $\varepsilon _{\mathbf{p}}^{\prime }=\varepsilon _{\mathbf{p}}-\mu
^{\prime }$ and $\varepsilon =2\pi \left( n+1/2\right) T$ and $\omega =2\pi
mT$ are fermionic and bosonic frequencies, respectively.

The expression in the second line of Eq.~(\ref{a25bc}) corresponds to the
solution of Eq.~(\ref{a25}) in the lowest order in $\phi $, which
demonstrates how the bosonic modes are obtained from the fermionic lines.

However, the first and the second lines in Eq.~(\ref{a25bc}) are different
at $\omega =0,\mathbf{q}=0.$ The first line gives the density of states at
the Fermi surface
\begin{equation}
\sum_{\mathbf{p}}\frac{1}{2T}\cosh ^{-2}\left( \frac{\varepsilon _{\mathbf{p}%
}^{\prime }}{2T}\right) =\frac{\partial n}{\partial \mu }  \label{a25bd}
\end{equation}%
but the expression in the second line is not defined. Of course, one could
speak about the limit $\mathbf{q}\rightarrow 0$ instead of just putting $%
\mathbf{q}=0$ but this is not justified for any finite physical system with
discrete energy levels.

The uncertainty of the expression in the second line of Eq.~(\ref{a25bc}) is
reflected in the existence of solutions [like those given by Eq.~(\ref{a25b}%
)] of the homogeneous equation corresponding to Eq.~(\ref{a25}).

We see that our bosonization scheme is not exactly equivalent to the initial
fermionic model and needs a regularization in order to avoid the
uncertainty. This uncertainty appears not only in the situation represented
in Fig. \ref{reg} (a), where the field $\phi $ enters with $\omega =0,%
\mathbf{q=}0$. The same problem is encountered when calculating the
contribution of, e.g., Fig. \ref{reg} (b), using the bosonization scheme
because this graph contains fermionic Green functions at coinciding momenta
and frequencies but the interaction lines may carry any momenta and
frequencies. We will come back to this point when discussing in detail the
diagrammatics of the bosonized theory.

The hint at a possible regularization is given by the form of Eq.~(\ref%
{a25bc}). If the momenta $\mathbf{q}$ were continuous rather then discrete
and we could simply neglect the contribution of the state with $\mathbf{q=}0$%
, the bosonization scheme would work in this order. This is not sufficient,
though, because the momenta and frequencies of the two horizontal fermion
lines in Fig. \ref{reg} (b) obtained after integration over the HS field $%
\phi $ coincide for any translationally invariant systems and one should
slightly violate the translational invariance in order to split the momenta.

This goal can be achieved if we assume that the system of the interacting
electrons considered here is imbedded in a \textquotedblleft
bath\textquotedblright . We introduce this bath considering a model of the
interacting electrons on a $d$-dimensional lattice with the total number of
sites $N_{d}^{total}$. However, we assume that the interaction $%
V_{r,r^{\prime }}$, Eq.~(\ref{a3}), entering the Hamiltonian $\hat{H}$, Eq.~(%
\ref{a1}), vanishes outside a subsystem consisting of a considerably smaller
number of sites $N_{d}$. As an example, we can suggest the following form of
the interaction $V_{r,r^{\prime }}$%
\begin{equation}
V_{r,r^{\prime }}^{total}=\left\{
\begin{array}{cl}
V_{r,r^{\prime }} & ,\; \left\vert r\right\vert ,\left\vert r^{\prime
}\right\vert <R_{0} \\
0 & ,\;\text{otherwise}%
\end{array}%
\right.  \label{a100}
\end{equation}%
which means that the interaction is finite inside the sphere of the radius $%
R_{0}$ containing $N_{d}$ sites and vanishes outside this sphere. In other
words, we attach metallic leads to the system of the interacting electrons
and assume that the inter-electron interaction vanishes in the leads.

It is clear that the leads cannot change physics of the system of the
interacting electrons if the latter is sufficiently large. However, this
model is reasonable even for a system with a small number sites $N_{d}$
(quantum dot), although properties of the systems with and without leads can
be different.

The form of the interaction $V_{r,r^{\prime }}^{total}$, Eq.~(\ref{a100}),
formally violates the translational invariance and the matrix elements
\begin{equation}
V^{total}\left( \mathbf{q,q}_{0}\right) =\sum_{\mathbf{r,r}^{\prime
}}V_{r,r^{\prime }}^{total}e^{-i\mathbf{q}_{0}\left( \mathbf{r+r}^{\prime
}\right) /2-i\mathbf{q}\left( \mathbf{r-r}^{\prime }\right) }  \label{a101}
\end{equation}%
are finite for $\mathbf{q}_{0}\neq 0$.

Of course, the matrix element with $\mathbf{q}_{0}=0$ remains finite and,
for such momenta, the two Green functions in Fig. \ref{reg} (b) still have
the same momenta and frequencies. However, we can neglect them because they
give a small contribution provided the number of the contributing momenta $%
\mathbf{q}_{0}$ is large. The latter is achieved for a large number of the
sites in the bath.

We emphasize, that neglecting the contribution of graphs containing several
electron Green functions with coinciding momenta and frequencies is possible
because their contribution is not singular. One can estimate the relative
error of this approximation as being proportional to $N_{d}/N_{d}^{total}$,
where $N_{d}^{total}$ is the total number of sites in the entire system
including the bath.

Following this idea we could simply exclude in the bosonization approach all
propagators with $\omega =0,$ $\mathbf{q}=0$ \textquotedblleft by hand".
However, this is possible only when doing a perturbation theory and such an
approach is not sufficient for non-perturbative and numerical studies.

The bosonic modes with $\omega =0,$ $\mathbf{q}=0$ correspond to solutions
of the homogeneous equation in Eq.~(\ref{a25}). In order to discard these
modes by a regular procedure we regularize Eq.~(\ref{a25}) slightly changing
it. This is not a trivial task because this slight modification should
guarantee the complete absence of the solutions of the homogeneous equation
corresponding to Eq.~(\ref{a25}).

We regularize the bosonized model replacing Eq.~(\ref{a25}) by a system of
two equations
\begin{equation}
\mathcal{H}_{r,r^{\prime }}\left( \tau \right) A_{a;r,r^{\prime }}\left(
z\right) =-u\sigma n_{r,r^{\prime },\sigma }\Phi _{r,r^{\prime };\sigma
}\left( \tau \right) B_{a},  \label{a102}
\end{equation}%
where $a=1,2$, $\mathcal{H}_{r,r^{\prime }}$ is a $2\times 2$ matrix
\begin{equation}
\mathcal{H}_{r,r^{\prime }}\left( \tau \right) =\Lambda _{1}M_{r,r^{\prime
}}\left( z\right) +i\Lambda _{2}\frac{\partial }{\partial \tau }+\Lambda
\gamma ,  \label{a103}
\end{equation}%
and
\begin{equation*}
A_{1;r,r^{\prime }}\left( z\right) =\left(
\begin{array}{c}
A_{1;r,r^{\prime }}^{\prime }\left( z\right) \\
A_{1;r,r^{\prime }}^{\prime \prime }\left( z\right)%
\end{array}%
\right) ,\quad B_{1}=\left(
\begin{array}{c}
0 \\
1%
\end{array}%
\right)
\end{equation*}%
\begin{equation*}
A_{2;r,r^{\prime }}\left( z\right) =\left(
\begin{array}{c}
A_{2;r,r^{\prime }}^{\prime \prime }\left( z\right) \\
A_{2;r,r^{\prime }}^{\prime }\left( z\right)%
\end{array}%
\right) ,\quad B_{2}=\left(
\begin{array}{c}
1 \\
0%
\end{array}%
\right)
\end{equation*}%
The functions $A_{a;r,r^{\prime }}\left( z\right) $ satisfy the bosonic
boundary conditions
\begin{equation}
A_{a;r,r^{\prime }}\left( \tau ,\sigma ,u\right) =A_{a;r,r^{\prime }}\left(
\tau +\beta ,\sigma ,u\right)  \label{a103a}
\end{equation}%
and $2\times 2$ Pauli matrices are used%
\begin{equation}
\Lambda _{1}=\left(
\begin{array}{cc}
0 & 1 \\
1 & 0%
\end{array}%
\right) ,\quad \Lambda _{2}=\left(
\begin{array}{cc}
0 & -i \\
i & 0%
\end{array}%
\right) ,\quad \Lambda =\left(
\begin{array}{cc}
1 & 0 \\
0 & -1%
\end{array}%
\right).  \label{a103b}
\end{equation}%
The real parameter $\gamma $ should be put to zero, $\gamma \rightarrow 0 $,
at the end of calculations.

With this modification we write the function $Z\left[ \phi \right] $ in the
form%
\begin{equation}
Z\left[ \phi \right] =Z_{0}\exp \left[ -\frac{1}{2}\sum_{r,\sigma
,a}\int_{0}^{\beta }\int_{0}^{1}\sigma \phi _{r\sigma }\left( \tau \right)
A_{a;r,r}^{\prime }\left( z\right) dud\tau \right]  \label{a104}
\end{equation}

The exponent in Eq.~(\ref{a104}) is an even function of $\gamma $. It is not
difficult to see that putting $\gamma =0$ in Eqs.~(\ref{a102}-\ref{a104})
one returns to Eqs.~(\ref{a24}, \ref{a25}). However, keeping in Eqs.~(\ref%
{a102}-\ref{a104}) the parameter $\gamma $ finite allows us to discard all
the solutions of the homogeneous equation in Eq.~(\ref{a25}).

This can be understood rather easily because the operator $\mathcal{H}$ is
hermitian and, therefore, its eigenvalues are real. Moreover, as will be
shown in the next subsection, they appear in pairs: if an eigenvalue $E$
exists, then the eigenvalue $-E$ exists, too. If, when changing parameters
of the operator $\mathcal{H}$, an eigenvalue $E$ with its counterpart $-E$
turned to zero at some point, this would mean that after crossing this point
they become imaginary. However, the latter is forbidden by the hermiticity
of the operator $\mathcal{H}$. A general proof of the absence of the
solutions of the homogeneous equation in Eq.~(\ref{a102}) or, in other
words, absence of zero eigenvalues $E$ of the operator $\mathcal{H}%
_{r,r^{\prime }}\left( \tau \right) $, Eq.~(\ref{a103}), is given in
Appendix \ref{solution}.

The absence of zero eigenvalues makes the operator $\mathcal{H}$ invertible.
Since all the parameters of Eqs.~(\ref{a102}-\ref{a103}) are real, the
solutions $A_{a;r,r^{\prime }}\left( z\right) $ are real, too. This means
that the exponent in Eq.~(\ref{a104}) is real and $Z\left[ \phi \right] $ is
real and positive. This property of the $Z\left[ \phi \right] $ can be very
important for numerical computations using the MC method.

Using the regularization with the parameter $\gamma $, Eqs.~(\ref{a102}-\ref%
{a104}), we can prove a stronger than Eq.~(\ref{a22c}) relation for $%
A_{r,r^{\prime }}\left( z\right) $,%
\begin{equation}
\sum_{r}A_{r,r}\left( z\right) =0  \label{a22ca}
\end{equation}
Eq.~(\ref{a22ca}) is fulfilled automatically for any solution of Eqs.~(\ref%
{a102}-\ref{a103}). This property can be proven putting in Eqs.~(\ref{a102}, %
\ref{a103}) $r=r^{\prime }$ and summing over $r$. Then, one obtains the
equations%
\begin{eqnarray}
\gamma I_{a}^{\prime }\left( z\right) +\partial I_{a}^{\prime \prime }\left(
z\right) /\partial \tau &=&0  \notag \\
\partial I_{a}^{\prime }\left( z\right) /\partial \tau +\gamma I_{a}^{\prime
\prime }\left( z\right) &=&0  \label{a105}
\end{eqnarray}%
where $I_{a}^{\prime }\left( z\right) =\sum_{r}A_{a;r,r}^{\prime }\left(
z\right) $, $I_{a}^{\prime \prime }\left( z\right)
=\sum_{r}A_{a;r,r}^{\prime \prime }\left( z\right) $.

Only the trivial solution $I_{a}^{\prime }\left( z\right) =I_{a}^{\prime
\prime }\left( z\right) =0$ satifies the boundary condition, Eq.~(\ref{a103a}%
), leading to Eq.~(\ref{a22ca}).

\subsubsection{Spectral expansion.}

In this subsection we discuss important properties of the solutions of Eqs.~(%
\ref{a102}, \ref{a103}) using spectral expansions in eigenfunctions of this
equation. In order to better understand the difference between solutions of
Eq.~(\ref{a25}) and Eqs.~(\ref{a102}, \ref{a103}), let us consider both the
equations.

The operator in the L.H.S. of Eq.~(\ref{a25}) is not hermitian, which
results in a double set of eigenfunctions $\left\{ v_{r,r^{\prime
}}^{K}\left( \tau \right) \right\} $ and $\left\{ \bar{v}_{r,r^{\prime
}}^{K}\left( \tau \right) \right\} $ satisfying the equations%
\begin{eqnarray}
\left( \frac{\partial }{\partial \tau }+M_{r,r^{\prime }}\right)
v_{r,r^{\prime }}^{K}\left( \tau \right) &=&\lambda ^{K}v_{r,r^{\prime
}}^{K}\left( \tau \right),  \label{a106} \\
\left( -\frac{\partial }{\partial \tau }+M_{r,r^{\prime }}\right) \bar{v}%
_{r,r^{\prime }}^{K}\left( \tau \right) &=&\lambda ^{K}\bar{v}_{r,r^{\prime
}}^{K}\left( \tau \right)  \notag
\end{eqnarray}%
with the boundary conditions $v_{r,r^{\prime }}^{K}\left( \tau \right)
=v_{r,r^{\prime }}^{K}\left( \tau +\beta \right) $, $\bar{v}_{r,r^{\prime
}}^{K}\left( \tau \right) =\bar{v}_{r,r^{\prime }}^{K}\left( \tau +\beta
\right) $. Two other equations can be written taking the complex conjugate
of Eqs.~(\ref{a106}),
\begin{eqnarray}
\left( \frac{\partial }{\partial \tau }+M_{r,r^{\prime }}\right)
v_{r,r^{\prime }}^{K\ast }\left( \tau \right) &=&\lambda ^{K\ast
}v_{r,r^{\prime }}^{K\ast }\left( \tau \right),  \label{a107} \\
\left( -\frac{\partial }{\partial \tau }+M_{r,r^{\prime }}\right) \bar{v}%
_{r,r^{\prime }}^{K\ast }\left( \tau \right) &=&\lambda ^{K\ast }\bar{v}%
_{r,r^{\prime }}^{K\ast }\left( \tau \right).  \notag
\end{eqnarray}%
The orthogonality conditions for the functions $v_{r,r^{\prime }}^{K}$ can
be written as%
\begin{equation*}
\sum_{r,r^{\prime }}\int_{0}^{\beta }\bar{v}_{r,r^{\prime }}^{K}\left( \tau
\right) v_{r,r^{\prime }}^{K^{\prime }}\left( \tau \right) d\tau =\delta
^{K,K^{\prime }}
\end{equation*}%
(the same for the complex conjugates).

The eigenvalues $\lambda ^{K}$ are generally complex and $\lambda ^{K}=0$ is
not excluded. For example, the function $A_{r,r^{\prime }}^{\left( 0\right)
} $, Eq.~(\ref{a25b}), corresponds to the zero eigenvalue for static $\phi
_{r\sigma }$.

All non-zero eigenvalues $\lambda ^{K}$ appear in pairs. If an eigenvalue $%
\lambda ^{K}$ with an eigenfunction $v_{r,r^{\prime }}^{K}\left( \tau
\right) $ exists, then the eigenvalue $-\lambda ^{K}$ with the eigenfunction
$\bar{v}_{r^{\prime },r}^{K}\left( \tau \right) $ exists, too. We see that
nothing prevents the eigenvalues $\lambda ^{K}$ from turning at some points
to zero and being complex, which is an unpleasant feature of the theory.

In contrast, the operator $\mathcal{H}_{r,r^{\prime }}\left( \tau \right) $
is hermitian. The structure of the two-component vectors $S_{r,r^{\prime
}}^{K}\left( \tau \right) $ can be represented as
\begin{equation*}
S^{K}=\left(
\begin{array}{c}
a^{K} \\
b^{K}%
\end{array}%
\right) ,\text{\quad }\left( S^{K}\right) ^{T}=\left(
\begin{array}{cc}
a^{K} & b^{K}%
\end{array}%
\right)
\end{equation*}
and they can be found from the equation%
\begin{equation}
\mathcal{H}_{r,r^{\prime }}\left( \tau \right) S_{r,r^{\prime }}^{K}\left(
\tau \right) =E^{K}S_{r,r^{\prime }}^{K}\left( \tau \right)  \label{a108}
\end{equation}%
supplemented by the boundary condition $S_{r,r^{\prime }}^{K}\left( \tau
\right) =S_{r,r^{\prime }}^{K}\left( \tau +\beta \right) $, where the
eigenvalues $E^{K}$ must be real.

The eigenstates with eigenfunctions $S_{r,r^{\prime }}^{K}\left( \tau
\right) $ and eigenvalues $E^{K}$ have their counterparts with the
eigenfunctions $\Lambda _{1}S_{r^{\prime },r}^{K}\left( \tau \right) $ and
eigenvalues $-E^{K}$.

The orthogonality condition for the eigenvectors $S_{r,r^{\prime
}}^{K}\left( \tau \right) $ follows from Eq.~(\ref{a108}) and can be written
as%
\begin{equation}
\sum_{r,r^{\prime }}\int_{0}^{\beta }\Big( S_{r,r^{\prime }}^{K}\left( \tau
\right) \Big)^{+}S_{r,r^{\prime }}^{K^{\prime }}\left( \tau \right) d\tau
=\delta ^{K,K^{\prime }}  \label{a109}
\end{equation}%
where the symbol $``+"$stands for the complex conjugation and transposition $%
``T"$.

The completeness of the system of the eigenvectors $S$ expressed by the
relation%
\begin{equation}
\sum_{K}S_{r,r^{\prime }}^{K}\left( \tau \right) \Big(S_{r_{1},r_{1}^{\prime
}}^{K}\left( \tau _{1}\right) \Big)^{+}=\delta _{r,r_{1}}\delta _{r^{\prime
},r_{1}^{\prime }}\delta \left( \tau -\tau ^{\prime }\right)  \label{a110}
\end{equation}%
allows us to write the spectral expansion for the $2\times 2$ matrix Green
function $\mathcal{G}_{r,r^{\prime };r_{1},r_{1}^{\prime }}\left( \tau ,\tau
_{1}\right) $ satisfying the equation%
\begin{equation}
\mathcal{H}_{r,r^{\prime }}\left( \tau \right) \mathcal{G}_{r,r^{\prime
};r_{1},r_{1}^{\prime }}\left( \tau ,\tau _{1}\right) =\delta
_{r,r_{1}}\delta _{r_{1},r_{1}^{\prime }}\delta \left( \tau -\tau _{1}\right)
\label{a111}
\end{equation}%
in the form
\begin{equation}
\mathcal{G}_{r,r^{\prime };r_{1},r_{1}^{\prime }}\left( \tau ,\tau
_{1}\right) =\sum_{K}\frac{S_{r,r^{\prime }}^{K}\left( \tau \right) \Big(%
S_{r_{1},r_{1}^{\prime }}^{K}\left( \tau _{1}\right) \Big)^{+}}{E^{K}}
\label{a112}
\end{equation}%
Since all eigenvalues $E^{K}$ are not equal to zero, the Green function $%
\mathcal{G}_{r,r^{\prime };r_{1},r_{1}^{\prime }}\left( \tau ,\tau
_{1}\right) $, Eq.~(\ref{a112}), is uniquely defined and does not have
singularities. Using the properties of the eigenvectors $S_{r,r^{\prime
}}^{K}\left( \tau \right) $ explained after Eq.~(\ref{a108}) we obtain a
symmetry property for these functions%
\begin{equation}
\mathcal{G}_{r,r^{\prime };r_{1},r_{1}^{\prime }}\left( \tau ,\tau
_{1}\right) =-\Lambda _{1}\mathcal{G}_{r^{\prime },r;r_{1}^{\prime
},r_{1}}\left( \tau _{1},\tau \right) \Lambda _{1}  \label{a113}
\end{equation}%
With the Green function $\mathcal{G}_{r,r^{\prime };r_{1},r_{1}^{\prime
}}\left( \tau ,\tau _{1}\right) ,$ Eqs.~(\ref{a111}, \ref{a112}), we write
the function $Z\left[ \phi \right] $, Eq.~(\ref{a104}), in the form
\begin{eqnarray}
&&Z\left[ \phi \right] =Z_{0}\exp \Big[\frac{1}{2}\sum_{r,r_{1},r_{1}^{%
\prime },\sigma }\int_{0}^{\beta }\int_{0}^{1}\phi _{r\sigma }\left( \tau
\right)  \notag \\
&&\times \left( \mathcal{G}_{r,r;r_{1},r_{1}^{\prime }}^{12}\left( \tau
,\tau _{1};u,\sigma \right) +\mathcal{G}_{r,r;r_{1},r_{1}^{\prime
}}^{21}\left( \tau ,\tau _{1};u,\sigma \right) \right)  \notag \\
&&\times un_{r_{1},r_{1}^{\prime }}\Phi _{r_{1},r_{1}^{\prime };\sigma
}\left( \tau \right) d\tau du\Big]  \label{a115}
\end{eqnarray}

Eqs.~(\ref{a112}, \ref{a115}) give an explicit unambiguous expression for
the function $Z\left[ \phi \right] $. It can be further integrated over $%
\phi _{r\sigma }$ using superfields, as it is done in the next Section, or
calculated numerically. In the latter case, it is more convenient to solve
directly Eq.~(\ref{a102}) and substitute the solution into Eq.~(\ref{a104}).

We check our regularization scheme in Appendix \ref{statics} on the model
with a static HS field $\phi _{r\sigma }$.

One can interpret the real function $A_{r,r^{\prime }}\left( z\right) $ as a
density matrix, while Eqs.~(\ref{a25}, \ref{a102}) are analogues of the von
Neumann equation. In this language, Eq.~(\ref{a22ca}) is the particle
conservation law. Using the density matrix in quantum mechanics implies a
connection of the system to an environment. Within our approach, we use the
regularization based on the introduction of the bath. This enforces in a
very natural way the analogy between the matrix $A$ and the density matrix
in quantum mechanics. One can say that our bosonization procedure
corresponds to replacing the description of the quantum mechanics in terms
of the wave functions by the description in terms of the density matrix.

In the classical limit, Eqs.~(\ref{a25}, \ref{a102}) play a role of the
kinetic equation and $A_{r,r^{\prime }}\left( z\right) $ (after Fourier
transforming in $r-r^{\prime }$) can be considered as the distribution
function for a particle moving in a fluctuating field in imaginary time.

We see that the original problem of calculating the partition function of
interacting fermions $Z$, Eq.~(\ref{a4}), has been reduced after the
regularization to solving Eqs.~(\ref{a102}, \ref{a103}) for a bosonic
function $A_{r,r^{\prime }}\left( z\right) $, substituting the solution into
Eq.~(\ref{a104}) and calculating the functional integral over $\phi
_{r\sigma }\left( \tau \right) $ in Eq.~(\ref{a10}). For an analytical
investigation, this integral over $\phi _{r\sigma }\left( \tau \right) $ can
be performed before doing any approximations and this is demonstrated in the
next Section.

\section{Superfield theory of interacting bosons and perturbation theory
\label{pt}}

\subsection{Superfield theory of interacting bosons}

In this subsection, we reduce the computation of the partition function $Z$,
Eq.~(\ref{a4}), to calculations with a superfield model describing
interacting bosonic excitations. Eq.~(\ref{a25}) resembles quasiclassical
equations written in Ref.~\onlinecite{aleiner}. The solution of these
equations was represented in terms of a functional integral over $48$%
-component supervectors, which allowed the authors to integrate over the HS
field. Unfortunately, the resulting Lagrangian and calculations with it were
rather cumbersome due to the large number of the components of the
supervectors.

Now we use another trick, known as the Becchi-Rouet-Stora-Tuytin (BRST)
transformation\cite{brst} (see also the book by Zinn-Justin\cite{justin}). A
similar transformation was used in the quantization of non-abelian gauge
theories\cite{faddeev}. In condensed matter physics, this trick has been
used for the first time in Ref.~\onlinecite{parisi}. Using this
transformation one can also represent the solution of an equation (not
necessarily linear) in terms of a functional integral over both conventional
and Grassmann anticommuting fields. Due to some special symmetries, both the
types of the fields can be unified in a superfields depending not only on
coordinates and times but also on additional anticommuting variables.

In order to avoid complicated formulas containing $2\times 2$ matrices that
appear as a result of the regularization, we restrict our consideration in
this Section by the thermodynamic limit. In this limit the spectrum is
continuous and the states with coinciding momenta and frequencies do not
give any essential contribution anyway. This means that we put the parameter
$\gamma =0.$ At the same time, we still assume that the system is imbedded
in the bath. This allows us to treat irregular expressions in a simple way
because the translational invariance of the subsystem (where the interaction
is finite) is broken and Green functions with coinciding momenta and
frequencies give a vanishing contribution in the thermodynamic limit
considered in this Section.

Let us explain this procedure in more details. Suppose, we have an equation%
\begin{equation}
F\left( A\right) =0,  \label{a30}
\end{equation}%
where $F\left( A\right) $ is a real matrix function of a real matrix
function $A$, and our task is to find the solution $A_{0}$ of this equation
and calculate a quantity $B\left( A_{0}\right) ,$ where $B\left( A\right) $
is another matrix function. Instead of proceeding in this way, one can write
$B\left( A_{0}\right) $ in a form of the integral%
\begin{equation}
B\left( A_{0}\right) =\int B(a)\delta \left[ F\left( a\right) \right]
\left\vert \mathrm{\det }\left( \frac{\partial F}{\partial a}\right)
\right\vert da  \label{a31}
\end{equation}%
over the variable $a$ of the same structure as $A$. The modulus of the
determinant in Eq.~(\ref{a31}) is included in the integrand in order to
normalize the $\delta $-function.

As the next step, we represent the $\delta $-function as%
\begin{equation}
\delta \left[ F\left( a\right) \right] =C\int \exp \left[ i\mathrm{Tr}\left(
f^TF\left( a\right) \right) \right] df  \label{a32}
\end{equation}%
where $f$ is a real matrix variable and the symbol ``$T$'' means the
transposition of $r$ and $r^{\prime }$, $f_{r,r^{\prime }}^{T}=f_{r^{\prime
},r}$ . $C$ is a normalization constant.

The determinant in Eq.~(\ref{a31}) can also be represented in a form of an
integral but now the variables of integration should be anticommuting
Grassmann variables $\eta $ and $\eta ^{+}$,%
\begin{equation}
\det \left( \frac{\partial F}{\partial a}\right) =\int \exp \left[ -\mathrm{%
Tr}\left( \eta ^{+}\frac{\partial F}{\partial a}\eta \right) \right] d\eta
^{+}d\eta .  \label{a33}
\end{equation}%
After unifying the fields $a$, $f^T$, $\eta $, and $\eta ^{+}$ in a
superfield~$\Psi$, the expression in the exponential can be written in a
compact way. We do not present here general formulas that can be found in
the book~\onlinecite{justin} but concentrate explicitly on Eqs.~(\ref{a24},%
\ref{a25}).

Eq.~(\ref{a25}) is linear and real and we can apply Eq.~(\ref{a31})
directly. Using the transformation of Eq.~(\ref{a32}) we obtain in the
exponent a quadratic form in terms of the variables $a$ and $f$. Then, we
write the determinant with the help of Eq.~(\ref{a33}) and obtain another
quadratic form in the exponent containing the anticommuting variables $\eta $
and $\eta ^{+}$.

At first glance, there should be a problem related to the presence of the
modulus in Eq.~(\ref{a31}) and absence of it in Eq.~(\ref{a33}).
Fortunately, the operator $\partial /\partial \tau + M _{r,r^{\prime
}}\left( z\right) $ is real and antisymmetric, which leads to an always
positive determinant. Therefore, the modulus in Eq.~(\ref{a31}) does not
play any role in the case under consideration.

Writing the quadratic forms in the exponent using the variables $a$, $f^{T}$%
, $\eta $, and $\eta ^{+}$ is not difficult and we do not write them in this
form. A more compact form of the functional integrals can be achieved
introducing the superfields $\Psi _{r,r^{\prime }}\left( R\right) $, $%
R=\left\{ z,\theta ,\theta ^{\ast }\right\} $, as follows%
\begin{equation}
\Psi _{r,r^{\prime }}\left( R\right) =a_{r,r^{\prime }}\left( z\right)
\theta +f_{r,r^{\prime }}^{T}\left( z\right) \theta ^{\ast }+\eta
_{r,r^{\prime }}\left( z\right) +\eta _{r,r^{\prime }}^{+}\left( z\right)
\theta ^{\ast }\theta  \label{a34}
\end{equation}%
where $a_{r,r^{\prime }}\left( z\right) $ and $f_{r,r^{\prime }}\left(
z\right) $ are real fields of the coordinates $r,r^{\prime }$ and the
variables $z=\left( \tau ,\sigma ,u\right) $. The hermitian conjugation
means complex conjugation $``\ast "$ supplemented by the transposition, such
that $\eta _{r,r^{\prime }}^{+}=\eta _{r,^{\prime }r}^{\ast }$. The
variables $\theta $, $\theta ^{\ast }$ are artificially introduced Grassmann
variables that help us to write the exponent in the compact form.

As a result, we write the functional $Z\left[ \phi \right] $, Eq.~(\ref{a24}%
), in the form
\begin{equation}
Z\left[ \phi \right] =Z_{0}\int \exp \left( -\mathcal{S}_{0}\left[ \Psi %
\right] -\mathcal{S}^{\left( u\phi \right) }\left[ \Psi \right] \right) D\Psi
\label{a35}
\end{equation}%
where $S_{0}\left[ \Psi \right] $ is the bare part of the action%
\begin{equation}
\mathcal{S}_{0}=\frac{i}{2}\sum_{r,r^{\prime }}\int \Big[\Psi _{r^{\prime
},r}\left( R\right) \left( \frac{\partial }{\partial \tau }+\left( \hat{%
\varepsilon}_{r}-\hat{\varepsilon}_{r^{\prime }}\right) \right) \Psi
_{r,r^{\prime }}\left( R\right) \Big]dR  \label{a36}
\end{equation}

The second term $\mathcal{S}^{\left( u\phi \right) }\left[ \Psi \right] $ in
the exponent in Eq.~(\ref{a35}) is linear with respect to the HS field $\phi
_{r\sigma }\left( \tau \right) $. Its explicit form reads%
\begin{eqnarray}
&&\mathcal{S}^{\left( u\phi \right) }\left[ \Psi \right] =-i\sum_{r,r^{%
\prime }}\int \phi _{r\sigma } \left( \tau \right) \Big[u\left( \Psi
_{r^{\prime },r}\left( R\right) -n_{r^{\prime },r}\theta \right)  \notag \\
&&\times \left( \Psi _{r,r^{\prime }}\left( R\right) -n_{r,r^{\prime
}}\theta \right) -i\sigma \delta _{r,r^{\prime }}\Psi _{r,r}\left( R\right)
\theta ^{\ast }\Big]dR  \label{a37}
\end{eqnarray}%
\newline
It is interesting to remark here that the superfield $\Psi \left( R\right) $
is anticommuting, which is a rather unusual feature of the field theory
considered here. However, it is very important that this field describes
bosons and not fermions. This follows from the periodic boundary condition $%
\Psi \left( \tau \right) =\Psi \left( \tau +\beta \right) $. It is this
condition that determines unambiguously the statistics of the particles. The
fact that $\Psi \left( R\right) $ is anticommuting is only a formal property.

The form of Eqs.~(\ref{a35}-\ref{a37}) allows us to average immediately over
the HS field $\phi _{r\sigma }\left( \tau \right) $. This integration is
Gaussian and is specified by Eq.~(\ref{a10}). The analytic supersymmetric
field theory written here relies crucially on the presence of a bath, whose
function is to break the translational symmetry of the system. In order to
take the bath into account, we model the interaction in the same way as in
Eq.~(\ref{a100}). The pair correlation of the fields with the distribution $%
W $ can be written as%
\begin{eqnarray}
&&\left\langle \sigma \sigma^\prime \phi _{r\sigma }\left( \tau \right) \phi
_{r^{\prime }\sigma ^{\prime }}^{\prime }\left( \tau ^{\prime }\right)
\right\rangle _{W} =U_{r,r^{\prime }}\left( R,R^{\prime }\right)  \label{a38}
\\
&&U_{r,r^{\prime }}\left( R,R^{\prime }\right) =\delta \left( \tau -\tau
^{\prime }\right) \left( \sigma \sigma ^{\prime }V_{0}^{total}\delta
_{r,r^{\prime }}+V_{r,r^{\prime }}^{\left( 1\right), total }\right) \ ,
\notag
\end{eqnarray}%
where $V_{0}^{total}$ and $V_{r,r^{\prime }}^{\left( 1\right), total }$ are
defined from $V_{0}$ and $V^{\left( 1\right) }$ according to Eqn. (\ref{a100}%
). The coupling constants $V_{0}$ and $V^{\left( 1\right) }$ contain the
constant $\bar{V}$ that has been introduced in a rather arbitrary manner
(see, below Eq.~(\ref{a4a}). However, the same constant enters the
renormalized chemical potential $\mu ^{\prime }$, Eq.~(\ref{a14}), and the
renormalized thermodynamical potential $\Omega $ written below Eq.~(\ref{a14}%
). Of course, the constant $\bar{V}$ should disappear from the final result
for the thermodynamical potential, which can serve as a check of any
computation. Actually, considering only perturbation theory in the
interaction there is no necessity to introduce this constant. In such
calculations the sign of $\tilde{V}_{r,r^{\prime }}$ in Eq.~(\ref{a4}) does
not play an important role because one could decouple the interaction term
with $\tilde{V}$ integrating over purely imaginary HS. We have added $\bar{V}
$ because we want to keep the HS fields real, keeping in mind a possibility
of applying the method to numerical investigations.

Using Eqs.~(\ref{a37},\ref{a38}) we easily integrate in Eq.~(\ref{a35}) over
the field $\phi _{r\sigma }\left( \tau \right) $ reducing the partition
function $Z$ to the form%
\begin{equation}
Z=Z_{0}\int \exp \left( -\mathcal{S}_{0}\left[ \Psi \right] -\mathcal{S}%
_{int}\left[ \Psi \right] \right) D\Psi  \label{a39}
\end{equation}%
with $\mathcal{S}_{0}\left[ \Psi \right] $ given by Eq.~(\ref{a36}) and
\begin{equation}
\mathcal{S}_{int}\left[ \Psi \right] =\mathcal{S}_{2}\left[ \Psi \right] +%
\mathcal{S}_{3}\left[ \Psi \right] +\mathcal{S}_{4}\left[ \Psi \right] .
\label{a40}
\end{equation}

The term $\mathcal{S}_{int}\left[ \Psi \right] $ in the action describes the
interaction between the fields $\Psi $. The terms $\mathcal{S}_{2}\left[
\Psi \right] $, $\mathcal{S}_{3}\left[ \Psi \right] $ and $\mathcal{S}_{4}%
\left[ \Psi \right] $ contain quadratic, cubic and quartic in $\Psi $ terms,
respectively. They can be written in the form,
\begin{subequations}
\begin{align}
& \mathcal{S}_{2}=\frac{1}{2}\int \sum_{r,r_{1}}\left\{ i\Psi
_{r,r}(R)\theta ^{\ast }-[\Psi \left( R\right) ,\hat{n}]_{rr}u\theta
\right\} U_{r,r_{1}}\left( R,R_{1}\right)  \notag \\
& \quad \times \left\{ i\Psi _{r_{1},r_{1}}(R_{1})\theta _{1}^{\ast }-[\Psi
\left( R_{1}\right) ,\hat{n}]_{r_{1},r_{1}}u_{1}\theta _{1}\right\} \
dRdR_{1},  \label{a41} \\
& \mathcal{S}_{3}=\int \sum_{r,r^{\prime },r_{1}}\Psi _{r^{\prime
},r}(R)\Psi _{r,r^{\prime }}(R)U_{r,r_{1}}(R,R_{1})  \notag \\
& \quad \times \left\{ i\Psi _{r_{1},r_{1}}(R_{1})\theta _{1}^{\ast }-[\Psi
\left( R_{1}\right) ,\hat{n}]_{r_{1},r_{1}}u_{1}\theta _{1}\right\} \
udRdR_{1},  \label{a42} \\
& \mathcal{S}_{4}=\frac{1}{2}\int \sum_{r,r^{\prime },r_{1},r_{1^{\prime
}}}\Psi _{r^{\prime },r}(R)\Psi _{r,r^{\prime }}(R)U_{r,r_{1}}(R,R_{1})
\notag \\
& \quad \times \Psi _{r_{1}^{\prime },r_{1}}(R_{1})\Psi
_{r_{1},r_{1}^{\prime }}(R_{1})u_{1}udRdR_{1}  \label{a43}
\end{align}%
where $\left[ ,\right] $ stands for the commutator. Integration over $R$ in
Eq.~(\ref{a41}-\ref{a43}) implies summation over $\sigma $ and integration
over $u,\tau ,\theta ,\theta ^{\ast }$. The bare action $\mathcal{S}_{0}$
and the interaction term~$\mathcal{S}_{4}$, being invariant under the
transformation of the fields $\Psi $
\end{subequations}
\begin{equation}
\Psi _{r,r^{\prime }}\left( \theta ,\theta ^{\ast }\right) \rightarrow \Psi
_{r,r^{\prime }}\left( \theta +\kappa ,\theta ^{\ast }+\kappa ^{\ast }\right)
\label{aSUSY}
\end{equation}%
($\kappa $ and $\kappa ^{\ast }$ being anticommuting variables) are fully
supersymmetric in the sense of Ref.~\onlinecite{justin}, whereas the terms $%
\mathcal{S}_{2}$ and $\mathcal{S}_{3}$ break this invariance. The invariance
under the transformation (\ref{aSUSY}) is stronger than the standard BRST
symmetry for stochastic field equations [invariance under the transformation
$\Psi \left( \theta ^{\ast }\right) \rightarrow \Psi \left( \theta ^{\ast
}+\kappa ^{\ast }\right) $]\cite{justin}, and reflects additional symmetries
of Eqs.~(\ref{a36},\ref{a43}).

Eqs.~(\ref{a39}-\ref{a43}) completely determine the new bosonic superfield
theory. This model can be studied using standard methods of field theory.
One can, e.g., expand in the interaction $U$ using the Wick theorem with
simple contraction rules following from the form of the bare action $%
\mathcal{S}_{0}$. In the next subsection, we will demonstrate how such
calculations can be carried out explicitly but now let us understand what
one obtains neglecting the cubic and quartic in $\Psi $ terms $\mathcal{S}%
_{3}$ and $\mathcal{S}_{4}$ in the action. In this approximation, one has a
purely quadratic action and, making the Fourier transform with respect to
time and space, we obtain easily for the partition function~$Z$ an RPA-like
expression,
\begin{eqnarray}
&&Z\simeq Z_{0}\exp \left[ -\frac{T}{2}\sum_{\omega }\int \frac{d^{d}\mathbf{%
k}}{\left( 2\pi \right) ^{d}}\ln K\right] ,  \label{aRPA} \\
&&K=1+V_{0}\int \frac{n_{\mathbf{p-k/}2}-n_{\mathbf{p+k/}2}}{i\omega
+\varepsilon _{\mathbf{p-k}/2}-\varepsilon _{\mathbf{p+k}/2}}\frac{d^{d}%
\mathbf{p}}{\left( 2\pi \right) ^{d}}.  \notag
\end{eqnarray}%
The same result can be obtained using Eqs.~(\ref{a10},\ref{a24}, \ref{a25})
and neglecting the field~$\phi _{r\sigma }(\tau )$ in the L.H.S. of Eq.~(\ref%
{a25}).

In Eq.~(\ref{aRPA}), $\left( K-1\right) $ is the contribution of
non-interacting bosonic excitations. Considering their interaction, one can
fully describe the initial fermionic system.

We note that the boson model, as formulated in this paper, does not appear
convenient for a standard analytical approach such as renormalization group
procedures, where the low lying modes should have been singled out in the
very beginning. However, the superfield model Eq.~(\ref{a40}) and a proper
low-energy formulation are formally quite similar. Therefore, this Section
does not only serve as check of the superfield model but also as a guide for
future analytical studies of low energy effects.

In the next subsections, we demonstrate how the contributions of the
conventional perturbation theory\cite{agd} in the original fermion language
are reproduced in the boson superfield representation. We perform this check
calculating the perturbation series for the thermodynamic potential~$\Omega $
up to the second order in the interaction.

\subsection{Wick theorem and diagrams}

In order to develop the perturbation theory in the interaction, we Fourier
transform the superfields
\begin{equation*}
\Psi _{rr^{\prime }}(R)=T\sum_{\omega }\int \Psi _{\mathbf{p}\mathbf{p}%
^{\prime },\omega }(\rho )e^{i(\mathbf{p}r-\mathbf{p}^{\prime }r^{\prime
})-i\omega \tau }\left( d\mathbf{p}d\mathbf{p}^{\prime }\right)
\end{equation*}%
where $\rho =(\sigma ,u,\theta ,\theta ^{\ast })$ and $\omega $ stands for
the bosonic Matsubara frequencies. The momentum integration~$(d\mathbf{p})=d%
\mathbf{p}/(2\pi )^{d}$ extends over the first Brillouin zone.

In the Fourier representation, the bare action reads
\begin{align}
\mathcal{S}_{0}& =\frac{1}{2i}T\sum_{\omega }\int \Psi _{\mathbf{p}^{\prime }%
\mathbf{p},-\omega }(\rho )\big\{i\omega -[\varepsilon _{\mathbf{p}%
}-\varepsilon _{\mathbf{p}^{\prime }}]\big\}\Psi _{\mathbf{p}\mathbf{p}%
^{\prime },\omega }(\rho )  \notag \\
& \qquad \qquad \times (d\mathbf{p}d\mathbf{p}^{\prime })d\rho  \label{b1}
\end{align}%
and similarly one can obtain the Fourier transform of the interaction, Eqs.~(%
\ref{a41}-\ref{a43}).

According to Eq.~(\ref{b1}), the bare propagator of the bosonized theory is
given by
\begin{align}
& \left\langle \Psi _{\mathbf{p},\mathbf{p}^{\prime };\omega }(\rho )\Psi _{%
\mathbf{p}_{1},\mathbf{p}_{1}^{\prime };\omega _{1}}(\rho _{1})\right\rangle
=(\theta -\theta _{1})(\theta ^{\ast }-\theta _{1}^{\ast })\delta (u-u_{1})
\notag \\
& \times \delta _{\sigma \sigma _{1}}(2\pi )^{2d}\delta (\mathbf{p}-\mathbf{p%
}_{1}^{\prime })\delta (\mathbf{p}^{\prime }-\mathbf{p}_{1})\frac{\delta
_{\omega ,-\omega _{1}}}{T}g(\omega ;\mathbf{p},\mathbf{p}^{\prime })
\label{bWick}
\end{align}%
wherein, $g(\omega ;\mathbf{p},\mathbf{p}^{\prime })$ is the Green function
in the momentum representation
\begin{equation*}
g(\omega ;\mathbf{p},\mathbf{p}^{\prime })=i\left[ i\omega -\big(\varepsilon
_{\mathbf{p}}-\varepsilon _{\mathbf{p}^{\prime }}\big)\right] ^{-1}.
\end{equation*}%
Using Wick's theorem and the pairing rule, Eq.~(\ref{bWick}), one can
calculate averages of $\Psi $-products arising in perturbation theory
expansions analytically in a standard manner.

For a convenient diagrammatical representation, the propagator can be
depicted in the form of an oriented double-line as shown in Fig.~\ref%
{fig:buildingblocks}~(a), the first momentum ~$\mathbf{p}$ being carried by
the right-facing upper line, the second momentum~$\mathbf{p}^{\prime }$ by
the left-facing lower line.

\begin{figure}[t]
\includegraphics[width = \linewidth]{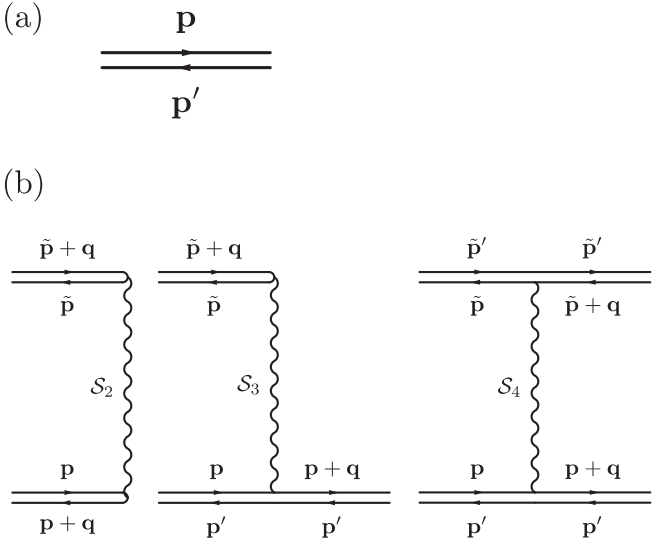}
\caption{Diagrammatic building blocks for perturbative calculations: (a)~the
boson propagator~$g(\protect\omega ;\mathbf{p},\mathbf{p}^{\prime })$, Eq.~(%
\protect\ref{bWick}), (b)~the interaction terms, Eqs.~(\protect\ref{a41}-%
\protect\ref{a43}).}
\label{fig:buildingblocks}
\end{figure}

In the same spirit, the interaction vertices, Eqs.~(\ref{a41}-\ref{a43}),
are diagrammatically represented as shown in Fig.~\ref{fig:buildingblocks}%
~(b). Depending on whether a leg is depicted as \textquotedblleft
out-going\textquotedblright\ [right-facing] or \textquotedblleft
incoming\textquotedblright\ [left-facing], the momenta and frequencies are
either reversed [momenta~$\mathbf{p}$ and~$\mathbf{p}^{\prime }$ are
interchanged and frequency~$\omega $ changes the sign] or not. The
interaction lines carrying the momentum~$\mathbf{q}$ correspond to the
propagator
\begin{equation}
U_{\mathbf{q},\omega }(\sigma ,\sigma _{1})=\sigma \sigma _{1}V_{0}+V_{%
\mathbf{q}}^{(1)}  \label{b1a}
\end{equation}%
and are attached according to the analytical structure of Eqs.~(\ref{a41}-%
\ref{a43}). In this way, the single lines within the double-line structure
obey evident momentum conservation rules, whereas the bosonic frequency of
each double-line is conserved in the usual way and, therefore, not labelled
in the diagram.

With the rules formulated, we are now ready to discuss the perturbation
theory for the thermodynamic potential~$\Omega $.

\subsection{First order of the perturbation theory}

As~$\mathcal{S}_{4}$ does not break the symmetry specified by Eq.~(\ref%
{aSUSY}), the first order contribution to~$\Omega $ in the superfield theory
comes only from $\mathcal{S}_{2}$. However, $T\langle \mathcal{S}_{2}\rangle
$ is not the only contribution to the first order correction~$\Delta \Omega
^{(1)}$ in the interaction potential. In addition, one should renormalize
the chemical potential~thus using $\mu ^{\prime }$, Eq.~(\ref{a14}), and
shift the thermodynamic potential~$\Omega \rightarrow \Omega +\left(
2n\right) ^{2}V^{\left( 1\right) }N_{d}/2$ in accordance with Eq.~(\ref{a13}%
). This leads to a trivial first order contribution to the thermodynamic
potential~$\Omega $ that can be written as
\begin{eqnarray}
\delta \Omega ^{(1)} &=&N_{d}\big[2n^{2}V^{\left( 1\right) }  \label{b2} \\
&&+\left( V_{r,r}+2\bar{V}\right) n-4n^{2}V^{\left( 1\right) }\big]  \notag
\end{eqnarray}%
where $V^{\left( 1\right) }=-\sum_{r^{\prime }}\tilde{V}_{r,r^{\prime }}+%
\bar{V}$ and $n$ is the (unperturbed) fermion density per one spin direction
at site~$r$. The first term in Eq.~(\ref{b2}) comes from the shift of the
thermodynamic potential, Eq.~(\ref{a13}). The second line originates from
the shift of the chemical potential, Eq.~(\ref{a14}), and is obtained using
the standard relation $\partial \Omega /\partial \mu =-2nN_{d}$ (factor $2$
is due to spin). So, the total first order contribution $\Delta \Omega
^{(1)} $ to the thermodynamic potential takes the form
\begin{equation}
\Delta \Omega ^{(1)}=T\langle \mathcal{S}_{2}\rangle +\delta \Omega ^{(1)}.
\label{b3}
\end{equation}%
\begin{figure}[t]
\includegraphics{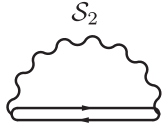}
\caption{First order diagram from~$\langle \mathcal{S}_{2}\rangle $.}
\label{bfig1}
\end{figure}
Diagrammatically, $T\langle \mathcal{S}_{2}\rangle $ is represented in Fig.~%
\ref{bfig1}. The correspondence of Fig.~\ref{bfig1} to the conventional
diagrammatic technique becomes evident if we associate the double-line with
two single lines representing the original fermionic propagators. Then, one
can easily imagine that the diagram in Fig. \ref{bfig1} should yield the
Fock contribution [Fig.~\ref{reg}~(a)]. Actually, this is not completely
correct because this diagram contains also some other terms. Calculating $%
T\langle \mathcal{S}_{2}\rangle $ with the help of the contraction rule (\ref%
{bWick}) and Eq.~(\ref{b1a}) we obtain in the first order
\begin{align}
& T\left\langle \mathcal{S}_{2}\right\rangle =T^{2}\sum_{\omega }\int
\left\langle \Psi _{\mathbf{p},\mathbf{p}+\mathbf{q};-\omega }(\rho )\Psi _{%
\mathbf{p}^{\prime }+\mathbf{q},\mathbf{p}^{\prime };\omega }(\rho
_{1})\right\rangle  \notag \\
& \;\times (-i)u_{1}U_{\mathbf{q};\omega }(\sigma ,\sigma _{1})\theta ^{\ast
}\theta _{1}[n_{\mathbf{p}^{\prime }+\mathbf{q}}^{0}-n_{\mathbf{p}^{\prime
}}^{0}]\ d\rho d\rho _{1}(d\mathbf{p}d\mathbf{p}^{\prime }d\mathbf{q})
\notag \\
=& -N_{d}T\sum_{\omega }\int \frac{n_{\mathbf{p}+\mathbf{q}}^{0}-n_{\mathbf{p%
}}^{0}}{i\omega -\varepsilon _{\mathbf{p}}+\varepsilon _{\mathbf{p}+\mathbf{q%
}}}\left[ V_{\mathbf{q}}-2V_{0}\right] (d\mathbf{p}d\mathbf{q})  \label{b3a}
\end{align}%
where $n_{\mathbf{p}}^{0}$ is the bare Fermi distribution.

We can represent Eq.~(\ref{b3a}) in a somewhat different form using the
identity
\begin{eqnarray}
&&T\sum_{\omega }\frac{n_{\mathbf{p}^{\prime }}^{0}-n_{\mathbf{p}}^{0}}{%
i\omega -\varepsilon _{\mathbf{p}}+\varepsilon _{\mathbf{p}^{\prime }}}=%
\frac{1}{2}\big(n_{\mathbf{p}^{\prime }}^{0}-n_{\mathbf{p}}^{0}\big)\coth
\frac{\varepsilon _{\mathbf{p}^{\prime }}-\varepsilon _{\mathbf{p}}}{2T}
\notag \\
&=&n_{\mathbf{p}}^{0}n_{\mathbf{p}^{\prime }}^{0}-\frac{1}{2}\big(n_{\mathbf{%
p}}^{0}+n_{\mathbf{p}^{\prime }}^{0}\big).  \label{b4}
\end{eqnarray}

Substituting Eq.~(\ref{b4}) into Eq.~(\ref{b3a}), we see that the term with $%
V_{\mathbf{q}}$ in Eq.~(\ref{b3a}), indeed, participates in forming the Fock
contribution $\Delta \Omega _{F}^{\left( 1\right) }$ to the thermodynamic
potential%
\begin{equation}
\Delta \Omega _{F}^{\left( 1\right) }=-N_{d}\int V_{\mathbf{q}}n_{\mathbf{p}%
}^{0}n_{\mathbf{p+q}}^{0}(d\mathbf{p}d\mathbf{q})  \label{b4a}
\end{equation}

However, there are other contributions $\delta ^{\prime }\Omega ^{\left(
1\right) }$ in Eq.~(\ref{b3a}) that should be added to that of Eq.~(\ref{b2})%
\begin{eqnarray}
&&\delta ^{\prime }\Omega ^{\left( 1\right) } =N_{d}\int \big[2n_{\mathbf{p}%
}^{0}n_{\mathbf{p+q}}^{0}\left( V_{r,r}+\bar{V}\right)  \notag \\
&&+\left( n_{\mathbf{p}}^{0}+n_{\mathbf{p+q}}^{0}\right) \left( V_{\mathbf{q}%
}/2-V_{r,r}-\bar{V}\right) \big]\left( d\mathbf{p}d\mathbf{q}\right)
\label{b4c}
\end{eqnarray}

The structure of the integrand in Eq.~(\ref{b4c}) allows one easily
integrate over $\mathbf{p}$ and $\mathbf{q}$. Then, adding Eqs.~(\ref{b2}, %
\ref{b4c}) to each other and using Eqs.~(\ref{a4a}-\ref{a4c}) we obtain
\begin{equation}
\Delta \Omega _{H}^{\left( 1\right) }=\delta \Omega ^{\left( 1\right)
}+\delta ^{\prime }\Omega ^{\left( 1\right) }=2N_{d}n^{2}V_{\mathbf{q}=0}
\label{b4d}
\end{equation}%
which is the standard Hartree contribution. The artificial interaction $\bar{%
V}$ introduced in the model drops out from the final formulas as it should.

Thus, we come in the first order to the standard expression for the
correction $\Delta \Omega ^{\left( 1\right) }$ to the thermodynamic potential%
\begin{equation}
\Delta \Omega ^{\left( 1\right) }=\Delta \Omega _{H}^{\left( 1\right)
}+\Delta \Omega _{F}^{\left( 1\right) }  \label{b4e}
\end{equation}
where $\Delta \Omega _{H}^{\left( 1\right) }$ and $\Delta \Omega
_{F}^{\left( 1\right) }$ are the Hartree, Eq.~(\ref{b4d}), and Fock, Eq.~(%
\ref{b4a}), contributions, respectively.

\subsection{Second order of the perturbation theory}

\begin{figure*}[tbp]
\includegraphics{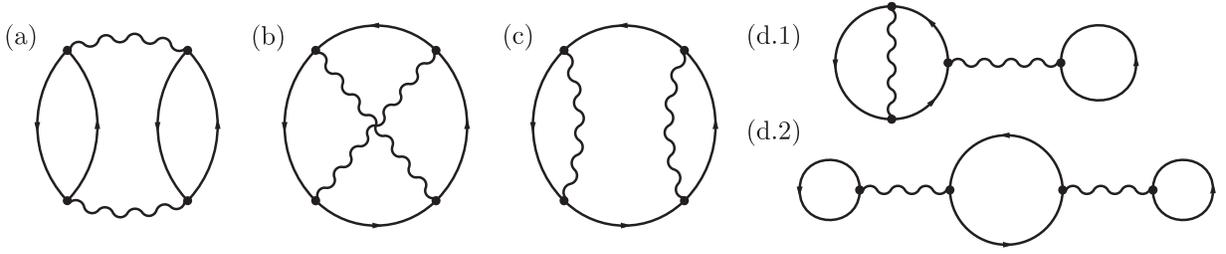}
\caption{Diagrams for the second order in the fermion language. Interesting
low-energy contributions are only due to diagrams~(a) and~(b) whereas
diagram~(c) just renormalize the chemical potential. To check the
equivalence of the bosonized model to the original fermion one, we must
check, however, that all fermion diagrammatic contributions are exactly
reproduced.}
\label{fig:conv2nd}
\end{figure*}
The perturbation expansion in the conventional field theory for interacting
fermions leads to the diagrams shown in~Fig.~\ref{fig:conv2nd}. Only the
diagrams in Fig.~\ref{fig:conv2nd}~(a) and~(b) are important for low-energy
physics, whereas the diagram~(c) merely modifies the chemical potential. The
diagrams~\ref{fig:conv2nd} (d.1) and~(d.2) contain the Hartree bubbles that
also result in a renormalization of the chemical potential.

Nevertheless, in order to understand details it is important to obtain \emph{%
all} contributions of the perturbation theory for the boson superfield
theory. Checking the exact correspondence between the original fermion and
the boson models is what this Section is devoted to.

\begin{figure*}[tbp]
\includegraphics{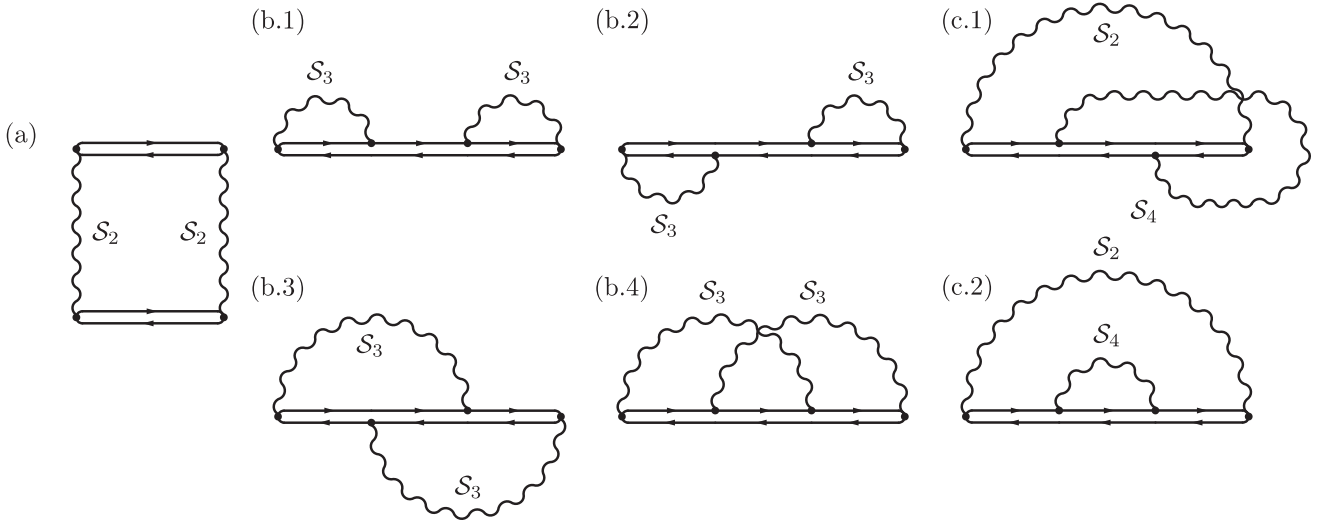}
\caption{Second order diagrams describing perturbation series of the
superfield theory. Contributions come from $\langle {\mathcal{S}_{2}}%
^{2}\rangle $ [(a)], $\langle {\mathcal{S}_{3}}^{2}\rangle $ [(b.1)-(b.4)],
and $\langle \mathcal{S}_{2}\mathcal{S}_{4}\rangle $ [(c.1),(c.2)].}
\label{fig:2nd}
\end{figure*}

Before presenting the details of the calculations, let us make some general
remarks. For~$\hat{H}_{int}^{(1)}$ [Eq.~(\ref{a5})], the Hartree-type
contributions \ref{fig:conv2nd}~(d.1) and~\ref{fig:conv2nd} (d.2) are in the
boson theory generally accounted for by the renormalization of the chemical
potential [Eq.~(\ref{a14})] and by the trivial contribution to the
thermodynamic potential in Eq.~(\ref{a13}). There is no contribution of
diagrams containing closed loops of boson propagators because they vanish
due to the superfield symmetry. The effective second order diagrams in the
bosonized theory are shown in Fig.~\ref{fig:2nd}.

Drawing parallels with the conventional fermion diagrams we interpret the
double-lines in Fig.~\ref{fig:2nd} as pairs of single fermion propagator
lines. This is similar to what we successfully did when discussing the first
order diagrams. In this spirit, inspection of the diagrams suggests that the
conventional RPA diagram Fig.~\ref{fig:conv2nd}~(a) is reproduced in the
boson language by Fig.~\ref{fig:2nd}~(a), the correlation diagram Fig.~\ref%
{fig:conv2nd}~(b) by Figs.~\ref{fig:2nd}~(b.4,c.1), and, finally, the
conventional diagram with the two Fock loops, Fig.~\ref{fig:conv2nd}~(c), by
Figs.~\ref{fig:2nd}~(b.1-b.3,c.2). However, for contributions coming from
the on-site interaction Hamiltonian~$\hat{H}_{int}^{(0)}$ [Eq.~(\ref{a5})],
this \textquotedblleft graphical picture\textquotedblright\ describes the
correspondence in a less strict way since the conventional diagrams~(a)
and~(b) as well as the conventional diagrams~(c), (d.1), and (d.2) in Fig.~%
\ref{fig:conv2nd} coincide in this case, thus making a parallel between just
Fig.~\ref{fig:conv2nd}~(a) and Fig.~\ref{fig:2nd}~(a) meaningless. In this
case one can speak only about a \textquotedblleft weaker\textquotedblright\
correspondence between Figs.~\ref{fig:conv2nd}~(a,b) and Figs.~\ref{fig:2nd}%
~(a,b.4,c.1,c.2).

Having established these parallels, the detailed calculations can be
performed in an organized manner, which is straightforward though a little
tedious.

Starting with $\langle {\mathcal{S}_{2}}^{2}\rangle $, the evaluation of the
diagram in Figs.~\ref{fig:2nd}~(a) yields
\begin{subequations}
\begin{align}
\langle {\mathcal{S}_{2}}^{2}\rangle & =-2N_{d}\sum_{\omega }\int \left( V_{%
\mathbf{q}}^{2}-2V_{0}V_{\mathbf{q}}+2V_{0}^{2}\right) [n_{\mathbf{p}+%
\mathbf{q}}-n_{\mathbf{p}}]  \notag \\
\times \;& [n_{\mathbf{p}^{\prime }+\mathbf{q}}-n_{\mathbf{p}^{\prime
}}]g(\omega ,\mathbf{p}+\mathbf{q},\mathbf{p})g(\omega ,\mathbf{p}^{\prime }+%
\mathbf{q},\mathbf{p}^{\prime })(d\mathbf{p}d\mathbf{p}^{\prime }d\mathbf{q}%
),  \notag \\
& =\ 2N_{d}T^{2}\sum_{\varepsilon \varepsilon ^{\prime }\omega }\int V_{%
\mathbf{q}}^{2}G_{\varepsilon ,\mathbf{p}+\mathbf{q}}^{0}G_{\varepsilon
+\omega ,\mathbf{p}}^{0}  \notag \\
\times \ & G_{\varepsilon ^{\prime },\mathbf{p}^{\prime }+\mathbf{q}%
}^{0}G_{\varepsilon ^{\prime }+\omega ,\mathbf{p}^{\prime }}^{0}(d\mathbf{p}d%
\mathbf{p}^{\prime }d\mathbf{q})  \label{b5a} \\
& +\ 2N_{d}T^{2}\sum_{\varepsilon \omega \omega ^{\prime }}\int (-2V_{0}V_{%
\mathbf{q}}+2V_{0}^{2})G_{\varepsilon ,\mathbf{p}}^{0}G_{\varepsilon +\omega
,\mathbf{p}+\mathbf{q}}^{0}  \notag \\
\times \ & G_{\varepsilon +\omega +\omega ^{\prime },\mathbf{p}+\mathbf{q}+%
\mathbf{q}^{\prime }}^{0}G_{\varepsilon +\omega ^{\prime },\mathbf{p}+%
\mathbf{q}^{\prime }}^{0}(d\mathbf{p}d\mathbf{q}d\mathbf{q}^{\prime })
\label{b5b}
\end{align}%
where Eq.~(\ref{a25bc}) has been applied and the notation $G_{\varepsilon ,%
\mathbf{p}}^{0}=(i\varepsilon -\xi _{\mathbf{p}})^{-1}$ for the fermion
Green function in the original electron language is used, $\xi _{\mathbf{p}%
}=\varepsilon _{\mathbf{p}}-\mu $. In the second term (\ref{b5b}), the
momenta in the Green functions have been shifted using the fact that at
least one of the involved interaction couplings~$V$ is independent of the
running momentum $\mathbf{q}$ (on-site interaction).

Inspection of the expressions obtained shows that the first term~(\ref{b5a})
is indeed identical to the conventional second order RPA contribution, Fig.~%
\ref{fig:conv2nd}~(a). This one-to-one correspondence of the $V_{\mathbf{q}%
}^{2}$-term confirms the validity of the \textquotedblleft graphical
correspondence\textquotedblright . The form of the second term resembles the
contribution of the diagram Fig.~\ref{fig:conv2nd}~(b) and therefore we
expect that it should be cancelled by those diagrams~of $\langle {\mathcal{S}%
_{3}}^{2}\rangle $ and $\langle \mathcal{S}_{2}\mathcal{S}_{4}\rangle $ that
we have already related to the conventional contribution of Fig.~\ref%
{fig:conv2nd}~(b). Since for the purely on-site interaction the diagrams
Fig.~\ref{fig:conv2nd}~(a) and~(b) are indistinguishable, there is no
surprise that $\langle {\mathcal{S}_{2}}^{2}\rangle $ also contributes to
the conventional diagram in Fig.~\ref{fig:conv2nd}~(b) due to the presence
of the $\hat{H}_{int}^{(0)}$-channel, Eq.~(\ref{a5}).

So, let us consider now the boson diagrams shown in Figs.~\ref{fig:2nd}%
~(b.4,c.1) which are of conventional type of Fig.~\ref{fig:conv2nd}~(b) in
terms of their graphical shape. Their evaluation yields
\end{subequations}
\begin{align}
& \langle \mathcal{S}_{int}^{2}\rangle ^{(\mathrm{b}.4,\mathrm{c}%
.1)}=\langle {\mathcal{S}_{3}}^{2}\rangle ^{(\mathrm{b}.4)}+\langle \mathcal{%
S}_{2}\mathcal{S}_{4}\rangle ^{(\mathrm{c}.1)}  \notag \\
=& -4N_{d}T\sum_{\omega \omega ^{\prime }}\int (2V_{0}-V_{\mathbf{q}%
})(2V_{0}-V_{\mathbf{q}^{\prime }})  \notag \\
& \times \ in_{\mathbf{p}}\ g(\omega +\omega ^{\prime },\mathbf{p}+\mathbf{q}%
+\mathbf{q}^{\prime },\mathbf{p})  \notag \\
& \times g(\omega ,\mathbf{p}+\mathbf{q}^{\prime },\mathbf{p})g(\omega
^{\prime },\mathbf{p}+\mathbf{q},\mathbf{p})\ (d\mathbf{p}d\mathbf{q}d%
\mathbf{q}^{\prime }).  \label{b5c}
\end{align}%
Recasting Eq.~(\ref{b5c}) into the original fermion language, we can
equivalently write
\begin{align}
& \langle \mathcal{S}_{int}^{2}\rangle ^{(\mathrm{b}.4,\mathrm{c}%
.1)}=-N_{d}T^{2}\sum_{\varepsilon \omega \omega ^{\prime }}\int
(4V_{0}^{2}-4V_{0}V_{\mathbf{q}}+V_{\mathbf{q}}V_{\mathbf{q}^{\prime }})
\notag \\
& \times G_{\varepsilon ,\mathbf{p}}^{0}G_{\varepsilon +\omega ,\mathbf{p}+%
\mathbf{q}}^{0}G_{\varepsilon +\omega ^{\prime },\mathbf{p}+\mathbf{q}%
^{\prime }}^{0}G_{\varepsilon +\omega +\omega ^{\prime },\mathbf{p}+\mathbf{q%
}+\mathbf{q}^{\prime }}^{0}\ (d\mathbf{p}d\mathbf{q}d\mathbf{q}^{\prime }).
\label{b6}
\end{align}

Inspecting Eq.~(\ref{b6}), we see that the term with $4V_{0}^{2}-4V_{0}V_{%
\mathbf{q}}$ exactly cancels the contribution (\ref{b5b}), while the term
with~$V_{\mathbf{q}}V_{\mathbf{q}^{\prime }}$ in Eq.~(\ref{b6}) is exactly
the contribution originating from the conventional fermionic diagram in Fig.~%
\ref{fig:conv2nd}~(b). Thus, we conclude that we have verified the
one-to-one correspondence of the conventional diagram family given by Fig.~%
\ref{fig:conv2nd}~(a,b) and their graphical counterparts in the boson
diagrammatics.

Now we address the remaining second order diagrams. The diagrams of Figs.~%
\ref{fig:2nd}~(b.1) and~(b.2) involve a formal propagator $g(0;\mathbf{p},%
\mathbf{p})\sim 1/0$ and we need a regularization scheme. This badly defined
boson propagator appears as a counterpart to the conventional diagrammatic
structure of two fermion lines which carry identical frequency and momentum,
cf. the discussion in Section~\ref{subsecIIb}. However, as has been
discussed previously, this irregular expression is healed as soon as the
interaction Hamiltonian is no longer translationally invariant.

As the main purpose of this Section is to check the perturbation series, it
is sufficient to perform the regularization by smearing the $\delta $%
-function in the interaction matrix element~$V_{\mathbf{q}}\delta _{\mathbf{q%
}\mathbf{q}^{\prime }}$. By this procedure, the momenta and frequencies in
the conventional diagram Fig.~\ref{fig:conv2nd}~(b) become in general
mutually different, while the central propagator in the boson diagrams in
Figs.~\ref{fig:2nd}~(b.1) and~(b.2) has two generally different momentum
arguments, e.g. $g(0;\mathbf{p},\mathbf{p}+\mathbf{q}^{\prime }-\mathbf{q})$%
. The smearing of the $\delta $-function is achieved by breaking the
translational symmetry of the interaction introducing the bath as it has
been done in the preceding Section.

Writing the proper analytical expressions we obtain a regularized boson
propagator $g(0;\mathbf{p},\mathbf{p}+\mathbf{q}^{\prime }-\mathbf{q})\simeq
1/\Delta \varepsilon $, where $\varepsilon _{\mathbf{p}+\mathbf{q}^{\prime }-%
\mathbf{q}}\simeq \varepsilon _{\mathbf{p}}+\Delta \varepsilon $ with small $%
\Delta \varepsilon $. Then, we should expand in $\Delta \varepsilon $ all
other terms. As the original fermionic contributions are not singular in the
limit $\Delta \varepsilon \rightarrow 0$, we expect that the small parameter
$\Delta \varepsilon $ should be compensated.

Calculating the diagrams in Fig.~\ref{fig:2nd}~(b.1) and~(b.2) with the help
of this regularization scheme, then evaluating the remaining diagrams~(b.3)
and~(c.2) that are regular by themselves and, finally, combining all these
contributions we come to the expression
\begin{align}
\langle \mathcal{S}_{int}^{2}& \rangle ^{(\mathrm{b}.1-3,\mathrm{c}%
.2)}=-2N_{d}T^{2}\sum_{\omega \omega ^{\prime }}\int (2V_{0}-V_{\mathbf{q}%
})(2V_{0}-V_{\mathbf{q}^{\prime }})  \notag \\
\times \Big\{& i\,n_{\mathbf{p}+\mathbf{q}}\,g^{2}(\omega ;\mathbf{p},%
\mathbf{p}+\mathbf{q})g(\omega ^{\prime };\mathbf{p}+\mathbf{q}^{\prime },%
\mathbf{p}+\mathbf{q})  \notag \\
+& i\,n_{\mathbf{p}+\mathbf{q}^{\prime }}\,g^{2}(\omega ^{\prime };\mathbf{p}%
,\mathbf{p}+\mathbf{q}^{\prime })g(\omega ;\mathbf{p}+\mathbf{q},\mathbf{p}+%
\mathbf{q}^{\prime })  \notag \\
-& i\,n_{\mathbf{p}}\,g^{2}(\omega ;\mathbf{p}+\mathbf{q},\mathbf{p}%
)g(\omega ^{\prime };\mathbf{p}+\mathbf{q}^{\prime },\mathbf{p})  \notag \\
-& i\,n_{\mathbf{p}}\,g(\omega ;\mathbf{p}+\mathbf{q},\mathbf{p}%
)g^{2}(\omega ^{\prime };\mathbf{p}+\mathbf{q}^{\prime },\mathbf{p})  \notag
\\
+& \frac{\partial n_{\mathbf{p}}}{\partial \mu }\,g(\omega ;\mathbf{p}+%
\mathbf{q},\mathbf{p})g(\omega ^{\prime };\mathbf{p}+\mathbf{q}^{\prime },%
\mathbf{p})\Big\}(d\mathbf{p}d\mathbf{q}d\mathbf{q}^{\prime }).
\end{align}%
We sum over the Matsubara frequencies and obtain using transformations
similar to those leading to Eq.~(\ref{b4})
\begin{subequations}
\begin{align}
\langle & \mathcal{S}_{int}^{2}\rangle ^{(\mathrm{b}.1-3,\mathrm{c}.2)}=-%
\frac{N_{d}}{2T}\int (2V_{0}-V_{\mathbf{q}})(2V_{0}-V_{\mathbf{q}^{\prime }})
\notag \\
& \qquad \times n_{\mathbf{p}}(n_{\mathbf{p}}-1)(2n_{\mathbf{p}+\mathbf{q}%
}-1)(2n_{\mathbf{p}+\mathbf{q}^{\prime }}-1)(d\mathbf{p}d\mathbf{q}d\mathbf{q%
}^{\prime })  \notag \\
=& \ 2N_{d}\int (d\mathbf{p}d\mathbf{q}d\mathbf{q}^{\prime })\Big\{V_{%
\mathbf{q}}V_{\mathbf{q}^{\prime }}\frac{\partial n_{\mathbf{p}}}{\partial
\mu }n_{\mathbf{p}+\mathbf{q}}n_{\mathbf{p}+\mathbf{q}^{\prime }}
\label{b8a} \\
& +(4V_{0}^{2}-2V_{0}V_{\mathbf{q}}-2V_{0}V_{\mathbf{q}^{\prime }})\frac{%
\partial n_{\mathbf{p}}}{\partial \mu }n_{\mathbf{p}+\mathbf{q}}n_{\mathbf{p}%
+\mathbf{q}^{\prime }}  \label{b8b} \\
& -\frac{1}{4}(2V_{0}-V_{\mathbf{q}})(2V_{0}-V_{\mathbf{q}^{\prime }})\frac{%
\partial n_{\mathbf{p}}}{\partial \mu }(2n_{\mathbf{p}+\mathbf{q}}+2n_{%
\mathbf{p}+\mathbf{q}^{\prime }}-1).  \label{b8c}
\end{align}

It is clear that all singularities have disappeared. Analyzing the
expressions above we find that the first summand~(\ref{b8a}) equals the
contribution of the conventional diagram Fig.~\ref{fig:conv2nd}~(c). Thus,
we have so far succeeded to show how the boson model reproduces all the
conventional diagrams without the Hartree bubbles.

Finally, we collect all trivial second order contributions arising from the
shift of the chemical potential, Eq.~(\ref{a14}), and the thermodynamic
potential, Eq.~(\ref{a13}). Note that in this procedure, we have to expand
also the Fermi functions in the contribution of the first order diagram,
Eq.~(\ref{b3a}). Combining these trivial second order contributions with the
contributions~(\ref{b8b}) and~(\ref{b8c}) yields the analytical expression
for the collection of the Hartree-type diagrams Fig.~\ref{fig:conv2nd}~(d.1)
and~(d.2). The calculation is completely analogous to the one for the first
order diagrams but is a little bit more tedious.

Thus, the calculations performed in this Section helped us to see that the
boson model yields up to the second order the same perturbation series as
the conventional theory. The perturbative calculations in the boson model
are somewhat more cumbersome than the standard perturbation theory. However,
our motivation for constructing the boson model is to develop a new tool for
studying both analytically and numerically the low temperature physics,
rather than considering lowest order diagrams exactly. The calculations
presented in this Section are merely a check of the bosonic approach.

We note that the regularization based on the breaking of translational
symmetry is crucial only for a class of diagrams which usually do not play
an important role in the framework of analytical investigation of low lying
excitations. Therefore, studying, e.g., the non-analytical corrections to
the specific heat of an interacting Fermi gas \cite{aleiner,chubukov}, the
inherent ambiguity of Eq.~(\ref{a25}) is rather harmless because the
important contributions do not require any additional regularization.
\end{subequations}
\begin{equation*}
\end{equation*}

\section{Discrete time representation and Monte Carlo simulations\label{mc}}

\subsection{Quantum Monte Carlo}

Using exact transformations, we have mapped in the previous Sections the
original fermion model, Eqs.~(\ref{a1}-\ref{a4}), onto the boson model,
Eqs.~(\ref{a102}-\ref{a104}), or Eqs. (\ref{a39}-\ref{a43}). We have shown
how one can treat the latter model diagrammatically making expansion in the
interaction terms and demonstrated in the first two orders in the
interaction that the perturbation theory agrees with the conventional
perturbation theory for the fermions.

In this Section we want to compare the two equivalent models from the point
of view of their suitability for Monte Carlo simulations. It is well known%
\cite{blankenbecler,hirsch,linden,dosantos,troyer} that studying models for
fermions one encounters the negative sign problem and the main advantages of
the MC get lost. At the same time, the sign in the MC sampling is always
positive for bosonic models. Therefore, it is quite natural to investigate
the suitability of the boson model, Eqs.~(\ref{a102}-\ref{a104}) or (\ref%
{a39}-\ref{a43}), to MC simulations and clarify the question about the sign
problem.

Before starting our investigation of the problem we would like to emphasize
that we do not attempt discussing here effectiveness of various MC
algorithms or compare them with each other. Our goal is to understand
semi-quantitatively where the problems come from, how one can overcome them
and how one can do computations in principle. We are aware of the fact that
the concrete schemes of the calculations we suggest at the end may be not
necessarily the most efficient ones, but writing explicit algorithms is
beyond the scope of the present paper.

The MC method is used for computation of the average of a quantity $A$ over
configurations $c$ in a space of these configurations

\begin{equation}
\left\langle A\right\rangle =Z^{-1}\sum_{c}A\left( c\right) p\left( c\right)
,\quad Z=\sum_{c}p\left( c\right)  \label{c1}
\end{equation}%
According to this approach one chooses a set of $M$ configurations $\left\{
c_{i}\right\} $, $i=1,2,...M$ with the probability~$Z^{-1}p\left(
c_{i}\right) $. Then, instead of calculating the entire sum in Eq.~(\ref{c1}%
), one computes the sample average%
\begin{equation}
\left\langle A\right\rangle \approx \frac{1}{M}\sum_{i=1}^{M}A\left(
c_{i}\right)  \label{c2}
\end{equation}%
This approach is very efficient because the time of a numerical computation
grows with the size of the sample in a power law, which contrasts
exponential growth when using other methods.

Of course, such an approximation is possible only provided all $p\left(
c\right) $ are not negative. In some cases the proper $p\left( c\right) $
can be both positive and negative and formally one may not use the
approximation (\ref{c2}). In this case one may use instead, e.g., $%
\left\vert p\left( c\right) \right\vert $ as the probability rewriting Eq.~(%
\ref{c1}) as%
\begin{equation}
\left\langle A\right\rangle =\frac{\sum_{c}A\left( c\right) s\left( c\right)
\left\vert p\left( c\right) \right\vert }{\sum_{c}\left\vert p\left(
c\right) \right\vert }\left( \frac{\sum_{c}s\left( c\right) \left\vert
p\left( c\right) \right\vert }{\sum_{c}\left\vert p\left( c\right)
\right\vert }\right) ^{-1},  \label{c3}
\end{equation}%
where $s\left( c\right) =\mathrm{sign}[p\left( c\right) ].$

One can perform the computations starting with Eq.~(\ref{c3}) but the
average sign $s\left( c\right) $ becomes in many important cases very small,
which leads to exponentially large computation time.

The traditional scheme \cite{blankenbecler,hirsch,linden,dosantos}of the
quantum MC simulations for the fermion models is based on the HS
transformation and using $Z\left[ \phi \right] $, Eq.~(\ref{a18}), as the
probability $p\left( c\right) $ for the MC sampling. The problem arising in
this approach is that for some important configurations of $\phi _{r\sigma
}\left( \tau \right) $ the functional $Z\left[ \phi \right] $ becomes
negative. There are some special cases like models with a half-filled band
or with attraction between the fermions when $Z\left[ \phi \right] $ is
necessarily positive. However, generically, the MC simulations are
inefficient for the fermionic models.

In principle, one could propose just to use the bosonized model, Eqs.~(\ref%
{a39}-\ref{a43}). However, it contains the superfields $\Psi ,$ and one
cannot apply the MC method immediately. This diffiuculty can be overcome by
decoupling the interaction with the help of HS transformation. Of course,
proceeding in this way we would come back to Eqs.~(\ref{a102}- \ref{a104}).
Now we can argue that the solution $A_{r,r^{\prime }}\left( z\right) $ of
Eq.~(\ref{a102}) is necessarily real and substituting it into Eq.~(\ref{a104}%
) one must obtain positive values of $Z\left[ \phi \right] $. We have
explained how one comes to Eqs. (\ref{a102}-\ref{a104}) using the
regularization based on introducing a bath and the small parameter $\gamma $%
. All the transformations leading to (\ref{a102}-\ref{a104}) were justified
and therefore one may enjoy computing using the scheme free of the sign
problem.

We think that this way of reasoning is correct, but it is definitely
somewhat artificial: one introduces superfields, integrates over the HS
fields but then returns to the equations (\ref{a102}-\ref{a104}). After
making these transformations the sign problem desappears but why this has
happened is not clear.

So, it is very instructive to try to understand in a more direct way how the
sign problem is overcome, what are the approximations made, etc. Of course,
one should discuss these things without using the superfields.

This is what the next subsections are devoted to.

\subsection{Discrete time representation\label{discrete}}

We start the discussion with a slight extension of the original fermionic
model, Eqs.~(\ref{a1}-\ref{a4}). The interaction between the electrons $%
V_{r,r^{\prime }}$, Eq.~(\ref{a3}), does not depend on time. As a result,
fluctuations of the HS field $\phi _{r\sigma }\left( \tau \right) $ are not
correlated in time, Eq.~(\ref{a38}). The function $\delta \left( \tau -\tau
^{\prime }\right) $ describing these correlations does not create
difficulties in analytic calculations but it is more convenient to slightly
smear it for our dicussion. So, we change the model replacing the weight $%
W_{0}\left( \phi \right) $ in Eq.~(\ref{a12}) by
\begin{equation}
W_{\alpha }\left[ \phi \right] =\exp \Big[-\frac{1}{2V_{0}}%
\sum_{r}\int_{0}^{\beta }\Big[\phi _{r}^{2}\left( \tau \right) +\alpha ^{2}%
\Big(\frac{\partial \phi _{r}\left( \tau \right) }{\partial \tau }\Big)^{2}%
\Big]d\tau \Big]  \label{c4}
\end{equation}%
To simplify the discussion we neglect the interaction $V^{\left( 1\right) }$
in Eq.~(\ref{a5}) thus restricting ourselves with the on-site repulsion.
According to the accepted regularization with the help of the bath we assume
that the interaction is equal to $V_{0}$ inside the sphere with the radius $%
R_{0}$ but vanishes outside this region (c.f. Eq.~(\ref{a100})). We are
interested in the limit $\alpha \rightarrow 0$ but let us take this limit at
the end of the calculations. We do not see any reason why this limit should
lead to results different from those for the original model.

The presence of the derivative in Eq.~(\ref{c4}) introduces correlations
between $\phi _{r}\left( \tau \right) $ at different times. We are mostly
interested in the low temperature limit when the sign problem is strongest.
In this limit, the pair correlation function between the fields $\phi
_{r}\left( \tau \right) $ inside the interacting region takes a simple form%
\begin{equation}
\left\langle \phi _{r}\left( \tau \right) \phi _{r^{\prime }}\left( \tau
^{\prime }\right) \right\rangle =V_{0}\delta _{r,r^{\prime }}\frac{\exp
\left( -\left\vert \tau -\tau ^{\prime }\right\vert /\alpha \right) }{%
2\alpha }  \label{c5}
\end{equation}%
showing the time correlations on the scale $\alpha $.

\bigskip

In order to proceed with the MC computations, one should discretize the time
and basic equations, which can be done in many different ways. The standard
procedure is to subdivide the interval $\left( 0,\beta \right) $ into slices
of the length $\Delta =\beta /N\ll 1$ and consider discrete times $\tau
_{l}=\Delta \left( l-1/2\right) $. Within this scheme one writes the
partition function $Z$ in the form%
\begin{equation}
Z=\lim_{\Delta \rightarrow 0}\frac{\int Z^{\prime }\left[ \phi \right]
W_{\alpha }^{\prime }\left[ \phi \right] D\phi }{\int W_{\alpha }^{\prime }%
\left[ \phi \right] D\phi },  \label{c6}
\end{equation}%
where
\begin{equation*}
Z^{\prime }\left[ \phi _{r,l}\right] =\mathrm{Tr_{\bar{c},c}}\Big[\exp
\left( -\hat{H}_{0}/T\right) \prod_{l=1}^{N}\exp \Big(\Delta \sum_{r,\sigma
}\sigma \phi _{r,l}\bar{c}_{r,l}c_{r,l}\Big)\Big],
\end{equation*}%
\begin{eqnarray*}
W_{\alpha }^{\prime }\left[ \phi \right] &=&\exp \Big(-\frac{\Delta }{2V_{0}}%
\sum_{r}\sum_{l=1}^{N}\Big(\phi _{r,l}^{2} \\
&&+\left( \alpha /\Delta \right) ^{2}\left( \phi _{r,l+1/2}-\phi
_{r,l-1/2}\right) ^{2}\Big)\Big),
\end{eqnarray*}%
and the subscript $l$ of $\phi $ and $c$ means that these functions are
taken at the times $\tau _{l}$.

The disadvantage of Eq.~(\ref{c6}) is that one cannot bring the trace of the
fermionic operators $c,\bar{c}$ into an useful form without making
additional approximations. Therefore, we use a slightly different
representation that allows us to calculate this trace exactly and derive the
bosonized form.

Instead of using Eq.~(\ref{c6}) we write the partition function $Z$ as%
\begin{equation}
Z=\lim_{\Delta \rightarrow 0}\frac{\int Z_{f}[\tilde{\phi}]\tilde{W}[\tilde{%
\phi}]D\tilde{\phi}}{\int \tilde{W}[\tilde{\phi}]D\tilde{\phi}}  \label{c8}
\end{equation}%
where $\tilde{\phi}_{r}\left( \tau \right) =\phi _{r,l}$ for $\left(
l-1\right) \Delta \leq \tau <l\Delta $ and the functional $Z_{f}[\tilde{\phi}%
]$ equals
\begin{eqnarray}
&&Z_{f}[\tilde{\phi}]=\mathrm{Tr_{\bar{c},c}}\Big[\mathrm{\exp }\left(
-\beta \hat{H}_{0}\right)  \label{c9} \\
&&\times T_{\tau }\exp \Big(-\int_{0}^{\beta }\sum_{r,\sigma }\sigma \tilde{%
\phi}_{r\sigma }\left( \tau \right) \bar{c}_{r\sigma }\left( \tau \right)
c_{r\sigma }\left( \tau \right) d\tau \Big)\Big],  \notag
\end{eqnarray}%
The weight $\tilde{W}[\tilde{\phi}]$ is chosen in the form
\begin{eqnarray}
\tilde{W}_{\alpha }[\tilde{\phi}] &=&\exp \Big\{-\frac{1}{2V_{0}}%
\sum_{r}\int_{0}^{\beta }\Big(\tilde{\phi}_{r}^{2}\left( \tau \right)
\label{c10} \\
&+&\left( \alpha /\Delta \right) ^{2}\left[ \tilde{\phi}_{r}\left( \tau
+\Delta \right) -\tilde{\phi}_{r}\left( \tau \right) \right] ^{2}\Big)d\tau %
\Big\}  \notag
\end{eqnarray}%
We assume, as for the continuous limit, that $\tilde{\phi}_{r}\left( \tau
\right) =\tilde{\phi}_{r}\left( \tau +\beta \right) .$ Of course, the
representations (\ref{c6}) and (\ref{c8}-\ref{c10}) are equivalent in the
continuous limit. In Eq.~(\ref{c9}) one integrates the exponent over each
slice instead of taking it in the middle of the slices as it is done in the
second of Eqs.~(\ref{c6}).

\begin{figure}[t]
\includegraphics{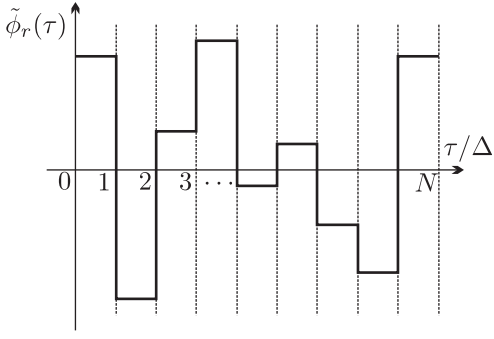}
\caption{Typical form of the function~$\tilde{\protect\phi}_{r}\left(
\protect\tau \right) .$}
\label{c:typicalphi}
\end{figure}
The function $\tilde{\phi}_{r}\left( \tau \right) $ is discontinuous and its
typical form is represented in Fig.~\ref{c:typicalphi}. The fluctuations of
this function are essentially dependent on the ratio $\alpha /\Delta $. In
the limit $\alpha \ll \Delta $, the typical average value $\langle \lbrack
\tilde{\phi}_{r}(\tau )]^{2}\rangle _{\tilde{W}}$ calculated with $\tilde{W}%
_{\alpha }$, Eq.~(\ref{c10}), equals%
\begin{equation}
\left\langle \left[ \tilde{\phi}_{r}\left( \tau \right) \right]
^{2}\right\rangle _{\tilde{W}}=V_{0}/\Delta ,  \label{c11}
\end{equation}%
whereas the correlation at neighboring slices is%
\begin{equation}
\left\langle \tilde{\phi}_{r}\left( \tau \right) \tilde{\phi}_{r}\left( \tau
+\Delta \right) \right\rangle _{\tilde{W}}=V_{0}\alpha ^{2}/\Delta ^{3}\ll
\left\langle \left[ \tilde{\phi}_{r}\left( \tau \right) \right]
^{2}\right\rangle _{\tilde{W}},  \label{c12}
\end{equation}%
which shows that $\langle \lbrack \tilde{\phi}_{r}(\tau )-\tilde{\phi}%
_{r}(\tau +\Delta )]^{2}\rangle _{\tilde{W}}$ and $\langle \lbrack \tilde{%
\phi}_{r}(\tau )]^{2}\rangle _{\tilde{W}}$ are of the same order.

In the opposite limit, $\alpha \gg \Delta ,$ the fluctuations of the
difference of the fields at different slices are strongly suppressed and one
can replace the difference in Eq.~(\ref{c10}) by the derivative. Then, we
can use Eq.~(\ref{c5}) for estimates to obtain%
\begin{equation}
\left\langle \left[ \tilde{\phi}_{r}\left( \tau \right) \right]
^{2}\right\rangle _{\tilde{W}}=V_{0}/\left( 2\alpha \right) ,  \label{c13}
\end{equation}%
while the average squared difference of $\tilde{\phi}$ at neighboring slices
equals
\begin{equation}
\left\langle \left( \tilde{\phi}_{r}\left( \tau \right) -\tilde{\phi}\left(
\tau +\Delta \right) \right) ^{2}\right\rangle _{\tilde{W}}=V_{0}\Delta
/\alpha ^{2}\ll \left\langle \left[ \tilde{\phi}_{r}\left( \tau \right) %
\right] ^{2}\right\rangle _{\tilde{W}}  \label{c14}
\end{equation}%
Eqs.~(\ref{c13}, \ref{c14}) show that in the limit $\alpha \gg \Delta $ the
field $\tilde{\phi}_{r}\left( \tau \right) $ changes slowly from slice to
slice.

Now we are prepared to discuss the origin of the sign problem and how it can
be avoided using the bosonization.

Although the HS field is now discontinuous, Eq.~(\ref{c9}) is a standard
continuous time representation for the partition function. This means that
one can use Eqs.~(\ref{a19}-\ref{a22}) replacing everywhere the function $%
\phi _{r}\left( \tau \right) $ by $\tilde{\phi}_{r}\left( \tau \right) $.
For example, the equation for the Green function takes the form%
\begin{eqnarray}
\left( -\frac{\partial }{\partial \tau }-\hat{h}[u\tilde{\phi}]\right)
G_{r,r^{\prime };\sigma }^{\left( u\phi \right) }\left( \tau ,\tau ^{\prime
}\right) &=&\delta _{r,r^{\prime }}\delta \left( \tau -\tau ^{\prime
}\right) ,  \label{c14a} \\
\hat{h}[u\tilde{\phi}] &=&\hat{\varepsilon}_{r}-\mu ^{\prime }-\sigma u%
\tilde{\phi}_{r}\left( \tau \right) .  \notag
\end{eqnarray}

Substituting the solution of Eq.~(\ref{c14a}) into Eq.~(\ref{a19}) one comes
to the fermionic determinant for $Z_{f}[\tilde{\phi}]$,
\begin{equation}
Z_{f}[\tilde{\phi}]=\det_{r,\sigma ,\tau }\left[ -\frac{\partial }{\partial
\tau }-\hat{h}[\tilde{\phi}]\right] .  \label{c14b}
\end{equation}

In the same way as in Ref.~\onlinecite{blankenbecler} we can reduce the
functional $Z_{f}[\tilde{\phi}]$ to the form%
\begin{align}
Z_{f}[\tilde{\phi}]& =\det_{r,\sigma }\left[ 1+T_{\tau }\exp \Big(%
-\int_{0}^{\beta }\hat{h}[\tilde{\phi}]d\tau \Big)\right]  \notag \\
& =\det_{r,\sigma }\Big[1+\prod_{l=1}^{N}\exp \left( -\left( \hat{\varepsilon%
}_{r}-\mu ^{\prime }-\sigma \phi _{r,l}\right) \Delta \right) \Big]
\label{c15}
\end{align}%
In Eq.~(\ref{c15}), the multipliers are ordered in $l$, such that the
multiplier with $l=1$ is on the right. The same representation is usually
used in the standard MC simulations [Actually, one replaces each multiplier
in the second line of Eq.~(\ref{c15}) by the product $\exp \left( -\left(
\hat{\varepsilon}_{r}-\mu ^{\prime }\right) \Delta \right) \exp \left(
\sigma \phi _{r,l}\Delta \right) ,$ which is clearly justified in the limit $%
\Delta \ll \beta $]. Eq.~(\ref{c15}) was obtained in Refs.~%
\onlinecite{blankenbecler,dosantos} as a result of discretizing the time. At
the same time, we see from Eqs.~(\ref{c14a}, \ref{c14b}) that only the HS
field $\tilde{\phi}_{r}\left( \tau \right) $ is discrete now, while the
derivative $\partial /\partial \tau $ is continuous as before. This allows
us to bosonize the fermion model even in this discrete time representation.
However, before making the proper transformations, let us discuss the origin
of negative signs of $Z_{f}[\tilde{\phi}]$.

\subsection{The negative sign problem for fermions \label{sign}}

In the static case, when $\phi _{r,l}$ does not depend on $l$, the product
in Eq.~(\ref{c15}) is trivially calculated because the multipliers commute
with each other. Then one comes to the standard expression for the partition
function of fermions in a static external po

tential, which is clearly positive.

If the function $\phi _{r,l}$ varies sufficiently slowly in time, the
adiabatic approximation can be used and the function $Z_{f}[\tilde{\phi}]$
takes the form \cite{blankenbecler}
\begin{equation}
Z_{f}[\tilde{\phi}]\approx \prod_{k}\Big[1+\exp \Big(-\Delta
\sum_{l=1}^{N}\lambda _{l}^{\left( k\right) }\Big)\Big]  \label{c16}
\end{equation}%
where $\lambda _{l}^{(k) }$ is the eigenenergy of $k$-state of the
Hamiltonian $\hat{h}[\tilde{\phi}]$ obtained on each slice $l$. This
expression is, again, positive.

Actually, the negative sign of $Z_{f}[\tilde{\phi}]$ may arise only when the
function $\tilde{\phi}_{r,l}$ strongly fluctuates in time.

Let us see how it happens. For a given $\tilde{\phi}(\tau) $, we consider a
matrix element $P_{nm}$ of the product in Eq.~(\ref{c15}) for one of the
spin directions
\begin{equation}
P_{nm}=\sum_{r}v_{r,1}^{\left( n\right) }\left[ \prod_{l=1}^{N}\exp \left( -%
\hat{h}_{l}\Delta \right) \right] v_{r,1}^{\left( m\right) }  \label{c18}
\end{equation}%
where $v_{r,l}^{\left( m\right) }$ is the $m$-th eigenfunction of the
Hamiltonian $\hat{h}_{l}( \tilde{\phi}) $ at the $l$-th slice [we assume
that the eigenfunctions $v_{r,l}^{\left( m\right) }$ are real and the
product in Eq. (\ref{c18}) is ordered in time from the right to the left].

Using the completeness of the sets of the wave functions $v_{r,l}^{\left(
k_{l}\right) }$ for each slice $l,$%
\begin{equation}
\sum_{k}v_{r,l}^{\left( k\right) }v_{r^{\prime },l}^{\left( k\right)
}=\delta _{rr^{\prime }},  \label{c18a}
\end{equation}%
we represent the matrix element $P_{nm}$ in the form%
\begin{equation}
P_{nm}=\sum_{\left\{ k_{l}\right\} }\delta _{n,k_{N}}\delta _{k_{1}m}\exp %
\Big(-\Delta \sum_{l=1}^{N}\lambda _{l}^{\left( k_{l}\right) }\Big)%
\prod_{l=1}^{N}\pi _{l}^{k_{l+1}k_{l}},  \label{c18c}
\end{equation}%
where

\begin{equation}
\pi _{l}^{k_{l+1}k_{l}}=\sum_{r}v_{r,l+1}^{\left( k_{l+1}\right)
}v_{r,l}^{\left( k_{l}\right) },  \label{c19}
\end{equation}%
are elements of orthogonal matrices $\pi _{l}$, and write the function $%
Z_{f}[\tilde{\phi}]$ as%
\begin{equation}
Z_{f}[\tilde{\phi}]=\det \left[ 1+P_{+}\right] \det \left[ 1+P_{-}\right] ,
\label{c20}
\end{equation}%
where $N_{d}^{total}\times N_{d}^{total}$ matrices $P_{\pm }$ are matrices
with real matrix elements $\left( P_{\pm }\right) _{mn}$, Eq. (\ref{c18}),
corresponding to spins \textquotedblleft up" and \textquotedblleft down" [$%
N_{d}^{total}$ is the number of the eigenfunctions corresponding to the
total number of the sites in the system]. The matrices $P_{\pm }$ are real
but not necessarily symmetric.

For fields $\tilde{\phi}\left( \tau \right) $ slowly varying in time, the
eigenfunctions $v_{r,l}^{\left( k_{l}\right) }$ on the neighboring slices
are almost equal to each other and the matrices $\pi _{l}$ are close to the
unity matrix. In this case, the matrices $\pi _{l}$ can be calculated using
the standard perturbation theory for wave functions%
\begin{equation}
v_{r,l+1}^{\left( m\right) }=v_{r,l}^{\left( m\right) }+\sum_{k\neq m}\frac{%
\left( \delta \phi _{r,l}\right) ^{km}}{\lambda _{l}^{\left( m\right)
}-\lambda _{l}^{\left( k\right) }}v_{r,l}^{\left( k\right) },  \label{c22}
\end{equation}%
where $\delta \phi _{r,l}=\phi _{r,l+1}-\phi _{r,l}$ is small. Then, the
off-diagonal elements of the matrices $\pi _{l}$ take in the main order in $%
\delta \phi $ the following form%
\begin{equation}
\pi _{l}^{mn}=\left\{
\begin{array}{cc}
\frac{\left( \delta \phi _{r,l}\right) ^{nm}}{\lambda _{l}^{\left( m\right)
}-\lambda _{l}^{\left( n\right) }}, & m\neq n \\
-\frac{1}{2}\sum_{k\neq n}\left[ \frac{\left( \delta \phi _{r,l}\right) ^{nk}%
}{\lambda _{l}^{\left( k\right) }-\lambda _{l}^{\left( n\right) }}\right]
^{2}, & m=n%
\end{array}%
\right. .\quad  \label{c23}
\end{equation}

In the limit $\Delta \ll \alpha ,$ the characteristic values of the
quantities $\left\vert \delta \phi _{r,l}\right\vert $ follow from Eqs. (\ref%
{c13}, \ref{c14})
\begin{equation}
\left\vert \phi _{r,l}\right\vert \sim \left( V_{0}/\alpha \right)
^{1/2},\quad \left\vert \delta \phi _{r,l}\right\vert \sim \left(
V_{0}\Delta /\alpha ^{2}\right) ^{1/2},  \label{c23c}
\end{equation}%
while the average values of $\phi _{r,l}$ and $\delta \phi _{r,l}$ vanish.
It is clear from Eq. (\ref{c23}) that $\left\vert \delta \phi
_{r,l}\right\vert \ll \left\vert \phi _{r,l}\right\vert $ and this justifies
using the perturbation theory for computation of the matrix elements $\pi
_{l}^{mn}$, Eq. (\ref{c23}).

As we assume that $\alpha ^{-1}\gg \left\{ t_{r,r^{\prime }},V_{0}\right\} $%
, the wave functions $\phi _{r,l}$ are almost localized on the sites $r_{n}$
and the states $n$ are characterized by the positions of centers of the
localization. Then, we estimate the denominators in Eq. (\ref{c23}) as%
\begin{equation}
\left\vert \lambda _{l}^{\left( m\right) }-\lambda _{l}^{\left( n\right)
}\right\vert =\left\vert \phi _{r_{m},l}-\phi _{r_{n},l}\right\vert \sim
\left( V_{0}/\alpha \right) ^{1/2}  \label{c22a}
\end{equation}%
Completely localized wave functions cannot give a contribution to
non-diagonal matrix elements $\left( \delta \phi _{r,l}\right) ^{mn}$ in Eq.
(\ref{c23}) and one should take into account their overlap at finite $%
t_{r,r^{\prime }}$ leading to finite values of the wave functions on sites
in the neighborhood of the center of the localization $r_{0}$. These values
can be estimated [considering the tunneling term $t_{r,r^{\prime }}$ as a
perturbation] as $t_{r,r_{0}}\left\vert \phi _{r,l}-\phi
_{r_{0},l}\right\vert ^{-1}$ and one comes to typical absolute values of
randomly fluctuating off-diagonal elements of the matrix $\pi _{l}$
\begin{equation}
\left\vert \pi _{l}^{mn}\right\vert \sim \left( t^{2}\Delta /V_{0}\right)
^{1/2},\quad m\neq n  \label{c22b}
\end{equation}%
The estimate (\ref{c22b}) is written under the assumption that all the
states are non-degenerate, which is most probable for randomly chosen fields
$\phi _{r,l}$. The matrix elements $\pi _{l}^{mn}$, Eq. (\ref{c22b}), are
small in the limit $\Delta \rightarrow 0$ but the matrix $P$, Eq. (\ref{c18c}%
), contains a product of $N=\beta /\Delta $ matrices $\pi _{l}$ and can
essentially be different from a diagonal one. Since the matrix elements $\pi
_{l}$ fluctuate randomly with a gaussian distribution, we estimate the
off-diagonal elements of the product of $N$ matrices entering Eq. (\ref{c18c}%
) as $\Big(\beta t_{r,r^{\prime }}^{2}/V_{0}\Big)^{1/2}$. This is a small
number in the limit of sufficiently high temperatures
\begin{equation}
T\gg \Big(t_{r,r^{\prime }}^{2}/V_{0}\Big)^{1/2}  \label{c22c}
\end{equation}%
and Eq. (\ref{c22c}) determines the adiabatic limit. In this limit one comes
to Eq. (\ref{c16}) and the fermionic determinant $Z_{f}[\tilde{\phi}]$ must
be positive. Alternatively, one could take lower temperatures but large $%
\alpha \sim \beta $, which would lead to the same result.

Let us discuss now what happens if $\alpha \ll \beta $ and the inequality (%
\ref{c22c}) is not fulfilled. Using Eq. (\ref{c18c}) we write $\det P$ for
one of the spin directions in the form%
\begin{equation}
\det P=\exp \Big(-\Delta \sum_{l=1}^{N}\sum_{k=1}^{N_{d}^{total}}\lambda
_{l}^{\left( k\right) }\Big)\prod_{l=1}^{N}\det \pi _{l}.  \label{c21}
\end{equation}%
As the matrices $\pi _{l}$ are orthogonal, one comes to values $\det \pi
_{l}=1$ for all $l$ and to positive values of $\det P$ in Eq. (\ref{c21}).
Of course, this does not necessarily mean that the function $Z_{f}[\tilde{%
\phi}]$, Eq. (\ref{c20}), is positive, too.

Matrix $P$ given by Eq. (\ref{c18c}) is real but not symmetric. Such
matrices cannot generally be diagonalized but one can always find a Jordan
normal form for them. Assuming that $P$ is an $M\times M$ real matrix one
writes
\begin{equation}
P=QJQ^{-1},  \label{c230}
\end{equation}%
where the block diagonal matrix $J$ can be written as%
\begin{equation*}
J=\left(
\begin{array}{ccccc}
J_{1} & 0 & 0 & 0 & 0 \\
0 & J_{2} & 0 & 0 & 0 \\
\cdot & \cdot & \cdot & \cdot & \cdot \\
0 & 0 & 0 & J_{n-1} & 0 \\
0 & 0 & 0 & 0 & J_{n}%
\end{array}%
\right)
\end{equation*}%
and

\begin{equation*}
J_{i}=\left(
\begin{array}{ccccc}
p_{i} & 1 & 0 & 0 & 0 \\
0 & p_{i} & 1 & 0 & 0 \\
\cdot & \cdot & \cdot & \cdot & \cdot \\
0 & 0 & 0 & p_{i} & 1 \\
0 & 0 & 0 & 0 & p_{i}%
\end{array}%
\right).
\end{equation*}%
The sizes $K_{i}$ of the blocks $J_{i}$ depend on the form of the matrix $P$
but
\begin{equation*}
\sum_{i=1}^{n}K_{i}=M.
\end{equation*}%
For diagonalizable matrices each block in the matrix $J$ is just a number
and all $K_{i}=1$, such that $n=M$. The numbers $p_{i}$ are solutions of the
equation%
\begin{equation}
\det \left( pI_{M}-P\right) =0,  \label{c231}
\end{equation}%
where $I_{M}$ is the unity matrix, and are not necessarily real. However, as
the matrix $P$ is real, for any solution $p_{i}$ of Eq. (\ref{c231}) its
complex conjugate $p_{i}^{\ast }$ is also a solution.

One can see easily from Eqs. (\ref{c230}, \ref{c231}) that
\begin{equation}
\det \left[ 1+P\right] =\prod_{i=1}^{n}\left( 1+p_{i}\right) ^{K_{i}},\quad
\det P=\prod_{i=1}^{n}p_{i}^{K_{i}}  \label{c232}
\end{equation}

As we have mentioned previously, in the adiabatic limit (\ref{c22c}), when
the function $\tilde{\phi}_{r}\left( \tau \right) $ is smooth in time, one
obtains%
\begin{equation*}
\prod_{l=1}^{N}\pi _{l}\approx 1,
\end{equation*}%
and one comes to inequalities $p_{i}>0$ for all $i$. Deforming the function $%
\tilde{\phi}_{r}\left( \tau \right)$, such that it can vary fast in the
interval $\left[ 0,\beta \right] $, makes the matrix $\prod_{l=1}^{N}\pi
_{l} $ essentially different from unity and some of the eigenvalues $p_{i}$
can become negative, which is a necessary condition for obtaining negative
values of $\det \left[ 1+P\right] $.

In the limit $\Delta \gg \alpha $, all the matrix elements of the matrices $%
\pi _{l}$ can be typically of order one and the changes of $p_{i}$ can be
large after a minimum change of a configuration of the field $\phi _{r,l}$.
In this case nothing prevents $p_{i}$ from jumping from positive to large
negative values and making $\det \left[ 1+P_{\pm }\right] $ negative.

In the opposite limit $\alpha \gg \Delta $, a smooth deformation in time of
the field $\phi _{r,l}$ results in a smooth motion of the eigenvalues $p_{i}$%
. If we start with $\phi _{r,l}$ slowly varying in the interval $[0,\beta ]$%
, all $p_{i}$ are real and positive. Deforming the function $\phi _{r,l}$
results in a motion of the parameters $p_{i}$ along the real axis. They
cannot cross zero because this would mean $\det P^{\uparrow ,\downarrow }=0,$
which is impossible because $\det \pi _{l}=1$ for all $l$.

However, two different values $p_{i}$ and $p_{k}$ can collide and move to
the complex plane forming a pair of complex conjugated values $p$ and $%
p^{\ast }$. Again, this cannot change the sign because these two eigenvalues
lead to a multiplier $\left\vert 1+p\right\vert ^{2}$ in the determinant. A
possibility to change the sign of the fermionic determinant $Z_{f}[\tilde{%
\phi}]$ first arises when the values $p$ and $p^{\ast }$ collide again on
the real axis at a point $p_{0}<0$. After this event one obtains again two
real negative eigenvalues $p_{i}$ and $p_{k}$. Then, they move separately
and one of them may cross the point $-1$ leading to a negative value of $%
\det \left[ 1+P_{\pm }\right] $. Still, this does not necessarily mean that
the full fermionic determinant, Eq. (\ref{c15}), becomes negative because
the product of the contributions corresponding to spin \textquotedblleft up"
and spin \textquotedblleft down" should be taken but the positivity of the
fermionic determinant is no longer guaranteed.

We see that, when deforming continuously the function $\tilde{\phi}_{r,l}$,
a chain of several events must happen before the fermionic determinant
becomes negative. Therefore, the probability of negative values of $Z_{f}[%
\tilde{\phi}]$ may be reduced in the continuous limit, $\Delta \ll \alpha ,$
with respect to the opposite one, $\Delta \gg \alpha $, but one should not
expect that it vanishes.

Of course, using these arguments we cannot check whether the sign of $Z_{f}[%
\tilde{\phi}]$ changes or not for a given function $\tilde{\phi}_{r,l}$
[unless it is sufficiently smooth in the entire interval $\left[ 0,\beta %
\right] $] and this can be done only numerically. Apparently, negative
values of $Z_{f}[\tilde{\phi}]$ are generally possible even in the limit $%
\Delta /\alpha \rightarrow 0$ but we do not intend to clarify this question
here. What we want to do is to replace the function $Z_{f}[\tilde{\phi}]$ by
\emph{another} function $Z_{b}[\tilde{\phi}]$ which is positive by
construction. In doing so we expect that, although these functions are
generally different for a given $\tilde{\phi}_{r,l}$, replacing $Z_{f}[%
\tilde{\phi}]$ by $Z_{b}[\tilde{\phi}]$ in Eq. (\ref{c8}) will not change
the result of the integration over $\tilde{\phi}$ in the limit of a large
number of sites in the system including both the main system and the bath.
In the next subsection, we present arguments justifying this expectation.

\subsection{And how it can be avoided in the bosonic representation}

\label{sign2}

We start our discussion by generalizing the model under consideration to a
model with an arbitrary complex electron-electron interaction writing
instead of Eqs. (\ref{a4}, \ref{a8})
\begin{equation}
Z^{s}=Tr\exp \left[ -\beta \hat{H}\left( s\right) \right] ,\quad \hat{H}=%
\hat{H}_{0}+\hat{H}_{int}^{\left( 0\right) }\left( s\right) ,  \label{c23a}
\end{equation}%
where the operator $\hat{H}_{0}$ is given by Eq. (\ref{a2}) with a properly
shifted chemical potential and the interaction $\hat{H}_{int}^{\left(
0\right) }\left( s\right) $, \textbf{\ }
\begin{equation}
\hat{H}_{int}^{\left( 0\right) }\left( s\right) =-\frac{sV_{0}}{2}%
\sum_{r}\left( c_{r+}^{+}c_{r+}-c_{r-}^{+}c_{r-}\right) ^{2},  \label{c24}
\end{equation}%
replaces $\hat{H}_{int}^{\left( 0\right) }$ in Eq. (\ref{a5}). The variable $%
s$ is an arbitrary complex variable and the original partition function $Z$,
Eq. (\ref{a4}), is obtained from Eq. (\ref{c23a}) by putting $s=1$.

For a finite system with a total number of sites $N_{d}$, the number of
fermions cannot exceed $2N_{d}$ and the function $Z^{s}$ is a finite sum of
exponential functions of $s$. Therefore, $Z^{s}$ is analytical everywhere in
the complex plane of $s$ except the infinite point. The function $Z^{s}$
decays in the left half-plane when $\mathrm{Re}\ s\rightarrow -\infty $ but
grows in the right half-plane when $\mathrm{Re}\ s\rightarrow \infty $.

If the function $Z_{f}[\tilde{\phi}]$ in the integral in Eq. (\ref{c8})
becomes negative for sufficiently many configurations $\tilde{\phi}_{r,l}$,
making approximations in the integrand is dangerous because the latter can
be a fast oscillating functional of $\tilde{\phi}_{r,l}$. A simpler task is
to calculate $Z^{s}$ approximately for real negative $s<0$ because the
fermionic determinant is strictly positive in this case. These values of $s$
correspond to an attraction between the fermions and the sign problem does
not exist in this case. As soon as the function $Z^{s}$ is known for real $%
s<0$ one can try to make an analytical continuation to $s=1$.

For practical computations it is more convenient to decouple the attraction
term by a HS transformation corresponding to an external \textquotedblleft
potential" rather than to a \textquotedblleft magnetic field" acting of the
spins used here because the latter enters the effective action with $\sqrt{s}
$, which is imaginary for real $s<0$, and such a representation is not very
convenient for numerics. However, we do not change the decoupling scheme for
the present discussion and simply replace $\tilde{\phi}_{r,l}\rightarrow
\sqrt{s}\tilde{\phi}_{r,l}$ in Eq. (\ref{c15}). Then, the fermionic
determinant $Z_{f}^{s}[\tilde{\phi}]$ can be written for real negative $s$
as
\begin{equation}
Z_{f}^{s}[\tilde{\phi}]=\left\vert \det_{r}\Big[1+\prod_{l=1}^{N}\exp \left(
-\left( \hat{\varepsilon}_{r}-\mu ^{\prime }-i\sqrt{-s}\tilde{\phi}%
_{r,l}\right) \Delta \right) \Big]\right\vert ^{2},  \label{c240}
\end{equation}%
and it is clearly positive.

Substituting Eq. (\ref{c240}) into Eq. (\ref{c8}) we obtain an integral with
a positive integrand. Absence of oscillations of the integrand makes
approximations more reliable and we apply the bosonization scheme for the
real negative $s$ simply substituting everywhere $\tilde{\phi}$ by $i\sqrt{-s%
}\tilde{\phi}$. As a result, we write a new bosonic partition function $%
Z_{b}^{s}[\tilde{\phi}]$ for real $s<0$ in the form

\begin{equation}
\frac{Z_{b}^{s}[\tilde{\phi}]}{Z_{0}}=\exp \left[ \frac{1}{2i}\sum_{r,\sigma
,a}\int_{0}^{\beta }\int_{0}^{1}\sigma \sqrt{-s}\tilde{\phi}_{r\sigma
}\left( \tau \right) A_{a;r,r}^{\prime }\left( z\right) dud\tau \right] ,
\label{c241}
\end{equation}%
where the function $A$ satisfies the equation
\begin{equation}
\mathcal{H}_{r,r^{\prime }}\left( \tau \right) A_{a;r,r^{\prime }}\left(
z\right) =-iu\sigma \sqrt{-s}n_{r,r^{\prime },\sigma }\tilde{\Phi}%
_{r,r^{\prime };\sigma }\left( \tau \right) B_{a}  \label{c242}
\end{equation}%
with%
\begin{eqnarray}
&&\mathcal{H}_{r,r^{\prime }}\left( \tau \right) =\Lambda _{1}\left( \hat{%
\varepsilon}_{r}-\hat{\varepsilon}_{r^{\prime }}-iu\sigma \sqrt{-s}\tilde{%
\Phi}_{r,r^{\prime };\sigma }\left( \tau \right) \right)  \notag \\
&&+i\Lambda _{2}\frac{\partial }{\partial \tau }+\Lambda \gamma  \label{c243}
\end{eqnarray}%
following from Eqs. (\ref{a102}, \ref{a103}).

One can easily see after summation over $\sigma $ that the exponent in Eq. (%
\ref{c241}) is real for real $s<0$ and the function $Z_{b}^{s}[\tilde{\phi}]$
is positive. As the transformation from the fermions to bosons is almost
exact and both the functions $Z_{f}^{s}[\tilde{\phi}]$ and $Z_{b}^{s}[\tilde{%
\phi}]$ are positive, we expect that they should be close to each other for
real $s<0$. The introduction of the finite parameter $\gamma $ and of the
bath is important for such $s$, too, because one should exclude
\textquotedblleft parasitic solutions" for $A_{a;r,r^{\prime }}\left( \tau
\right) $.

So, we have bosonized the fermionic system with attraction for which the
sign problem does not exist anyway. Integrating over $\tilde{\phi}_{r}\left(
\tau \right) $ in Eq. (\ref{c8}) with $Z_{f}^{s}[\tilde{\phi}]$ and $%
Z_{b}^{s}[\tilde{\phi}]$ we thus obtain two partition functions $Z_{f}^{s}$
and $Z_{b}^{s}$ on the entire semi-axis of real $s<0$ and argue that they
should be close to each other because $Z_{f}^{s}[\tilde{\phi}]$ and $%
Z_{b}^{s}[\tilde{\phi}]$ are. The functionals $Z_{f}^{s}[\tilde{\phi}]$ and $%
Z_{b}^{s}[\tilde{\phi}]$ are positive and a small difference between them
should not result in a considerable difference between the functions $%
Z_{f}^{s}$ and $Z_{b}^{s}$.

In order to obtain the partition functions $Z_{f}$ and $Z_{b}$ corresponding
to the original model under consideration we perform an analytical
continuation of the functions $Z_{f}^{s}$ and $Z_{b}^{s}$ in the complex
plane of $s$ from real negative $s<0$ to $s=1$. In order to analytically
continue $Z_{f}^{s}$, one can use Eqs. (\ref{c23a}, \ref{c24}) giving a
finite sum of exponentials of $s$ and the analytical continuation is
performed just putting $s=1$ everywhere. Then, one obtains immediately all
standard formulas for the fermionic system with repulsion and $Z_{f}[\tilde{%
\phi}]$ derived from these formulas can be both positive and negative
depending on $\tilde{\phi}_{r}\left( \tau \right) $, which leads to the sign
problem.

Having constructed the function $Z_{b}^{s}[\tilde{\phi}]$ for real negative $%
s,$ Eqs. (\ref{c241}-\ref{c243}), one should integrate over $\tilde{\phi}%
_{r}\left( \tau \right) $ using Eq. (\ref{c8}) and obtain the partition
function $Z_{b}^{s}$. After this function has been found for all real $s<0$
one can find uniquely an analytical continuation $\tilde{Z}_{b}^{s}$ for
complex $s$ coinciding with $Z_{b}^{s}$ for real $s<0$. This means that one
can represent $\tilde{Z}_{b}^{s}$ in a form of a convergent series in powers
of the complex $s$ in a certain region of complex $s$ including the negative
real half-axis.

Below are some arguments for the analyticity of $Z_{b}^{s}$. As we have
seen, the difference between the terms of this expansion and those for the
fermionic partition function $Z_{f}^{s}$ arises from a different treatment
of two or more electronic Green functions at coinciding Matsubara
frequencies and momenta. This difference should be small in the limit of a
large bath and small parameter $\gamma $ and therefore the Taylor expansion
of the function $Z_{b}^{s}$ should converge for all complex $s$ in the same
way as the expansion for $Z_{f}^{s}$ does. In Section \ref{pt} we have
checked explicitly the agreement between the partition functions $Z_{f}$ and
$Z_{b}$ in the first two orders of the expansion in the interaction. The
calculation can be repeated without changes for arbitrary complex $s$
including real $s<0$. Thus, the function $Z_{b}^{s}$ is analytic for all
complex $s$ except the infinite point and the physical partition function $%
Z_{b}$ can be obtained putting $s=1$ in this series. This result allows us
to expect that the function $Z_{b}$ is a good approximation for $Z_{f}$.

Having put $s=1$ everywhere in the series for $Z_{b}^{s}$ we can represent
it again in a form of an expansion of the integral over $\tilde{\phi}$ with $%
Z_{b}[\tilde{\phi}]$, Eq. (\ref{a104}), thus coming back to the bosonization
formulas obtained in the previous sections. This means, that replacing the
original fermionic model with the repulsion by the bosonic one, we expect a
precise agreement between the results obtained by integration over $\tilde{%
\phi}$ of the functional $Z_{f}[\tilde{\phi}]$ and $Z_{b}[\tilde{\phi}]$ in
the limit $\gamma \rightarrow 0$ and $N_{d}\rightarrow \infty $. Although
both the functionals can equally be used for analytical calculations, the
use of $Z_{b}[\tilde{\phi}]$ may be preferable for the MC method because it
is always positive in contrast to $Z_{f}[\tilde{\phi}]$.

The analyticity in the the complex plane is evident for the partition
function $Z^{s}$, Eq. (\ref{c23a}), and our arguments justifying the
approximation done in the bosonization procedure have been based on this
property. However, the analyticity of $Z_{f}^{s}$ and $Z_{b}^{s}$ does not
automatically imply the analyticity of the functions $Z_{f}^{s}[\tilde{\phi}%
] $ and $Z_{b}^{s}[\tilde{\phi}]$ for \emph{any given }$\tilde{\phi}%
_{r}\left( \tau \right) $ and one should check this property using explicit
formulas. These functionals are close to each other for real negative $s$
and would remain always close for other complex $s$ including $s=1$ if they
were analytical in the complex plane of $s$ for all $\tilde{\phi}_{r}\left(
\tau \right) $, which would imply a convergent series in $s$. If this is not
so one can expect that the functional $Z_{b}^{s}[\tilde{\phi}]$ does not
well approximate the functional $Z_{f}^{s}[\tilde{\phi}]$ for certain $%
\tilde{\phi}\left( \tau \right) $ in some regions of complex $s$ including $%
s=1$. In this case, the functional $Z_{b}[\tilde{\phi}]$ used for our
explicit calculations can be considerably different from the analytical
continuation of $Z_{b}^{s}[\tilde{\phi}]$ from real $s<0$ to $s=1$ and,
hence, from $Z_{f}^{s}[\tilde{\phi}]$. In particular, the functionals $Z_{b}[%
\tilde{\phi}] $ and $Z_{f}^{s}[\tilde{\phi}]$ may have different signs for
certain functions $\tilde{\phi}\left( \tau \right) $ when $Z_{f}[\tilde{\phi}%
]$ becomes negative.

It is not easy to consider the analyticity properties of the functionals $%
Z_{f}^{s}[\tilde{\phi}]$ and $Z_{b}^{s}[\tilde{\phi}]$ analytically
continued from Eq. (\ref{c240}) and Eqs. (\ref{c241}-\ref{c243}),
respectively. It seems that $Z_{f}^{s}[\tilde{\phi}]$ is still analytical
everywhere in the complex plane of $s$ because the fermionic determinant is
a finite product of exponentials and therefore can be expanded in series in
integer powers of $s$. However, the same cannot be said about $Z_{b}^{s}[%
\tilde{\phi}],$ Eqs. (\ref{c241}-\ref{c243}), and we think that this
quantity cannot be analytical everywhere in the complex plane of $s$ for any
$\tilde{\phi}_{r}\left( \tau \right) $.

A simple way to see this is to try to repeat the regularization of Appendix %
\ref{solution} for complex $s$. The only difference with respect to $s=1$ in
the derivation is that the eigenvalue in Eqs. (\ref{a107}) written for
complex $M_{r,r^{\prime }}$ is not just $\lambda ^{K\ast }$, where $\lambda
^{K}$ is is taken from Eq. (\ref{a106}), but generally another value $\tilde{%
\lambda}^{K\ast }$. It coincides with $\lambda ^{K\ast }$ only for real $%
M_{r,r^{\prime }}$ corresponding to positive real $s$.

As a result, one should replace in Eqs. (\ref{e102}, \ref{e104}) the value $%
\lambda ^{\ast }$ by $\tilde{\lambda}^{\ast }.$ Then, one cannot exclude
that for some functions $\tilde{\phi}_{r}\left( \tau \right) $ and some
complex $s$ the eigenvalue $E$ turns to zero. The subsequent integration
over $u$ in Eq. (\ref{c241}) would lead to cuts in the complex plane of $s.$

In this situation, the functional $Z_{b}[\tilde{\phi}]$ obtained by just
putting $s=1$ in Eqs. (\ref{c241}-\ref{c243}) can differ from the analytical
continuation from real $s<0$ to $s=1$. Then, the functionals $Z_{f}[\tilde{%
\phi}]$ and $Z_{b}[\tilde{\phi}]$ can essentially be different from each
other for certain functions $\tilde{\phi}_{r}\left( \tau \right) $ and can
even have different signs. At the same time, the physical partition
functions $Z_{f}$ and $Z_{b}$ can still be close to each other because the
functions $Z_{f}^{s}$ and $Z_{b}^{s}$ are analytical everywhere in the
complex plane.

This discussion may help the reader to understand why replacing the not
necessarily positive fermionic determinant $Z_{f}[\tilde{\phi}]$ by the
always positive functional $Z_{b}[\tilde{\phi}]$ we still expect the correct
result for the physical partition function $Z$ obtained after integration
over all functions $\tilde{\phi}_{r}\left( \tau \right) $.

The arguments of this subsection are strongly based on the assumption that
the system is finite, which allowed us to speak about the analyticity of the
partition function $Z^{s}$ in the interaction constant $sV$. This
analyticity can be lost at low temperatures and real $s<0$ in the limit of
an infinite sample when the system becomes a superconductor but this does
not seem to create problems. A good quantitative agreement with exact
results is expected for large systems implying that they include the bath.

Although the arguments presented in the last subsections may help to
understand the general scenario of what happens in the process of the
bosonization, they are not rigorous and their final validity is to be
checked numerically.

\subsection{Some remarks about the regularization}

\label{sign3} In this subsection, we want to emphasize that one can select
the solution of Eq.~(\ref{a25}) in many different ways but only one of them,
Eq.~(\ref{a102}), gives positive real $Z_{b}[\tilde{\phi}]$ for all $\tilde{%
\phi}_{r}\left( \tau \right) $. In principle, in the limit of a large number
of sites in the bath all these solutions should lead to the same result and
we simply choose the most convenient one.

Of course, just taking the original electron Green function $G_{r,r^{\prime
};\sigma }^{\left( u\phi \right) }\left( \tau ,\tau ^{\prime }\right) $ from
Eq.~(\ref{a21}) we would satisfy Eq.~(\ref{a25}) for $\tau ^{\prime }=\tau
+\delta $ but we understand that using this solution, one encounters the
sign problem. There are infinitely many other solutions of Eq.~(\ref{a25})
leading to non-positive weights in the MC procedure.

To see this, let us carry out the regularization keeping small but finite $%
\delta $ for $\tau ^{\prime }=\tau +\delta $. Although the parameter $\delta
$ is infinitesimally small, its presence can be very important for choosing
the proper solution of Eq.~(\ref{a25}). As we consider discontinuous
functions $\tilde{\phi}_{r}\left( \tau \right) $, taking the proper limit is
especially important because $\tau $ and $\tau +0$ can belong to different
slices leading to completely different values $\tilde{\phi}_{r}\left( \tau
\right) $ and $\tilde{\phi}_{r}\left( \tau +0\right) $. So, we write Eq.~(%
\ref{a22a}) in the form
\begin{equation}
\left( \frac{\partial }{\partial \tau }-\tilde{M}_{r,r^{\prime }}^{\left(
+\right) }\left( z\right) \right) G_{r,r^{\prime };\sigma }^{\left( u\phi
\right) }\left( \tau ,\tau +\delta \right) =0  \label{c23ay}
\end{equation}%
with the operator $\tilde{M}_{r,r^{\prime }}^{\left( +\right) }\left(
z\right) ,$ \textbf{\ }
\begin{equation}
\tilde{M}_{r,r^{\prime }}^{\left( +\right) }\left( z\right) =\hat{\varepsilon%
}_{r}-\hat{\varepsilon}_{r^{\prime }}-u\sigma \tilde{\Phi}_{r,r^{\prime
};\sigma }^{\left( +\right) }\left( \tau \right) ,  \label{c24y}
\end{equation}%
where%
\begin{equation}
\tilde{\Phi}_{r,r^{\prime };\sigma }^{\left( +\right) }\left( \tau \right) =%
\tilde{\phi}_{r}\left( \tau \right) -\tilde{\phi}_{r^{\prime }}\left( \tau
+\delta \right)  \label{c240a}
\end{equation}%
and $\delta \rightarrow +0$.

We can repeat the transformations of subsection \ref{subsecIIb} keeping a
finite $\delta >0$ and introducing the regularization with the parameter $%
\gamma $ and the bath. As a result, we come instead of Eqs.~(\ref{a102}, \ref%
{a103}) to the following equations
\begin{equation}
\mathcal{\tilde{H}}_{r,r^{\prime }}^{\left( +\right) }\left( \tau \right)
A_{a;r,r^{\prime }}^{\left( +\right) }\left( z\right) =-u\sigma
n_{r,r^{\prime },\sigma }\tilde{\Phi}_{r,r^{\prime };\sigma }^{\left(
+\right) }\left( \tau \right) B_{a},  \label{c240y}
\end{equation}%
where $a=1,2$, and $\mathcal{\tilde{H}}_{r,r^{\prime }}^{\left( +\right) }$
is a $2\times 2$ matrix
\begin{equation}
\mathcal{\tilde{H}}_{r,r^{\prime }}^{\left( +\right) }\left( \tau \right)
=\Lambda _{1}\tilde{M}_{r,r^{\prime }}^{\left( +\right) }\left( z\right)
+i\Lambda _{2}\frac{\partial }{\partial \tau }+\Lambda \gamma .
\label{c241y}
\end{equation}%
As we have seen previously, just putting $\delta =0$ and keeping the
parameter $\gamma $ finite in Eqs.~(\ref{c240}, \ref{c241}) results in
strictly positive weights $Z_{b}[\tilde{\phi}]$ because the solutions of the
homogeneous equation in Eq.~ (\ref{c240y}) do not exist for finite $\gamma $
and one comes to a real solution for $A_{a;r,r^{\prime }}\left( z\right) $.
Putting $\gamma =0$ for finite $\delta $ we can choose the solution for $%
A_{a;r,r^{\prime }}\left( z\right) $ corresponding to the initial fermionic
problem and thus obtain the functional $Z_{f}[\tilde{\phi}]$ that can be
both positive and negative.

In principle, one could obtain many other solutions of Eq.~(\ref{c240y}, \ref%
{c241y}) keeping both $\gamma $ and $\delta $ finite. Using the spectral
expansion, Eq.~(\ref{a112}), it is clear that the singularities in the
function $A_{r,r^{\prime }}\left( z\right) $ can originate only from the
zero eigenvalues $E^{K}=0$ of the operator in the L.H.S. of Eq.~(\ref{a25}).
In order to avoid this problem we have introduced a regularization that
discards such solutions. Actually, the crucial step in this regularization
is taking equal times in the function $\tilde{\Phi}_{r,r^{\prime }}\left(
\tau \right) $ (putting $\delta =0$ in Eq.~(\ref{c240a})).

On the other hand, keeping a small $\delta >0$ makes the eigenvalues $E^{K}$
complex with an imaginary part of order $\delta $. Although this imaginary
part is infinitesimally small in the limit $\delta \rightarrow 0$, it
determines the imaginary part of the integral over $u$ in Eq.~(\ref{a104})
arising near the poles corresponding to the zero eigenvalues $E^{K}=0$.

We calculate explicitly the imaginary part of the eigenvalue $E^{K}$ in
Appendix \ref{equaltimes} assuming that the eigenvalue $E^{K}$ is very close
to zero and show that it is finite for finite $\delta $ but vanishes in the
limit $\delta /\gamma \rightarrow 0$.

Thus, our bosonization scheme should give different results depending on how
we treat Eqs.~(\ref{c240y}, \ref{c241y}). The presence of the bath makes the
partition function insensitive to the choice of the solution of this
equation and we take the limit $\delta /\gamma \rightarrow 0$ leading us to
the the real positive $Z_{b}[\tilde{\phi}]$.

In the next subsection, we discuss the basic properties of $Z_{b}[\tilde{\phi%
}]$ and demonstrate that there is no reason to encounter the sign problem
when using this functional.

\subsection{Properties of the bosonic action and its suitability for
numerical investigations\label{num}}

\subsubsection{Bosonic action}

In this subsection we want to demonstrate that using $Z_{b}[\tilde{\phi}]$
is convenient for computations and there are no singularities or bad
features, like the negative sign, that would create problems. Final formulas
will be brought to a form that can immediately be used for explicit
computations. We do not see any problems in taking the limit $\Delta
\rightarrow 0$, $\alpha /\Delta \rightarrow 0$ in this representation and
put $\alpha =0$ in the beginning of the discussion. For convenience, we
display in one place the formulas that may serve as the basis of
computational schemes for the case of the on-site repulsion.

We consider the system in a bath, which means that the HS field $\tilde{\phi}%
_{r}\left( \tau \right) $ is not equal to zero only in a region restricted
by a radius $R_{0}$ (c.f. Eq.~(\ref{a100})). In the rest of the sample the
particles do not interact with each other and the field $\tilde{\phi}%
_{r}\left( \tau \right) $ vanishes there. The interacting part of the free
energy functional should be proportional to the number of the sites $N_{d}$
located inside the sphere with the radius $R_{0}$. In order to obtain a good
precision one needs to consider a large number of sites in the bath, such
that the total number of the sites $N_{d}^{total}$ should considerably
exceed $N_{d}$. At the same time, considering the bath should not demand a
large computation time because the HS field is identically zero in that
region.

We write the exact partition function $Z$ of the original model in the form
\begin{equation}
Z=\lim_{\Delta \rightarrow 0}\frac{\int Z_{b}[\tilde{\phi}]W_{\Delta }[%
\tilde{\phi}]\prod_{l=1}^{N}d\phi _{r,l}}{\int W_{\Delta }[\tilde{\phi}%
]\prod_{l=1}^{N}d\phi _{r,l}}  \label{c25}
\end{equation}%
where $\beta =N\Delta $, $\tilde{\phi}_{r}\left( \tau \right) =\phi _{r,l}$
for $\left( l-1\right) \Delta \leq \tau <l\Delta $ (see Fig. \ref%
{c:typicalphi}), and $V_{0}$ is the on-site repulsion. The values $\phi
_{r,l} $ satisfy the periodicity conditions $\phi _{r,l}=\phi _{r,l+N}$. The
weight $W_{\Delta }[\tilde{\phi}]$ has the form%
\begin{equation}
W_{\Delta }[\tilde{\phi}]=\exp \Big[-\frac{\Delta }{2V_{0}}%
\sum_{r}\sum_{l=1}^{N}\phi _{r,l}^{2}\Big].  \label{c25ab}
\end{equation}%
in the region under the radius $R_{0}$. Outside this region the HS field is
equal to zero identically.

In order to write the correct functional $Z_{b}[\tilde{\phi}]$ in the
presence of the bath, the equations of Section~\ref{model} have to be
adjusted as concerns the shift of the chemical potential by the amount $%
-V_{0}/2$, Eq.~(\ref{a14}). We assume that main system is located in a
potential well and that the height of the walls equals $V_{0}/2$. In other
words, we shift the chemical potential $\mu $ everywhere in the space by $%
V_{0}/2$, although the electron-electron interaction is present only in the
main system, using the value
\begin{equation}
\mu ^{\prime }=\mu -V_{0}/2  \label{c25abc}
\end{equation}

This homogeneous shift is convenient because at $\mu ^{\prime }=0$ the
presence of the bath does not violate the particle-hole symmetry. Then, the
functional $Z_{b}[\tilde{\phi}]$ equals
\begin{equation}
Z_{b}[\tilde{\phi}]=Z_{0}\exp \Big[-\frac{1}{2}\sum_{r,\sigma
,a}\int_{0}^{\beta }\int_{0}^{1}\sigma \tilde{\phi}_{r}(\tau
)A_{a;r,r}^{\prime }\left( z\right) dud\tau \Big].  \label{c26}
\end{equation}%
where
\begin{equation}
Z_{0}=\prod_{k,\sigma }\Big[1+\exp \Big(-\beta \epsilon _{k}\Big)\Big]
\label{c26a}
\end{equation}%
and
\begin{equation}
\epsilon _{k}=-t\left( \mathbf{k}\right) -\mu +V_{0}/2.  \label{c26b}
\end{equation}%
In Eq.~(\ref{c26b}), $\mu $ is the chemical potential of the original
fermion model, Eq.~(\ref{a2}). The functions $A_{a;r,r}^{\prime }\left(
z\right) $ should be found from the linear equations
\begin{eqnarray}
&&\left(
\begin{array}{cc}
\gamma & \mathcal{M}_{r,r^{\prime }}+\frac{\partial }{\partial \tau } \\
\mathcal{M}_{r,r^{\prime }}-\frac{\partial }{\partial \tau } & -\gamma%
\end{array}%
\right) A_{a;r,r^{\prime }}\left( z\right)  \notag \\
&=&-un_{r,r^{\prime }}\sigma \tilde{\Phi}_{r,r^{\prime }}\left( \tau \right)
B_{a},  \label{c100}
\end{eqnarray}%
where $a=1,2$ and
\begin{equation*}
A_{1;r,r^{\prime }}\left( z\right) =\left(
\begin{array}{c}
A_{1;r,r^{\prime }}^{\prime }\left( z\right) \\
A_{1;r,r^{\prime }}^{\prime \prime }\left( z\right)%
\end{array}%
\right) ,\ B_{1}=\left(
\begin{array}{c}
0 \\
1%
\end{array}%
\right) ,
\end{equation*}%
\begin{equation*}
A_{2;r,r^{\prime }}\left( z\right) =\left(
\begin{array}{c}
A_{2;r,r^{\prime }}^{\prime \prime }\left( z\right) \\
A_{2;r,r^{\prime }}^{\prime }\left( z\right)%
\end{array}%
\right) ,\ B_{2}=\left(
\begin{array}{c}
1 \\
0%
\end{array}%
\right) .
\end{equation*}%
The operator $\mathcal{M}_{r,r^{\prime }}$ and the function $\tilde{\Phi}%
_{r,r^{\prime }}\left( \tau \right) $ are defined as%
\begin{eqnarray}
\mathcal{M}_{r,r^{\prime }}\left( z\right) &=&\hat{\varepsilon}_{r}-\hat{%
\varepsilon}_{r^{\prime }}-u\sigma \tilde{\Phi}_{r,r^{\prime }}(\tau ),
\notag \\
\tilde{\Phi}_{r,r^{\prime }}\left( \tau \right) &=&\tilde{\phi}_{r}\left(
\tau \right) -\tilde{\phi}_{r^{\prime }}\left( \tau \right) ,  \label{c101a}
\end{eqnarray}%
Finally, $n_{r,r^{\prime }}$ is the Fermi distribution
\begin{eqnarray}
n_{r,r^{\prime }} &=&\sum_{k}n_{k}e^{ik(r-r^{\prime })},  \label{c27a} \\
n_{k} &=&\left[ \exp \left\{ \beta \epsilon _{k}\right\} +1\right] ^{-1}.
\notag
\end{eqnarray}%
Both the functions $\tilde{\phi}_{r}\left( \tau \right) $ and $%
A_{r,r^{\prime }}\left( z\right) $, $z=\left( \tau ,\sigma ,u\right) $,
should be periodic in time: $\tilde{\phi}_{r}\left( \tau \right) =\tilde{\phi%
}_{r}\left( \tau +\beta \right) $ and
\begin{equation}
A_{a;r,r^{\prime }}\left( \tau ,\sigma ,u\right) =A_{a;r,r^{\prime }}\left(
\tau +\beta ,\sigma ,u\right)  \label{c102}
\end{equation}%
The real parameter $\gamma $ should be taken as small as necessary to be
sure that the final result is insensitive to its value.

In order to calculate the partition function $Z$, one should solve Eq.~(\ref%
{c100}-\ref{c102}) for a set of functions $\phi _{r,l}$ defined on the
slices and a small parameter $\gamma $, substitute the solution $%
A_{a;r,r^{\prime }}^{\prime }\left( z_{l}\right) $ into Eq.~(\ref{c26}) and
find the function $Z_{b}[\tilde{\phi}]$. This function is positive because
the solution $A_{a;r,r^{\prime }}(z_{l})$ must be real. Then, one can
calculate $Z$ in Eq.~(\ref{c25}) using $Z_{b}[\tilde{\phi}]$ as the
probability in the MC sampling for finite $\Delta $ diminishing this length
as much as necessary. We emphasize here that there should not be any
principle limitation of the computational accuracy provided a sufficiently
large system is considered and sufficiently small length of the time slices
is used.

These arguments are based on the fact that the solutions $A_{r,r^{\prime
}}\left( z\right) $ are real and non-singular, such that integrating over $u$
does not generate imaginary parts. This property of the solution follows
from the spectral expansion, Eqs.~(\ref{a112}, \ref{a115}) supplemented by
the proof that the real eigenvalues $E$ entering the sum in Eq.~(\ref{a112})
in pairs $E$ and $-E$ cannot be equal to zero (see Appendix \ref{solution}).
Moreover, due to the fact that the eigenstates enter in pairs with the
eigenvalues $E$ and $-E,$ the entire exponent in Eqs.~(\ref{a115}, \ref{c26}%
) does not become singular even for very small $E$ due to the mutual
compensation of the divergencies in the pairs. This follows also from the
symmetry property of the Green function, Eq.~(\ref{a113}). As all the
parameters of the equations are real, there is no danger to obtain an
imaginary part in the exponent in Eq.~(\ref{a115}) and therefore the weight $%
Z_{b}[\tilde{\phi}]$ is always positive.

In the limit of a very weak interaction $V_{0}$, typical $\tilde{\phi}%
_{r}\left( \tau \right) $ are small and one can neglect them in the
expression for $\mathcal{M}_{r,r^{\prime }}$, Eq.~(\ref{c101a}). Then,
eigenvalues $\lambda _{k}$ and eigenfunctions $v_{r}^{k}$ of this operator
do not depend on the number of the slices and can easily be obtained in the
form%
\begin{equation}
v_{r}^{k}=c_{\mathbf{k}}\left( \cos \mathbf{kr,\sin kr}\right) \text{,\quad }%
\lambda _{k}=\varepsilon _{\mathbf{k}},  \label{d10}
\end{equation}%
where $c_{k}$ are normalization coefficients.

In this limit, one can solve Eq.~(\ref{c100}) by Fourier transforming both
the sides in space and time. This allows one to write the Fourier
transformed solution as
\begin{equation}
A_{a}^{\prime }\left( \mathbf{k,k}^{\prime },\omega \right) =\frac{\sigma
u\left( \varepsilon _{\mathbf{k}}-\varepsilon _{\mathbf{k}^{\prime
}}-i\omega \right) \left( n_{\mathbf{k}}-n_{\mathbf{k}^{\prime }}\right)
\phi _{\mathbf{k-k}^{\prime },\omega }}{\omega ^{2}+\left( \varepsilon _{%
\mathbf{k}}-\varepsilon _{\mathbf{k}^{\prime }}\right) ^{2}+\gamma ^{2}}
\label{c103}
\end{equation}%
Substituting Eq.~(\ref{c103}) into Eqs.~(\ref{c25}, \ref{c26}) and taking
the limit $\gamma \rightarrow 0$ and $\Delta \rightarrow 0$ we come to the
RPA result, Eq.~(\ref{aRPA}). The integral over $u$ and summation over $%
\sigma $ are trivial in this calculation.

The presence of the regularizing parameter $\gamma $ results in the
vanishing of the contribution of the state with $\mathbf{k=k}^{\prime },$ $%
\omega =0$. However, the error is small because the sum extends over many
states with different $\mathbf{k}$ that inevitably exist due to the bath.

\subsubsection{Integral form of the equation}

The differential with respect to time equation (\ref{c100}) can be written
in the integral form with the help of the bare Green function $\mathcal{G}%
_{r,r^{\prime };r_{1},r_{1}^{\prime }}^{0}\left( \tau ,\tau _{1}\right) $
introduced as the solution of the equation
\begin{eqnarray}
&&\left(
\begin{array}{cc}
\gamma & \hat{\varepsilon}_{r}-\hat{\varepsilon}_{r^{\prime }}+\frac{%
\partial }{\partial \tau } \\
\hat{\varepsilon}_{r}-\hat{\varepsilon}_{r^{\prime }}-\frac{\partial }{%
\partial \tau } & -\gamma%
\end{array}%
\right) \mathcal{G}_{r,r^{\prime };r_{1},r_{1}^{\prime }}^{0}\left( \tau
,\tau _{1}\right)  \notag \\
&=&\delta _{r,r_{1}}\delta _{r^{\prime },r_{1}^{\prime }}\delta \left( \tau
-\tau _{1}\right)  \label{c104}
\end{eqnarray}
The function $\mathcal{G}_{r,r^{\prime };r_{1},r_{1}^{\prime }}^{0}\left(
\tau ,\tau _{1}\right) $ is a $2\times 2$ matrix and is uniquely defined by
the boundary conditions%
\begin{eqnarray}
\mathcal{G}_{r,r^{\prime };r_{1},r_{1}^{\prime }}^{0}\left( \tau ,\tau
_{1}\right) &=&\mathcal{G}_{r,r^{\prime };r_{1},r_{1}^{\prime }}^{0}\left(
\tau +\beta ,\tau _{1}\right)  \notag \\
&=&\mathcal{G}_{r,r^{\prime };r_{1},r_{1}^{\prime }}^{0}\left( \tau ,\tau
_{1}+\beta \right)  \label{c105}
\end{eqnarray}
The corresponding homogeneous equation does not have solutions due to the
presence of the parameter $\gamma $.

The bare Green function $\mathcal{G}_{r,r^{\prime };r_{1},r_{1}^{\prime
}}^{0}\left( \tau ,\tau _{1}\right) $ can easily be found by Fourier
transformation and can be written explicitly as%
\begin{eqnarray}
&&\mathcal{G}_{r,r^{\prime };r_{1},r_{1}^{\prime }}^{0}\left( \tau ,\tau
_{1}\right)  \label{d13} \\
&=&\frac{T}{\left( N_{d}^{total}\right) ^{2}}\underset{\mathbf{k,k}^{\prime
},\omega }{{\sum }}\frac{e^{i\mathbf{k}\left( \mathbf{r-r}_{1}\right) -i%
\mathbf{k}^{\prime }\left( \mathbf{r}^{\prime }\mathbf{-r}_{1}^{\prime
}\right) -i\omega \left( \tau -\tau _{1}\right) }}{\omega \Lambda
_{2}+\left( \varepsilon _{\mathbf{k}}-\varepsilon _{\mathbf{k}^{\prime
}}\right) \Lambda _{1}+\gamma \Lambda }  \notag
\end{eqnarray}%
where the Pauli matrices $\Lambda ,$ $\Lambda _{1},$ $\Lambda _{2}$ are
specified in Eq.~(\ref{a103b}).

Eq.~(\ref{d13}) is written for the entire system including the bath and the
momenta $\mathbf{k,k}^{\prime }$ correspond to this enlarged system.
Accordingly, the total number of sites $N_{d}^{total}$ enters the
normalization factor in Eq.~(\ref{d13}). The state with $\omega =0$, $%
\varepsilon _{\mathbf{k}}=\varepsilon _{\mathbf{k}^{\prime }}$ is included
in the sum in Eq.~(\ref{d13}).

One can perform summation over the bosonic frequencies $\omega =2\pi Tn$ in
Eq.~(\ref{d13}) using the formula%
\begin{equation}
T\sum_{\omega }\frac{\exp \left( -i\omega \left( \tau -\tau _{1}\right)
\right) }{-i\omega +a}=\frac{sgn\left( \tau -\tau _{1}\right) \exp \left(
-a\left( \tau -\tau _{1}\right) \right) }{1-\exp \left( -a\beta sgn\left(
\tau -\tau _{1}\right) \right) }  \label{d9a}
\end{equation}%
where the symbol \textquotedblleft $sgn$" stands for the sign and $a$ is a
number. This gives the following expression for the Green function $\mathcal{%
G}_{r,r^{\prime };r_{1},r_{1}^{\prime }}^{0}\left( \tau ,\tau _{1}\right) $
for arbitrary coordinates and times
\begin{eqnarray}
&&\mathcal{G}_{r,r^{\prime };r_{1},r_{1}^{\prime }}^{0}\left( \tau ,\tau
_{1}\right) =g\left[ \frac{\exp \left[ b\left( \tau -\tau _{1}\right) \hat{f}%
_{r_{1},r_{1}^{\prime }}\right] }{2\sinh \left[ \beta \hat{f}%
_{r_{1},r_{1}^{\prime }}/2\right] }\right] \delta _{r,r_{1}}\delta
_{r^{\prime },r_{1}^{\prime }},  \notag \\
&&g=\left(
\begin{array}{cc}
0 & -1 \\
1 & 0%
\end{array}%
\right) ,\quad \hat{f}_{r,r^{\prime }}=\left( \hat{\varepsilon}_{r}-\hat{%
\varepsilon}_{r^{\prime }}\right) \Lambda -\gamma \Lambda _{1}\quad
\label{d15a}
\end{eqnarray}%
where
\begin{equation}
b\left( \tau -\tau _{1}\right) =\,sgn\left( \tau -\tau _{1}\right) \beta
/2-\left( \tau -\tau _{1}\right) ,  \label{d9c}
\end{equation}%
In Eq.~(\ref{d15a}) the operators $\hat{\varepsilon}_{r_{1}}$ and $\hat{%
\varepsilon}_{r_{1}^{\prime }}$ act on the corresponding variables entering
the $\delta $-functions on the right. The function $\mathcal{G}_{r,r^{\prime
};r_{1},r_{1}^{\prime }}^{0}\left( \tau ,\tau _{1}\right) $, Eq.~(\ref{d15a}%
), is explicitly real and free of singularities.

Using the Green function $\mathcal{G}_{r,r^{\prime };r_{1},r_{1}^{\prime
}}^{0}\left( \tau ,\tau _{1}\right) ,$ Eq.~(\ref{d15a}), we reduce Eq.~(\ref%
{c100}) to the form%
\begin{align}  \label{d12}
\Big[-& \sum_{r_{1},r_{1}^{\prime }}\int_{0}^{\beta }\mathcal{G}%
_{r,r^{\prime };r_{1},r_{1}^{\prime }}^{0}\left( \tau ,\tau _{1}\right)
u\sigma \tilde{\Phi}_{r_{1},r_{1}^{\prime }}(\tau _{1})\Lambda _{1} & &
\notag \\
& \qquad +\delta \left( \tau -\tau _{1}\right) \delta _{r,r_{1}}\delta
_{r^{\prime }r_{1}^{\prime }}\Big]A_{a;r_{1},r_{1}^{\prime }}\left( \tau
_{1},\sigma ,u\right) d\tau _{1} & &  \notag \\
& =-\sum_{r_{1},r_{1}^{\prime }}\int \mathcal{G}_{r,r^{\prime
};r_{1},r_{1}^{\prime }}^{0}(\tau ,\tau _{1})un_{r_{1},r_{1}^{\prime
}}\sigma \tilde{\Phi}_{r_{1},r_{1}^{\prime }}\left( \tau _{1}\right)
B_{a}d\tau _{1} &
\end{align}%
Eq.~(\ref{d12}) can serve as the basis of many numerical approaches. One
can, e.g., solve this equation recursively neglecting in the zeroth
approximation the function $\tilde{\Phi}_{r_{1},r_{1}^{\prime }}\left( \tau
_{1}\right) $ in the L.H.S. and then calculating all orders of $\tilde{\Phi}%
_{r_{1},r_{1}^{\prime }}\left( \tau _{1}\right) $ or one can just invert the
operator acting on $A_{a;r,r^{\prime }}\left( z\right) $ in the L.H.S. of
the equation. Due to the regularization, the zero eigenvalues of the
operator in the L.H.S. are discarded. Therefore all these operations are
well defined and one should not encounter any singularity. In particular,
one can invert the operator acting on $A_{a;r,r^{\prime }}$ in the L.H.S. of
Eq.~(\ref{d12}).

As a result, after integration over $u$ the exponent in the function $Z_{b}[%
\tilde{\phi}]$ remains real and $Z_{b}[\tilde{\phi}]$ positive. This
demonstrates that this bosonization method of computations is free of the
sign problem. Substituting the solution obtained by inversion of the L.H.S.
operator into Eq.~(\ref{c26}), the function $Z_{b}[\tilde{\phi}]$ is given
after discretization by
\begin{widetext}
\begin{align}
Z_b[\tilde{\phi}]=Z_0 \exp \Big[\frac{1}{2}\int_0^1\Delta\sum_{l,l_1=1}^N\sum_{r,r_1,r_1';\sigma}\sigma\phi_{r}(\tau)n_{r_1,r_1'}\;
Tr\left[
\left\{1-u\Delta\ \mathcal{G}^0\sigma\tilde{\Phi}\Lambda_1\right\}^{-1}
\right]_{\{r,r\},\{r_1,r_1'\}}^{l,l_1}\ du \Big]
.\label{d17}
\end{align}
\end{widetext}In Eq.~(\ref{d17}) the symbol $Tr$ stands for the trace of the
$2\times 2$ matrices introduced in the regularization, Eq.~(\ref{a103}). The
expression entering the square brackets under $Tr$ is a $2\times 2$ matrix
in this \textquotedblleft regularization space" and an $N_{d}^{2}\times
N_{d}^{2}\times N\times N$ matrix with the space $\left\{ r,r^{\prime
}\right\} ,$ $\left\{ r_{1},r_{1}^{\prime }\right\} $ and time $\tau ,\tau
_{1}$ matrix elements. This matrix is obtained from the matrix $\mathcal{G}%
^{0}\sigma \tilde{\Phi}\Lambda _{1}$ using the conventional rules of matrix
operations. The only peculiarity is that the coordinates enter in pairs $%
\left\{ r,r^{\prime }\right\} $ and $\left\{ r_{1},r_{1}^{\prime }\right\} .$
Matrix elements can be written explicitly as
\begin{equation}
\left[ \mathcal{G}^{0}\sigma \tilde{\Phi}\Lambda _{1}\right] _{\left\{
r,r^{\prime }\right\} ,\left\{ r_{1},r_{1}^{\prime }\right\} }^{\tau ,\tau
_{1}}=\ \mathcal{G}_{r,r^{\prime };r_{1},r_{1}^{\prime }}^{0}\left( \tau
,\tau _{1}\right) \sigma \tilde{\Phi}_{r_{1},r_{1}^{\prime }}(\tau
_{1})\Lambda _{1}.  \label{d18}
\end{equation}

Eq.~(\ref{d17}) can directly be used for computations. The integral over~$u$
cannot lead to singularities, which is guaranteed by the absence of the zero
eigenvalues of the operator in the L.H.S. of Eq.~(\ref{d12}) written for an
arbitrary $u$. This property is valid for an arbitrarily large $\tilde{\Phi}$%
.

In principle, one could easily perform the integration over~$u$ already
analytically before doing numerics, which gives in the continuous time limit
\begin{eqnarray}
&&Z_{b}[\tilde{\phi}]=Z_{0}\exp \Big[-\frac{1}{2}\sum_{r,r_{1},r_{1}^{\prime
}}\int_{0}^{\beta }\int_{0}^{\beta }d\tau d\tau _{1}\tilde{\phi}_{r}\left(
\tau \right) n_{r_{1},r_{1}^{\prime }}  \notag \\
&\times &Tr\Big[(\mathcal{G}^{0}\tilde{\Phi}\Lambda _{1})^{-1}\ln \Big(1-(%
\mathcal{G}^{0}\tilde{\Phi}\Lambda _{1})^{2}\Big)\Big]_{\left\{ r,r\right\}
,\left\{ r_{1},r_{1}^{\prime }\right\} }^{\tau ,\tau _{1}}\Big]\ ,  \notag \\
&&  \label{d17a}
\end{eqnarray}%
The integrand in Eq.~(\ref{d17a}) is free of singularities as well. At small
$\tilde{\phi}_{r}\left( \tau \right) $ one can expand the logarithm and take
the first order of the expansion. This leads us again to the RPA formula,
Eq.~(\ref{aRPA}). However, Eq. (\ref{d17}) looks more convenient for MC
computations because it allows one to peform updating more easily.

Note that in the present paper, we decoupled the interaction by the Gaussian
integration over $\tilde{\phi}_{r}\left( \tau \right) $. For the on-site
interaction, one can also use the \textquotedblleft Ising spin" auxiliary
field~$s_{r,l}$ \cite{hirsch}. All the steps of the derivation of the
functional $Z_{b}[\tilde{\phi}]$ can be repeated with this field. The final
expression, Eq.~(\ref{c25}), should be replaced by the following sum over
all configurations of $s_{r,l}$%
\begin{equation}
Z=\left( \frac{1}{2}\right) ^{N_{d}N}\sum_{s_{r,l}}Z_{b}[\tilde{\phi}],
\label{c28b}
\end{equation}%
where $N_{d}$ is the number of the sites in the system. The functional $%
Z_{b}[\tilde{\phi}]$ for the Ising spin field has the same form as
previously with the field $\tilde{\phi}_{r}\left( \tau \right) =\bar{\lambda}%
s_{r,l}/\Delta ,$ $\cosh \bar{\lambda}=\exp \left( V_{0}\Delta /2\right) $,
for $\left( l-1\right) \Delta \leq \tau <l\Delta $. Using the
\textquotedblleft Ising spin" auxiliary field can further improve the
efficiency of the computational schemes.

\subsubsection{A numerical test}

As we have discussed previously the final formulas, Eqs.~(\ref{d12}, \ref%
{c26}) or Eq.~(\ref{d17}), are different from the formula of the original
fermionic model, Eqs.~(\ref{c14}-\ref{c15}). These deviations are due to the
introduced bath and states effectively discarded by introducing the
parameter $\gamma $. So, generally speaking there is no reason to expect
that the functions $Z_{f}[\tilde{\phi}]$ and $Z_{b}[\tilde{\phi}]$ are close
to each other. As we have argued, they should be considerably different for,
e.g., piece-wise functions with big jumps. The agreement between the
partition functions $Z$ is possible only after integration over $\tilde{\phi}%
_{r}$ and only in the limit of a large number of the sites $N_{d}^{total}$
in the system.

However, what one can expect is the agreement between the values of $Z_{f}[%
\tilde{\phi}]$ and $Z_{b}[\tilde{\phi}]$ for static $\tilde{\phi}$ (one
slice in time) for large $N_{d}$. Although we have demonstrated this
agreement analytically in Appendix \ref{statics}, it is instructive to check
this property numerically using Eqs.~(\ref{d12}, \ref{c26}) or Eq.~(\ref{d17}%
). Such a test can also give a feeling of how well the numerical procedure
based on using these equations can work.

\begin{figure}[t]
\includegraphics[width = 0.8\linewidth]{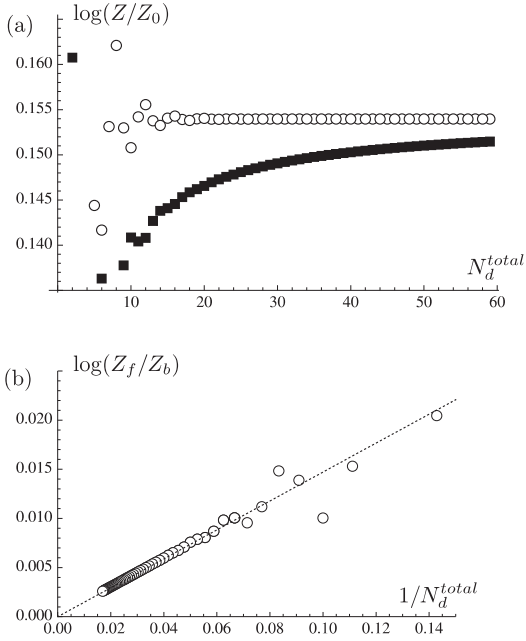}
\caption{ (a) Numerical evaluation of $\log(Z_{f}/Z_{0})$ (upper curve) and $%
\log(Z_{b}/Z_{0})$ (lower curve) in the case of two sites Hubbard model
imbedded in a bath of $N_{d}^{total}-2$ sites in the static case (one time
slice). The hopping parameter is taken as $t=4.0$, the inverse temperature $%
\protect\beta =1.0$, the chemical potential $\protect\mu' =0.1$, and the
regularization parameter $\protect\gamma =10^{-2}$. The HS fields on sites $%
1 $ and $2$ are respectively $\protect\phi _{r=1}=-1.0$ and $\protect\phi %
_{r=2}=1.0$. The entire system is a one dimensional ring in which the field $%
\protect\phi $ can be finite only on two neighboring sites. (b) The
deviation $\log(Z_{f}/Z_{b})$ as a function of the inverse number of the
sites in the system for the same parameters showing the $1/N_{d}^{total}$
rate of convergence.}
\label{test}
\end{figure}

We present in Fig.~\ref{test} an example of the convergence between $Z_{f}$
and $Z_{b}$ in a bath for the static case of one time slice. We consider a
two sites Hubbard model embedded in a bath of $N_{d}^{total}-2$ sites with
periodic boundary conditions. The HS fields $\phi _{r}$ are not equal to
zero on the sites $r=\{1,2\}$ but are equal to zero elsewhere, such that
only the sites $1$ and $2$ are interacting. Both analytical and numerical
integration over $u$ in Eq. (\ref{d17}) are suitable and we use both of
them. Integrating over $u$ in Eq.~(\ref{d17}) analytically makes the
computation faster but we have found a good agreement between the results
obtained using both the approaches.

In Fig.~\ref{test}~(a) the convergence between $\log\left(
Z_{f}/Z_{0}\right) $ (upper curve) and $\log\left( Z_{b}/Z_{0}\right) $
(lower curve) is demonstrated in the static case, as the number of the sites
in the bath is increased. Although the difference between these quantities
is fairly small, the complete convergence is not very fast. Therefore, in
order to understand the dependence of the deviation $\log\left(
Z_{f}/Z_{b}\right) $ on the size of the system we plot in Fig. \ref{test}
(b) this quantity as a function of the inverse total number of the sites $%
N_{d}^{total}$ in the system. As previously, the HS is non-zero on two sites
only and all the parameters are the same.

Fig. \ref{test} (b) shows that the deviation $\log \left( Z_{f}/Z_{b}\right)
$ decays with the size of the sample inversely proportional to the total
number of the sites $N_{d}^{total}$. This dependence is natural because the
regularization with the parameter $\gamma $ cuts out $N_{d}^{total}$ of $%
\left( N_{d}^{total}\right) ^{2}$ states in the system. This example of the
convergence completes the analytical proof of Appendix \ref{statics}, and
shows that the bath is necessary in order to equate $Z_{f}$ and $Z_{b}$. It
shows as well that for any finite size system, $Z_{f}$ and $Z_{b}$ are
different, $Z_{b}$ converging asymptotically towards $Z_{f}$ within an
infinite bath.

\begin{figure}[t]
\includegraphics[width = 0.8\linewidth]{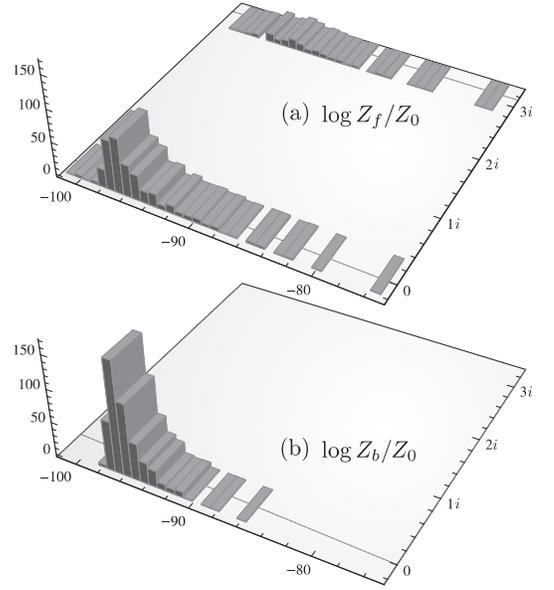}
\caption{ Distribution of (a) $\log (Z_{f}/Z_{0})$ and (b) $\log
(Z_{b}/Z_{0})$ for $500$ configurations for $\protect\phi _{r}(\protect\tau )
$ randomly chosen according to the Gaussian distribution, Eq.~(\protect\ref%
{a12}), for a system containing two interacting and one bath site in a ring
geometry with parameters $t=1.0$, $\protect\mu =2.0$, $V_{0}=3.0$, $\protect%
\beta =14$, and $\protect\gamma =10^{-3}$ and for three time slices. Only $%
\log (Z_{f}/Z_{0})$ can have the imaginary part leading to the negative sign
of the fermionic determinant. }
\label{test_spreading}
\end{figure}

The numerical procedure carried out here can be extended in a
straightforward way to the case of the HS field $\tilde{\phi}_{r}\left( \tau
\right) $ varying in time, which corresponds to the case of many slices $N$.
According to our analytical consideration there should not be principal
problems with this general case and the function $Z_{b}[\tilde{\phi}]$ must
remain real and positive. This is demonstrated in Fig. \ref{test_spreading}

For the calculation of the bosonic function~$Z_{b}[\tilde{\phi}]$, the
integral over~$u$ in Eq.~(\ref{d17}) has been calculated analytically (see
Eq. (\ref{d17a})). The resulting matrix entering $\log (Z_{b}[\tilde{\phi}]/Z_0)$
was evaluated using numerical diagonalization. This quantity remains real to
a high precision [Imaginary parts due to numerical inaccuracy are of order $%
10^{-13}$] showing the positivity of $Z_{b}[\tilde{\phi}]$, whereas the
imaginary part of $\log Z_{f}[\tilde{\phi}]$ jumps frequently between $0$
and $\pi $ showing the sign fluctuations of $Z_{f}[\tilde{\phi}]$. It is
interesting to note that $\log Z_{f}$ is spread over a broader region of the
real axis than $\log Z_{b}[\tilde{\phi}]$.

The form of the distribution of $\log Z_{b}[\tilde{\phi}]$, Fig. \ref%
{test_spreading}, shows that the sign problem is avoided in the bosonized
representation. Moreover, the distribution is narrower than that of $\log
Z_{f}[\tilde{\phi}]$ and we hope that the computational scheme may be
efficient.

\section{Discussion\label{dis}}

We have mapped models of interacting fermions onto models describing
collective bosonic excitations. This was performed by decoupling the
interaction between the fermions by a Hubbard-Stratonovich transformation
and considering non-interacting fermions in the fluctuating HS field. The
standard step of tracing out the fermions, which led to the fermionic
determinant, was followed by converting this determinant into an integral
containing in the integrand the solution of a linear differential equation.
The latter step resembles writing quasiclassical equations of Ref.~%
\onlinecite{aleiner} but is now exact in the thermodynamic limit, which has
not been anticipated previously.

The basic equation is actually an analogue of the von Neumann equation for
the density matrix and our transformation can be interpreted by analogy with
the replacement of the description of quantum mechanics in terms of wave
functions by the description with the density matrix. The solution of the
equation $A_{r,r^{\prime }}\left( \tau \right) $ plays the role of the
density matrix. It obeys the bosonic boundary conditions $A_{r,r^{\prime
}}\left( \tau \right) =A_{r,r^{\prime }}\left( \tau +\beta \right) $ and is
a bosonic field.

In order to regularize the obtained equations we assumed that the model of
the interacting electrons is imbedded in a bath. The latter is a part of the
sample where the electrons do not interact with each other. In other words,
the sample we consider is in a contact with metallic leads. The introduction
of the environment can serve as an additional support of the the analogy of
our field $A_{r,r,}\left( \tau \right) $ with the density matrix.

We have checked the suitability of the new bosonic model for both analytical
and numerical computations.

As concerns the analytical approach, we have expressed the solution of the
equation for $A_{r,r^{\prime }}\left( \tau \right) $ in terms of a
functional integral over superfields $\Psi $ using the well-known in field
theory Becchi-Rouet-Stora-Tyutin transformation \cite{faddeev,brst}. This
allowed us to integrate out the HS field $\phi $ and reduce the original
fermionic model to a model of interacting bosons. One can work with this
model using expansions in the interaction and we have compared the results
of the perturbation theory with those for the initial fermionic model up to
the second order in the interaction. This comparison demonstrated the exact
agreement.

In a short future, a renormalization group scheme analogous to the one
suggested previously in the quasiclassical approach \cite{aleiner} will be
developed and applied for studying anomalous contributions to
thermodynamical quantities. More complicated models for strongly correlated
systems can also be studied analytically using the bosonization approach
developed here.

As concerns using the developed scheme for the MC calculations, the
superfield theory obtained after averaging over $\phi $ with the help of the
superfields $\Psi $ is not directly suitable for this purpose. Since we do
not know how to reach the final form without introducing integrals over
Grassmann variables, the only possibility for numerics is to calculate using
the bosonic model in the fluctuating HS field. There is a very strong
motivation to try to study the bosonic model using the MC method. It is
well-known that using the MC method for fermion models suffers generically
from the famous negative sign problem. Since we have mapped the fermionic
model onto a bosonic one, one might have a hope that the sign problem will
not show up.

We have analyzed this idea and come to the conclusion that the MC
simulations for the bosonized model shall be free of the sign problem.
Practically, one can simply forget how the bosonic model has been obtained
from the initial fermionic one and start investigating it by discretizing
time. Then one can make a HS transformation with a discrete field $\tilde{%
\phi}_{r}\left( \tau \right) $ and obtain the formulas of subsection \ref%
{num}. This would be logically sufficient for justifying our scheme.

However, we have considered this issue in great detail in subsections \ref%
{discrete} and \ref{sign}. When deriving the basic equation for the bosonic
field $A_{r,r^{\prime }}\left( \tau \right) $, the fact that only the first
order derivative with respect to $\tau $ enters the equations for the
electron Green functions, Eqs. (\ref{a21}, \ref{a22}), is crucial. Numerical
schemes are based on discretizing the time but just replacing the derivative
by a finite difference would make the reduction to the bosonic field $%
A_{r,r^{\prime }}\left( \tau \right) $ impossible. We suggested a procedure
of time discretization that enabled us to use a piece-wise HS field $\tilde{%
\phi}_{r,l}$ keeping the time continuous and carry our the bosonization. A
crucial step in the transformation of the fermions into the bosons is the
regularization based on the introduction of the bath and of the small
parameter $\gamma $.

The importance of this regularization can be understood considering the
conventional diagrammatic expansion for fermions. In this language, the
bosonization means converting pairs of the fermion Green functions into a
propagator of bosonic excitations. However, it is not possible to do this
transformation if two or more Green functions have coinciding Matsubara
frequencies and momenta.

The regularization based on introducing a small parameter $\gamma $ removes
such states and the presence of the bath makes their contribution small in
the limit of a large number of sites in the bath. Our mapping can be used
for any temperature, interaction and dimension of the system and, in this
sense, the mapping is exact. At the same time, the presence of a
sufficiently large bath is necessary to provide the agreement between the
boson and fermion models.

We have discussed in subsections \ref{sign} and \ref{sign2} the origin of
the negative sign in the fermionic determinant $Z_{f}[\tilde{\phi}]$ and
demonstrated that the corresponding functional $Z_{b}[\tilde{\phi}]$, being
generally different from $Z_{f}[\tilde{\phi}]$, is always positive. We have
done this representing the solution for the differential equation for the
bosonic field $A_{r,r^{\prime }}\left( \tau \right) $ in terms of a sum
containing eigenfunctions and eigenvalues, Eqs.~(\ref{a112}, \ref{a115}).
With the regularization scheme developed in Section \ref{model}, all the
quantities entering these equations are real. In the language of Eqs. (\ref%
{a112}, \ref{a115}), one would encounter the sign problem if an eigenvalue $%
E^{K}$ could turn to zero at certain $u$. Then, the integration over $u$
near such poles might generate an imaginary part and make the function $%
Z_{b}[\phi ]$ not necessarily real and positive. However, we have
demonstrated (see Appendix \ref{solution}) that the eigenvalues $E^{K}$
cannot turn to zero and there are no singularities in this spectral
expansion. This allowed us to demonstrate that the distribution $Z_{b}[%
\tilde{\phi}]$ that should be used for MC sampling is always real and
positive, hence showing that there is a chance to overcome the negative sign
problem by the bosonization.

Although the functionals $Z_{f}[\tilde{\phi}]$ and $Z_{b}[\tilde{\phi}]$ may
even have opposite signs for certain functions $\tilde{\phi}$, we present
arguments that after integration over $\tilde{\phi}$ the partition function $%
Z_{b}$ obtained in the bosonic model has to be a good approximation to the
partition function $Z_{f}$ of the original fermion model provided both the
models are taken with the bath and the regularizing parameter $\gamma $ is
small. The absence of the sign oscillations in $Z_{b}[\tilde{\phi}]$ should
lead to smaller variations of the modulus of this quantity and, hence, to a
better convergence in the MC simulations.

The logics of all these manipulations can be illustrated calculating two
different integrals%
\begin{equation}
I_{1}=\sqrt{2}\int_{0}^{\infty }\cos ax^{2}dx\text{,\quad }%
I_{2}=\int_{0}^{\infty }e^{-ax^{2}}dx  \label{e1}
\end{equation}%
Calculating the integral $I_{1}$ analytically, one can reduce it by turning
the contours of the integration to the integral $I_{2}$ and show exactly
that $I_{1}=I_{2}$ for any $a$.

At the same time, we can discretize the variable $x$, replace the integrals
by sums and calculate these sums by the MC method. Then, one can see that
difficulties in the calculation of these sums are quite different. The
integrand of $I_{2}$ is positive and the sum converges very fast. In
contrast, the integrand of $I_{1}$ oscillates faster and faster with growing
$x$ and one cannot use it as the weight in the MC sampling. Of course, one
could replace the integrand by its modulus and use the latter as the weight
but then one would calculate the average sign of $\cos ax^{2}$ as in Eq.~(%
\ref{c3}), which is not a very pleasant task.

Our scheme of the replacement of the fermionic determinant $Z_{f}[\tilde{\phi%
}]$ by the bosonic functional $Z_{b}[\tilde{\phi}]$ resembles the analytical
reduction of the integral $I_{1}$ to the integral $I_{2}$. This
transformation can easily be carried out in the continuous limit by using
the powerful theory of complex variables but making the same for finite
length of slices is difficult and cannot be done exactly. It is clear that
the discrete versions of the integrals $I_{1}$ and $I_{2}$ are different and
the only justification for a replacement of $I_{1}$ by $I_{2}$ is that they
must be equal to each other in the continuous limit.

A nice example relating the occurrence of the sign problem to the
representation frame of a physical problem is discussed on p.~103 of Ref.~%
\onlinecite{linden}. In this example, the model of a single spin in an
external magnetic field is discussed. Although the corresponding Hamiltonian
$H=-\mathbf{h}\cdot \bm{\sigma}$ can be solved in a simple way, tackling the
problem using the path integral approach can either lead to sums containing
both positive and negative terms [choosing $\mathbf{h}=(h_{x},h_{y},0)$ to
lie in the $xy$-place], or lead to purely positive matrix elements [after
rotating~$\mathbf{h}$ into the $xz$-plane, $\mathbf{h}=(h_{x},0,h_{z})$].
While the result of the former choice resembles the fermionic sign problem,
this sign problem is apparently \textquotedblleft solved\textquotedblright\
by switching to the other reference frame.

This illuminating example shows that in some cases the sign problem can be
overcome by a simple analytical transformation. Our approach for solving the
fermionic sign problem looks similar: we make a rotation in a
\textquotedblleft generalized space\textquotedblright\ of bosons and
fermions from the \textquotedblleft fermion plane\textquotedblright\ to the
\textquotedblleft boson plane\textquotedblright . Actually, it is not
surprising that one does not encounter the sign problem in a bosonic model,
but the fact that there can be an exact mapping between a fermion and boson
model has not been anticipated previously.

The sign problem is a long standing problem of quantum MC computations for
fermionic systems. There have been many attempts to overcome this problem
writing sophisticated algorithms but it has not generally been solved until
now. It has even been asserted \cite{troyer} that the sign problem belonged
to the class of NP-hard problems \cite{cook}, which would imply that chances
to solve it are very low. However, we do not think that the arguments
presented in Ref. \onlinecite{troyer} can be considered as a rigorous proof.
Even if our bosonization scheme will really work for the fermion models, we
do not know yet how to apply it for any NP problem.

The functional $Z_{b}[\tilde{\phi}]$ can be computed for any $\tilde{\phi}%
_{r}\left( \tau \right) $ using directly Eq. (\ref{d17}). One can perform
integration over $u$ either numerically or analytically. The numerical
integration over $u$ looks more convenient because one can easily carry out
the MC updating. Then, one can use the standard MC scheme for integration
over the field $\tilde{\phi}_{r}\left( \tau \right) $. One can also use the
\textquotedblleft Ising spin" auxiliary field \cite{hirsch} instead of the
Gaussian field $\phi $, which can make the computation faster.

In order to make an independent check and get a feeling of how one could
compute using Eq. (\ref{d17}) we have calculated the function $Z_{b}[\tilde{%
\phi}]$ for a static field $\phi _{r}$ integrating over $u$ both
analytically and numerically and found a good agreement with the
corresponding function $Z_{f}[\tilde{\phi}]$ of the original fermionic
model. We do not expect essential difficulties in extending the computation
to time-dependent HS fields $\tilde{\phi}_{r}\left( \tau \right) $ and hope
that our bosonization scheme will be checked numerically in the nearest
future.

\section*{Acknowledgements}

We thank Transregio 12 of DFG, the French ANR for financial support and the
Aspen Center for Physics where part of this work was completed. We
acknowledge useful discussions with Y. Alhassid, A. Bulgac, D. Galanakis, M.
Jarrell, S. Kettemann, M.Yu. Kharitonov, O. Parcollet, M. Troyer, Ph. Werner
and S-X. Yang.

\appendix

\section{Absence of zero eigenvalues in the regularized model}

\label{solution}

Here, we demonstrate that the operator $\mathcal{H}_{r,r^{\prime }}\left(
\tau \right) $, Eq.~(\ref{a103}), does not have zero eigenvalues $E^{K}$.
This property is guaranteed by the presence of the regularizer $\gamma $. If
we put $\gamma =0$ in Eq.~(\ref{a103}), zero eigenvalues $E$ exist.

Putting $\gamma =0$ we come to equations (\ref{a106}, \ref{a107}). As the
operators in the L.H.S. are not hermitian, the eigenvalues $\lambda ^{K}$
are complex and can turn to zero for certain fields $\phi _{r}\left( \tau
\right) $.

Suppose that the modulus of an eigenvalue $\lambda $ becomes very small or
can even turn to zero and let us show that, nevertheless, the corresponding
eigenvalue $E$ remains finite for a finite $\gamma $.

Let $v_{r,r^{\prime }}\left( \tau \right) $ be the eigenfunction
corresponding to the eigenvalue $\lambda $ in the first equation in (\ref%
{a106}). Another function $\bar{v}_{r,r^{\prime }}^{\ast }\left( \tau
\right) $ should correspond to the eigenvalue $\lambda ^{\ast }$ in the
second equation in (\ref{a107}).

The small values of $\left\vert \lambda \right\vert $ correspond to small
values of the eigenvalues $E$ of Eq.~(\ref{a108}) of a state with an
eigenvector $S_{r,r^{\prime }}\left( \tau \right) $. At $\gamma =0$ and $%
\lambda =0$ the solution of Eq.~(\ref{a108}) is very simple. One obtains
immediately $E=0$, while the solution $S_{r,r^{\prime }}\left( \tau \right) $
can be written in the form%
\begin{equation}
S_{r,r^{\prime }}\left( \tau \right) =c_{1}S_{r,r^{\prime }}^{a}\left( \tau
\right) +c_{2}S_{r,r^{\prime }}^{b}\left( \tau \right)  \label{e100}
\end{equation}%
with the vectors $S_{r,r^{\prime }}^{a}\left( \tau \right) $ and $%
S_{r,r^{\prime }}^{b}\left( \tau \right) $ given by
\begin{equation}
S_{r,r^{\prime }}^{a}\left( \tau \right) =\left(
\begin{array}{c}
\bar{v}_{r,r^{\prime }}^{\ast }\left( \tau \right) \\
0%
\end{array}%
\right) ,\ S_{r,r^{\prime }}^{b}\left( \tau \right) =\left(
\begin{array}{c}
0 \\
v_{r,r^{\prime }}\left( \tau \right)%
\end{array}%
\right)  \label{e101}
\end{equation}%
and arbitrary coefficients $c_{1}$ and $c_{2}$. This means that the state
with $E=0$ is degenerate.

The case of small $\lambda $ and $\gamma $ can be considered using quantum
mechanical perturbation theory. As the state with $E=0$ is degenerate we
seek for the solution $S_{r,r^{\prime }}\left( \tau \right) $ writing it in
the form of Eq.~(\ref{e100}). Following the standard scheme other states are
neglected in this expansion.

Substituting Eq.~(\ref{e100}) into Eq.~(\ref{a108}) we multiply both sides
of the equation first by $S_{r,r^{\prime }}^{a\ast }\left( \tau \right) ,$
then by $S_{r,r^{\prime }}^{b\ast }\left( \tau \right) $, sum over $%
r,r^{\prime }$ and integrate over $\tau $. This gives us a system of two
equations for the coefficients $c_{1}$ and $c_{2}$,%
\begin{eqnarray}
\left( E-\gamma \right) x_{1}c_{1}-c_{2}\lambda &=&0  \label{e102} \\
-c_{1}\lambda ^{\ast }+\left( E+\gamma \right) x_{2}c_{2} &=&0  \notag
\end{eqnarray}%
where
\begin{eqnarray}
x_{1} &=&\sum_{r,r^{\prime }}\int_{0}^{\beta }\left\vert \bar{v}%
_{r,r^{\prime }}\left( \tau \right) \right\vert ^{2}d\tau ,  \label{e103} \\
x_{2} &=&\sum_{r,r^{\prime }}\int_{0}^{\beta }\left\vert v_{r,r^{\prime
}}\left( \tau \right) \right\vert ^{2}d\tau .  \notag
\end{eqnarray}%
Non-trivial solutions of Eq.~(\ref{e102}) exist provided the determinant
equals to zero. This condition gives us the eigenvalues $E$%
\begin{equation}
E=\pm \sqrt{\gamma ^{2}+\frac{\left\vert \lambda \right\vert ^{2}}{x_{1}x_{2}%
}}.  \label{e104}
\end{equation}%
Eq.~(\ref{e104}) demonstrates explicitly that the eigenvalues cannot turn to
zero even in the situation when $\lambda =0$. This is a well known effect of
level repulsion. The absence of the zero eigenvalues $E$ justifies our
method of the regularization.

\section{Checking the regularization for a static Hubbard-Stratonovich field}

\label{statics}

In this Appendix we demonstrate that using the spectral expansion, Eqs.~(\ref%
{a112}, \ref{a115}), we come in the case of static HS fields back to Eq.~(%
\ref{a19}). This can serve as a check of the transformations we have carried
out in order to derive Eqs.~(\ref{a112}, \ref{a115}) and can help the reader
to visualize our regularization scheme.

Assuming that the HS field $\phi _{r}$ does not depend on time we write time
independent solutions $S_{r,r^{\prime }}^{K}$ of Eq.~(\ref{a108}) in the form%
\begin{equation}
S_{r,r^{\prime }}^{K}=\left(
\begin{array}{c}
S_{1;r,r^{\prime }}^{K} \\
S_{2;r,r^{\prime }}^{K}%
\end{array}%
\right) =\left(
\begin{array}{c}
a \\
b%
\end{array}%
\right) w_{r}^{k}w_{r^{\prime }}^{k^{\prime }}  \label{e105}
\end{equation}%
where $a$ and $b$ are coefficients and the functions $v_{r}^{k}$ are the
eigenfunctions of the operator $\hat{h}_{r},$ Eq.~(\ref{a21})%
\begin{equation}
\hat{h}_{r}w_{r}^{k}=\lambda ^{k}w_{r}^{k}  \label{e106}
\end{equation}

In the static case, only time independent solutions contribute into the
exponent in Eq.~(\ref{a115}) and for such states $K=\left\{ k,k^{\prime
}\right\} $.

Substituting Eq.~(\ref{e105}) into Eq.~(\ref{a108}) we come to a system of
two linear equations%
\begin{eqnarray}
a\left( \lambda ^{k}-\lambda ^{k^{\prime }}\right) -b\gamma &=&E^{K}b
\label{e107} \\
a\gamma +b\left( \lambda ^{k}-\lambda ^{k^{\prime }}\right) &=&E^{K}a  \notag
\end{eqnarray}%
Solution of these equations leads to the eigenvalue%
\begin{equation}
E^{K}=\pm \sqrt{\gamma ^{2}+\left( \lambda ^{k}-\lambda ^{k^{\prime
}}\right) ^{2}}  \label{e108}
\end{equation}%
Substituting Eq.~(\ref{a112}) into Eq.~(\ref{a115}) we have to calculate the
sum over the eigenstates $S_{r,r^{\prime }}^{K}$ and thus find the solutions
for $a$ and $b$ from Eq.~(\ref{e107}) using the eigenvalues $E^{K}$, Eq.~(%
\ref{e108}). The function $Z\left[ \phi \right] $ can be written as%
\begin{eqnarray}
&&Z\left[ \phi \right]=Z_{0}\exp \Big[-\frac{\beta }{2}\sum_{r,\sigma
}\int_{0}^{1}\phi _{r\sigma }\left( \phi _{r_{1}\sigma }-\phi
_{r_{1}^{\prime }\sigma }\right) n_{r_{1},r_{1}^{\prime }}  \notag \\
&&\times \sum_{K}\frac{\left( S_{1;r,r^{\prime }}^{K\ast
}S_{2;r_{1},r_{1}^{\prime }}^{K}+S_{2;r,r^{\prime }}^{K\ast
}S_{1;r_{1},r_{1}^{\prime }}^{K}\right) }{E^{K}}udu\Big]  \label{e109a}
\end{eqnarray}%
The subscripts $1$ and $2$ in Eq.~(\ref{e109a}) correspond to two different
components of the vector $S_{r,r^{\prime }}^{K}$, Eq.~(\ref{e105}).

One can easily obtain from Eqs.~(\ref{e107}, \ref{e108}) that the product $%
ab $ entering Eq.~(\ref{e109a}) takes the form%
\begin{equation}
ab=\pm \frac{1}{2}\frac{\left( \lambda ^{k}-\lambda ^{k^{\prime }}\right) }{%
\sqrt{\left( \lambda ^{k}-\lambda ^{k^{\prime }}\right) ^{2}+\gamma ^{2}}}
\label{e110}
\end{equation}%
and we come to the following form of the function $Z\left[ \phi \right] $%
\begin{eqnarray}
&&Z\left[ \phi \right] =Z_{0}\exp \Big[-\beta \sum_{r,r_{1},r_{1}^{\prime
},\sigma ,k,k^{\prime }}\phi _{r\sigma }\left( \phi _{r_{1}\sigma }-\phi
_{r_{1}^{\prime }\sigma }\right) n_{r_{1},r_{1}^{\prime }}  \notag \\
&&\times w_{r}^{k}w_{r}^{k^{\prime }}w_{r_{1}}^{k}w_{r_{1}^{\prime
}}^{k^{\prime }}\frac{\lambda ^{k}-\lambda ^{k^{\prime }}}{\left( \lambda
^{k}-\lambda ^{k^{\prime }}\right) ^{2}+\gamma ^{2}}udu\Big]  \label{e111a}
\end{eqnarray}

We make a further transformation of Eq.~(\ref{e111a}) using Eq.~(\ref{a22a})
and a corresponding equation written in the absence of the HS field $\phi
_{r\sigma }$. Subtracting these equations from each other we obtain%
\begin{eqnarray}
&-&u\sigma \left( \phi _{r\sigma }-\phi _{r^{\prime }\sigma }\right)
n_{r,r^{\prime }}=\Big(\hat{\varepsilon}_{r}-u\sigma \phi _{r}  \notag \\
&-&\hat{\varepsilon}_{r^{\prime }}+u\sigma \phi _{r^{\prime }}\Big)\Big(%
G_{r,r^{\prime };\sigma }^{\left( u\phi \right) }\left( \tau ,\tau +0\right)
-G_{r,r^{\prime }}^{\left( 0\right) }\left( \tau ,\tau +0\right) \Big)
\notag \\
&&  \label{e112}
\end{eqnarray}%
Substituting Eq.~(\ref{e112}) into Eq.~(\ref{e111a}) we reduce the function $%
Z\left[ \phi \right] $ to the following form
\begin{eqnarray}
&&Z\left[ \phi \right] =Z_{0}\exp \Big[\sum_{r,r_{1},r_{1}^{\prime },\sigma
,k,k^{\prime }}\int_{0}^{1}\int_{0}^{\beta }dud\tau \sigma \phi _{r\sigma
}w_{r}^{k}w_{r}^{k^{\prime }}w_{r_{1}}^{k}w_{r_{1}^{\prime }}^{k^{\prime }}
\notag \\
&&\times \frac{\left( \lambda ^{k}-\lambda ^{k^{\prime }}\right) ^{2}u}{%
\left( \lambda ^{k}-\lambda ^{k^{\prime }}\right) ^{2}+\gamma ^{2}}\Big(%
G_{r_{1},r_{1}^{\prime }}^{\left( u\phi \right) }\left( \tau ,\tau +0\right)
-G_{r_{1},r_{1}^{\prime }}^{\left( 0\right) }\left( \tau ,\tau +0\right) %
\Big)\Big]  \notag \\
&&  \label{e114}
\end{eqnarray}%
In the limit $\gamma \rightarrow 0$ the state with $k=k^{\prime }$ drops out
from the sum and we obtain finally using the orthogonality of the functions $%
w_{r}^{k}$%
\begin{eqnarray}
&&Z\left[ \phi \right] =Z_{0}\exp \Big[\int_{0}^{\beta }\int_{0}^{1}\Big[%
\sum_{r,\sigma }\sigma \phi _{r\sigma }\Big(G_{r,r;\sigma }^{\left( u\phi
\right) }\left( \tau ,\tau +0\right)  \notag \\
&&-G_{r,r}^{\left( 0\right) }\left( \tau ,\tau +0\right) \Big)%
-\sum_{r,r_{1},r_{1}^{\prime },\sigma ,k}\sigma \phi _{r\sigma
}w_{r}^{k}w_{r}^{k}w_{r_{1}}^{k}w_{r_{1}^{\prime }}^{k}  \notag \\
&&\Big(G_{r_{1},r_{1}^{\prime };\sigma }^{\left( u\phi \right) }\left( \tau
,\tau +0\right) -G_{r_{1},r_{1}^{\prime }}^{\left( 0\right) }\left( \tau
,\tau +0\right) \Big)\Big]\Big]d\tau du  \label{e115}
\end{eqnarray}

We see that, containing the second sum, Eq.~(\ref{e115}) is somewhat
different from Eq.~(\ref{a19}). At this point we should recall that the
system we consider contains a bath making the number of states large. The
normalized eigenfunctions $w_{r}^{k}$ are proportional to $\left(
N_{d}^{total}\right) ^{-1/2}$. Therefore, the second sum in the exponent in
Eq.~(\ref{e115}) is much smaller than the first term and can be neglected
provided the total number of the sites in the system $N_{d}^{total}$ is
large. This example of the static HS field demonstrates very well how the
regularization works.

\section{Splitting times in the equation for $A_{r,r^{\prime }}\left(
z\right) $.}

\label{equaltimes}

In this appendix we demonstrate that keeping in Eqs.~(\ref{c23a}-\ref{c241})
the parameter $\delta $ finite one can obtain imaginary part in the
eigenvalues $\tilde{E}^{K}$ of the operator $\mathcal{\tilde{H}}%
_{r,r^{\prime }}^{\left( +\right) }\left( \tau \right) $, Eq.~(\ref{c241}).
The finite $\delta $ appears because the times $\tau $ and $\tau ^{\prime }$
entering the Green function $G_{r,r^{\prime }}^{\left( u\phi \right) }\left(
\tau ,\tau ^{\prime }\right) $ should be slightly different, $\tau ^{\prime
}=\tau +\delta $. Although the value of $\delta $ can be infinitesimally
small, the imaginary part of $\tilde{E}^{K}$ can be very important in
situations when the eigenvalue $\tilde{E}^{K}$ is close to zero. The results
derived in this appendix are an extension of those obtained in appendix \ref%
{solution} to the case of finite $\delta $.

Assuming that $\delta $ is small we expand the field $\tilde{\phi}%
_{r^{\prime }}\left( \tau +\delta \right) $ entering Eqs.~(\ref{c23a}-\ref%
{c241}) in this parameter and represent the eigenvalue problem of the
operator $\mathcal{\tilde{H}}_{r,r^{\prime }}^{\left( +\right) }\left( \tau
\right) $, Eq.~(\ref{c241}), in the form%
\begin{equation}
\mathcal{\tilde{H}}_{r,r^{\prime }}^{\left( +\right) }\left( \tau \right)
S_{r,r^{\prime }}\left( z\right) =\tilde{E}S_{r,r^{\prime }}\left( z\right) ,
\label{f1}
\end{equation}%
\begin{equation*}
\mathcal{\tilde{H}}_{r,r^{\prime }}^{\left( +\right) }\left( \tau \right) =%
\mathcal{H}_{r,r^{\prime }}\left( z\right) +\hat{\delta}_{r,r^{\prime
}}\left( z\right) ,
\end{equation*}%
where the hermitian operator $\mathcal{H}_{r,r^{\prime }}\left( z\right) $
is defined in Eq.~(\ref{a103}) and the matrix $\hat{\delta}_{r,r^{\prime
}}\left( z\right) $ can be written as%
\begin{equation}
\hat{\delta}_{r,r^{\prime }}\left( z\right) =\Big(%
\begin{array}{cc}
0 & \sigma u\delta \frac{\partial \phi _{r^{\prime }}\left( \tau \right) }{%
\partial \tau } \\
-\sigma u\delta \frac{\partial \phi _{r}\left( \tau \right) }{\partial \tau }
& 0%
\end{array}%
\Big)  \label{f2}
\end{equation}%
Eq.~(\ref{f1}) is a direct generalization of Eq.~(\ref{a108}) to the case of
the split times. We assume that the eigenvalue $\tilde{E}$ is close to zero.
The regularization with the finite $\gamma $ leads to level repulsion. All
eigenvalues $E^{K}$ calculated for $\delta =0$ remain finite and real.

Considering finite $\delta $ drastically changes the situation even if $%
\delta $ is infinitesimally small. This is a consequence of the fact that
the matrix $\hat{\delta}_{r,r^{\prime }}\left( z\right) $ is not hermitian
and therefore the operator $\mathcal{\tilde{H}}_{r,r^{\prime }}^{\left(
+\right) }\left( \tau \right) $ acting on $S_{r,r^{\prime }}\left( z\right) $
in the L.H.S. of Eq.~(\ref{f1}) is not hermitian, too. This means that the
eigenvalues $\tilde{E}$ can now be complex.

In order to calculate the eigenvalue $\tilde{E}$ we use, as in appendix \ref%
{solution} the degenerate perturbation theory considering the matrices $\hat{%
\delta}_{r,r^{\prime }}\left( z\right) $ and $\gamma \Lambda $ as a
perturbation.

We represent the solution of Eq.~(\ref{f1}) in the form of Eq.~(\ref{e100}, %
\ref{e101}) and multiply both sides of Eq.~(\ref{f1}) by $S_{r,r^{\prime
}}^{a\ast }\left( \tau \right) $ and then by $S_{r,r^{\prime }}^{b\ast
}\left( \tau \right) ,$summing after that over $r,r^{\prime }$and
integrating over $\tau $. As a result, we obtain equations for the
coefficients $c_{1}$ and $c_{2}$%
\begin{eqnarray}
\left( \tilde{E}-\gamma \right) x_{1}c_{1}-c_{2}\left( \lambda +\tilde{%
\lambda}_{1}\delta \right) &=&0  \label{f3} \\
-\left( \lambda ^{\ast }-\tilde{\lambda}_{2}\delta \right) c_{1}+\left(
\tilde{E}+\gamma \right) x_{2}c_{2} &=&0  \notag
\end{eqnarray}%
with positive numbers $x_{1}$ and $x_{2}$ given by Eq.~(\ref{e103}). The
complex parameters $\tilde{\lambda}_{1}$ and $\tilde{\lambda}_{2}$ have the
form
\begin{eqnarray}
\tilde{\lambda}_{1} &=&\sigma u\sum_{r,r^{\prime }}\int_{0}^{\beta }\bar{v}%
_{r,r^{\prime }}\left( \tau \right) \frac{\partial \phi _{r^{\prime }}\left(
\tau \right) }{\partial \tau }v_{r,r^{\prime }}\left( \tau \right) d\tau ,
\label{f4} \\
\tilde{\lambda}_{2} &=&\sigma u\sum_{r,r^{\prime }}\int_{0}^{\beta }\bar{v}%
_{r,r^{\prime }}^{\ast }\left( \tau \right) \frac{\partial \phi _{r}\left(
\tau \right) }{\partial \tau }v_{r,r^{\prime }}^{\ast }\left( \tau \right)
d\tau  \notag
\end{eqnarray}%
where the functions $v_{r,r^{\prime }}\left( \tau \right) $ and $\bar{v}%
_{r,r^{\prime }}\left( \tau \right) $ and the complex eigenvalues $\lambda $
are introduced in Eqs.~(\ref{a106}, \ref{a108}).

In the limit $\left\vert \lambda \right\vert \rightarrow 0$ the functions $%
v_{r,r^{\prime }}\left( \tau \right) $ and $\bar{v}_{r,r^{\prime }}\left(
\tau \right) $ are related to each other as
\begin{equation}
v_{r,r^{\prime }}\left( \tau \right) \simeq \bar{v}_{r^{\prime },r}\left(
\tau \right)  \label{f5}
\end{equation}%
and we obtain
\begin{equation}
\tilde{\lambda}_{2}^{\ast }\simeq \tilde{\lambda}_{1}=\tilde{\lambda}
\label{f6}
\end{equation}%
Substituting Eq.~(\ref{f6}) into Eqs.~(\ref{f3}) and solving the latter we
obtain the eigenvalue $\tilde{E}$%
\begin{equation}
\tilde{E}=\pm \sqrt{\gamma ^{2}+\frac{\left\vert \lambda \right\vert
^{2}+2\delta iIm\left( \tilde{\lambda}\lambda ^{\ast }\right) }{x_{1}x_{2}}}
\label{f7}
\end{equation}

Eq.~(\ref{f7}) demonstrates that the value of $\tilde{E}$ near the point $%
\lambda =0$ strongly depends on the ratio $\gamma /\delta $. In the limit $%
\left\vert \lambda \right\vert \gg \delta \gg \gamma $ we obtain
\begin{equation}
\tilde{E}=\pm \left[ \left\vert \lambda \right\vert +i\delta \frac{Im\left(
\tilde{\lambda}\lambda ^{\ast }\right) }{\left\vert \lambda \right\vert }%
\right] \frac{1}{\left( x_{1}x_{2}\right) ^{1/2}}  \label{f8}
\end{equation}%
We see from Eqs.~(\ref{f7}, \ref{f8}) that a finite value of $\delta $ that
determines the splitting of the times makes the eigenvalues $\tilde{E}$
complex. The imaginary part, being infinitesimally small is very important
for small $\lambda $ generating a finite imaginary part of the integral over
$u$ in Eq.~(\ref{a115}). Working in the limit of $\gamma \ll \delta $ one
would inevitably encounter the sign problem. In contrast, putting $\delta =0$
and working with small but finite $\gamma $ gives strictly positive $Z\left[
\phi \right] $ and one avoids the sign problem.

\end{document}